\documentclass[twocolumn, prx, reprint, amsfonts, amsmath, amssymb, superscriptaddress, longbibliography, nofootinbib]{revtex4-2}
\usepackage{graphicx}
\usepackage{footnote}
\usepackage{booktabs}
\usepackage{float}
\usepackage{braket}
\usepackage{bbold}
\usepackage{tikz-cd}
\usepackage{pgfplots}
\usepackage{bm}
\usepackage{enumitem}

\usepackage{amsthm}
\newtheorem*{JW*}{Jordan Wigner-Flux attachment equivalence}

\usepackage[graphicx]{realboxes}

\usepackage{tikz}
\usepackage{multirow}
\def\checkmark{\tikz\fill[scale=0.4](0,.35) -- (.25,0) -- (1,.7) -- (.25,.15) -- cycle;}
\usepackage{tikz,tikz-3dplot}
\tdplotsetmaincoords{80}{45}
\tdplotsetrotatedcoords{-90}{180}{-90}
\usetikzlibrary{decorations.markings}
\def\NrLines{7}

\pgfmathsetmacro{\UnitSegment}{1/(\NrLines+1)}
\usepackage[title]{appendix}

\usepackage{extarrows} 
\usepackage[colorlinks,linkcolor=blue,anchorcolor=red,citecolor=green]{hyperref}
\usepackage{cancel}
\tikzset{labl/.style={anchor=south, rotate=90, inner sep=.5mm}}
\usepackage[]{mdframed}
\usepackage[caption=false]{subfig}
\pgfplotsset{compat=1.18} 
\usepackage[compat=1.1.0]{tikz-feynman}

\begin{document}


\title{Jordan-Wigner Composite-fermion liquids in 2D quantum spin-ice
}

\author{Leonardo Goller}
\affiliation{SISSA, Via Bonomea 265, 34136 Trieste, Italy}

\author{Inti Sodemann Villadiego}
\affiliation{Institut für Theoretische Physik, Universität Leipzig, Brüderstraße 16, 04103, Leipzig, Germany}
\affiliation{Max-Planck-Institut für Physik komplexer Systeme, Nöthnitzer Straße 38, 01187 Dresden, Germany}
\date{\today}

\begin{abstract}
The Jordan-Wigner map in 2D is as an exact lattice regularization of the $2 \pi$-flux attachment to a hard-core boson (or spin-$1/2$) leading to a composite-fermion particle. When the spin-$1/2$ model obeys ice rules this map preserves locality, namely, local Rohkshar-Kivelson models of spins are mapped onto local models of Jordan-Wigner/composite-fermions. Using this composite-fermion dual representation of RK models, we construct spin-liquid states by projecting Slater determinants onto the subspaces of the ice rules. Interestingly, we find that these composite-fermions behave as ``dipolar" partons for which the projective implementations of symmetries are very different from standard ``point-like" partons. We construct interesting examples of these composite-fermi liquid states that respect all microscopic symmetries of the RK model. In the six-vertex subspace, we constructed a time-reversal and particle-hole-invariant state featuring two massless Dirac nodes, which is a composite-fermion counterpart to the classic $\pi$-flux state of Abrikosov-Schwinger fermions in the square lattice. This state is a good ground state candidate for a modified RK-like Hamiltonian of quantum spin-ice. In the dimer subspace, we construct a state fearturing a composite Fermi surface but with nesting instabilities towards ordered phases such as the columnar state. We have also analyzed the low energy emergent gauge structure. If one ignores confinement, the system would feature a $U(1) \times U(1)$ low energy gauge structure with two associated gapless photon modes, but with the composite fermion carrying gauge charge only for one photon and behaving as a gauge neutral dipole under the other. These states are examples of pseudo-scalar $U(1)$ spin liquids where mirror and time-reversal symmetries act as composite-fermion particle-hole conjugations, and the emergent magnetic fields are even under such time-reversal or lattice mirror symmetries.
\end{abstract}




\maketitle

\tableofcontents

\section*{Introduction}

The appearance of fermionic particles in systems whose microscopic building blocks are spins or bosons is a remarkable example of the emergence of non-local excitations in quantum states of matter. An exact and powerful map that allows to understand this phenomenon in one-dimension is the Jordan-Wigner transformation, which provides a re-writing of one-dimensional spin-$\frac{1}{2}$ models in terms of fermions. In two-dimensions the Jordan-Wigner transformation is closely related to another celebrated statistical transmutation procedure known as flux attachment \cite{Fra89, Sem88, ES92, Azz93, PRB15,derzhko2001jordan}. More specifically, as we will review in detail in Sec. \ref{Thermodynamics}, a standard Jordan-Wigner Fermion constructed by ordering spin-$\frac{1}{2}$ operators in a 2D lattice is exactly equivalent to a hard-core boson carrying a fictitious solenoid of $2\pi$-flux. In this sense, the 2D Jordan-Wigner Fermion is an exact lattice regularized version of the composite Fermion particle that is commonly used to understand certain quantum Hall states emerging from  microscopic bosons, such as those making the bosonic composite Fermi liquid state at filling $\nu=1$ \cite{Read, Pas, Dong}.

This 2D Jordan-Wigner/flux-attachment has been exploited in several studies of non-trivial quantum disordered states (``spin liquids'') and their competition with traditional ordered phases  \cite{Azz93, Wang91, Wang912, Wang92, LGF94, Yan94, Yan97, YZC01, KSF14, SGK17}. 
One of the central challenges with the models investigated in these previous studies is that the Jordan-Wigner/flux-attachment map in 2D does not preserve space locality, in the sense that not all local spin-$\frac{1}{2}$ operators appearing in the Hamiltonian are mapped onto local fermionic operators. This sharply contrasts with the situation in 1D, where simply imposing a global parity symmetry guarantees that local spin Hamiltonians map onto local fermionic Hamiltonians. In most of the 2D studies this difficulty is dealt with in a non-systematic manner by adding background magnetic fields that account for the relation between particle density and flux in an average fashion, similarly to how it is  done in mean-field treatments of composite fermions in quantum Hall states \cite{halperin1993theory, Fra, Wen, Jai07}.

Recently, however, it has been emphasized that another kind of exact Jordan-Wigner-like maps in 2D are possible which in some sense preserve space locality \cite{Chen, Pozo, Peng, PI}. This is achieved by imposing local conservation of certain $\mathbb{Z}_2$-valued operators and thus endowing the Hilbert space of spin-$\frac{1}{2}$ with a $\mathbb{Z}_2$ lattice gauge theory structure. The gauge invariant spin operators (namely those commuting with the $\mathbb{Z}_2$ conservation laws) can then be mapped exactly into bilinears of fermion operators. The single fermion creation operator remains non-local and can be explicitly constructed as a Jordan-Wigner-like string operator in 2D \cite{Chen, Pozo, Peng, PI}. This construction realizes an exact lattice regularization of a different kind of flux attachment, namely the one associated with a mutual Chern-Simons theory comprised of two U(1) gauge fields and the following $K$-matrix:

\begin{equation*}
       K =   \begin{pmatrix}
       0 & 2\\
       2 & 0
    \end{pmatrix}, 
   \end{equation*}

which is the Chern-Simons description of the topological order associated with Kitaev's Toric code model \cite{Chen, Pozo, Peng, PI}, and the Jordan-Wigner fermions are the $\epsilon$ particles, while the operator associated with the local $\mathbb{Z}_2$  conservation law measures the parity of one of the other self-bosonic anyons, e.g. the $e$ or $m$ particle. For related constructions see also \cite{KITAEV20062, PhysRevB.76.193101, Chen_2008,  PhysRevLett.104.020402, Cobanera_2011,  Nussinov_2012, PhysRevB.107.045114}.

Motivated by these precedents, our current work investigates systems with a different kind of local conservation law that allows to preserve the locality of the usual Jordan-Wigner map in 2D associated with attaching $2\pi$-flux to a hard-core boson. Our local conservation laws will be associated with a two dimensional spin ``ice rule" with a correspondingly conserved local operator that generates a U(1) gauge group, and the models of interest will be the classic 2D Rokshar-Kivelson-like (RK) Hamiltonians \cite{RK88, Moessner_2004, PhysRevLett.108.247210, PhysRevLett.113.027204, Sode, PhysRevB.69.220403, PhysRevB.65.024504,  Banerjee_2013,tschirsich2019phase,ran2023fully,stornati2023crystalline,banerjee2021quantum,biswas2022scars}\footnote{See Ref.\cite{PhysRevB.106.L041115} for a recently proposed interesting variant.}. Similar to the situation in 1D, these models will remain local under Jordan-Wigner maps, however the price we will have to pay for this is that the resulting fermionic model will be necessarily interacting and endowed with local conservation laws. Despite this, the advantage of re-writing the RK models in terms fermionic variables is that they are much more flexible degrees of freedom to construct non-trivial quantum disordered states than the original spin degrees of freedom. For example, simple fermionic Slater determinant states would serve already as a mean-field approximation to describe quantum spin liquid states. There is however a crucial caveat to this mean field approach, which is that generically free fermion Slater determinant states will not obey the local spin-ice rules. In other words, the naive free fermion states would break the local gauge invariance and violate Elitzur's theorem  \cite{PhysRevD.12.3978}. To remedy this, we will project the free-fermion states onto the susbpace of the Hilbert space satisfying the ice rules, in an analogous fashion to how the Gutzwiller projection is employed in parton constructions of spin liquid states \cite{WenA02}. Despite the similarity of spirit, we have encountered that these states differ in crucial aspects from the standard parton constructions, such as the Abrikosov-Schwinger fermions \cite{PhysRevB.37.580, PhysRevB.37.3774,  PhysRevLett.54.966}. 


One of the key distinctions, is that the Abrikosov-Schwinger fermion of standard parton constructions behaves as a point-like object under the parton gauge transformations, whereas the Jordan-Wigner fermion behaves as a dipole-like object under the local $U(1)$ gauge transformations of the RK models. This is because the Abrikosov-Schwinger fermion operator at a given lattice site only transforms non-trivially under the parton gauge transformations acting on its site, but it transforms trivially under gauge transformations of different sites. In other words, the creation of a single Abrikosov-Schwinger fermion would violate the constraint defining the physical Hilbert space only at a single lattice site, and in this sense it is point-like. The Jordan-Wigner fermion violates the spin-ice rule of two neighboring vertices, in such a way so as to create a dipole under of the $U(1)$ Gauss' law of the RK type Hamiltonians. 

Because of the above, we will refer to our construction of spin-ice projected Slater determinants of the Jordan-Wigner fermions as an ``extended parton construction". These differences between extended vs point-like partons lead to crucial physical differences between for the states constructed from them. Some of these differences will manifest as unconventional implementations of lattice symmetries. For example, we will show that $\pi/2$ rotation symmetries\footnote{These are $\pi/2$ rotations that will be denoted by $R_{\frac{\pi}{2}}$.} do not admit an ordinary fermion implementation for the Jordan-Wigner fermions, but need to be dressed by a unitary transformation that is not part of the $U(1)$ lattice gauge group. 

But the most remarkable difference we have found between the extended partons and the point-like partons, is the nature of the gauge fluctuations around their mean field Slater determinant states. According to the principles of the projective symmetry group constructions for ordinary point-like partons, like Abrikosov-Schwinger fermions, a Slater determinant state which conserves the global particle number fermions will describe a $U(1)$ spin liquid state, whenever it is stable against gauge confinement. The deconfined state has therefore an emergent $U(1)$ photon gauge field, and the fermionic parton will carry charge under this field. As we will see, however, the a Slater determinant of the Jordan-Wigner extended partons will feature a $U(1) \times U(1)$ gauge structure, namely two distinct gapless photons. The Jordan-Wigner fermion will carry a net gauge charge under one of these two photons, but it will be gauge neutral under the other photon, for which it will only carry a gauge dipole. 

Moreover despite the fact that the Jordan-Wigner fermion is a composite fermion that can be viewed as a boson attached to $2\pi$ flux, we will see that the expected action of the two $U(1) \times U(1)$ gauge fields is an ordinary Maxwell-like action with no Chern-Simons terms, as a result of the enforcement of time reversal and microscopic mirror symmetries of the models in question. This is interesting because it demonstrates an explicit instance of the existence of a composite Fermi liquid-like states arising from flux attachment, for which the emergent gauge structure does not feature an explicit Chern-Simons term. This feature is somewhat reminiscent of the Dirac composite fermion theories of the half-filled Landau level \cite{PhysRevX.5.031027, DAMSON,  Geraedts_2016}, and of some of the more sophisticated composite Fermi liquid theories of bosons at $\nu=1$ \cite{Read, Pas, Dong}, which contrast from the more traditional explicit forms of flux attachment in the HLR description of composite fermions \cite{PhysRevB.47.7312}.

We will also construct interesting explicit examples of mean-field spin-liquid states for RK-like 2D quantum spin-ice Hamiltonians. As we will see, the sectors defined by different values of the spin-ice rules will correspond to different fillings of the Jordan-Wigner/composite-fermion bands. For example the sector with zero spin, which maps to the quantum six-vertex model \cite{Sode, PhysRevB.69.220403}, will correspond to half-filling of a two-band model. We will construct an explicit mean-field state that satisfies all the space symmetries of the lattice, time-reversal and particle-hole transformations, and that features two gapless linearly dispersing Dirac fermion modes at low energy, which can be pictures as a composite fermion counterpart to the classic $\pi$-flux state of Abrikosov-Schwinger fermions \cite{PhysRevB.37.3774,PhysRevB.37.3664}. Ignoring compactification-driven instabilities (see below however), this would be therefore a specific kind of Dirac composite Fermi liquid state, with an emergent low energy $U(1)\times U(1)$ gauge structure with two massless photons, with the fermions carrying charge under only one $U(1)$ photon and neutral under the other $U(1)$ photon. 

Field theories of masless Dirac fermions coupled to a single $U(1)$ compact gauge field are known to remain deconfined at low energies in the limit of large-$N$ number of Dirac fermions flavors \cite{Bor02, HSFLNW04} and to also avoid spontaneous chiral symmetry breaking \cite{PhysRevLett.60.2575,  PhysRevLett.62.3024}. However, understanding the ultimate infrared fate of these field theories at finite $N$ has remained challenging \cite{PhysRevD.105.085008,  PhysRevD.100.054514, Chester_2016}. In our case we have $N=2$ Dirac fermions and two photons (with the fermions carrying charge under only one of these photons).
We will not address systematically the impact of gauge compactification, but we expect that at least the photon under which the fermions are neutral will undergo Polyakov-like confinement \cite{POLYAKOV197582,  POLYAKOV1977429}, which will remove it from low energies, leaving possibly only two masless Dirac fermions coupled to a single $U(1)$ photon at low energies, analogously to $\text{QED}_3$ with $N=2$ (to the extend that this theory avoids confinement and other instabilities at low energies).

On the other hand, we will see that for the subspace of spin-ice that maps onto the quantum dimer model \cite{Sode}, the band structure of the Jordan-Wigner/composite-fermions will be at quarter-filling leading to the appearance of a composite Fermi-surface state. Moreover, the state arising when the composite fermions only hop between nearest neighbor sites will display a perfectly nested Fermi surface. This nesting is accidental in the sense that it can be removed by adding symmetry-allowed further-neighbor hopping terms. Nevertheless, such strong tendency for perfect nesting can be viewed as related to the tendency of the quantum dimer model systems to have ordinary gauge confined ground states (such as the resonant plaquette or the columnar phases \cite{PhysRevB.40.5204,  PhysRevB.54.12938, PhysRevB.73.245105, PhysRevLett.100.037201, PhysRevB.90.245143, PhysRevB.98.064302}), which would appear if the Fermi surface is fully gapped via a composite fermion particle-hole pair condensation. This nested state could be therefore useful as a mother state to understand the descending competing broken symmetry states of the quantum dimer model and perhaps help understand the strong tendency towards the columnar phase of the classic RK model, which has been advocated in recent studies to take over the complete phase diagram on the side of the RK point where a unique ground state exists ($v/t < 1$) \cite{PhysRevB.90.245143, PhysRevB.98.064302}.

\begin{table}
    \centering
\caption{Dictionary spin-$\frac{1}{2}$ to hard-core-boson language.}
\label{Traduzione}
\vspace{0.05in}
\begin{tabular}{c|c}
\toprule
\text{Spin-$\frac{1}{2}$} & Boson   \\
\midrule
  $\ket{\uparrow}$ &$\ket{0}$  \\
   $\ket{\downarrow}$ &$\ket{1}$  \\
   $\sigma^+$ &$b$  \\
   $\sigma^-$ &$b^{\dag}$  \\
    $\frac{1-\sigma^z}{2}$ & $n$\\
\bottomrule
\end{tabular}
\end{table}

Our manuscript is organized as follows: Chapter \ref{Thermodynamics} reviews  the one-dimensional Jordan-Wigner transformation and its interpretation as flux-attachment in the 2D square lattice. Chapter \ref{MFCF} applies this construction to 2D quantum spin-ice models, and introduces the general extended parton construction of mean-field states. Then we apply this to the specific cases of Quantum Six Vertex and Quantum Dimer Models, and construct the mean-field states with two Dirac cones 
and a Fermi Surface for each of these models respectively. Chapter \ref{gauge} develops a description of the gauge field fluctuations, and discusses the derivation of the effective low energy theories for these states, demonstrating the appearance of two U(1) gauge fields with two associated gapless photons, with the fermions being charged under only one of the U(1) fields. We close then with a summary and discussion where we also comment on future research directions. 

\section{Equivalence of Jordan-Wigner Transformation and flux attachment in 2D}
\label{equivJWCF}

\label{Thermodynamics}

\begin{figure}
  \centering
   \includegraphics[trim={0cm 4cm 0cm 4cm}, clip, width=0.5\textwidth]{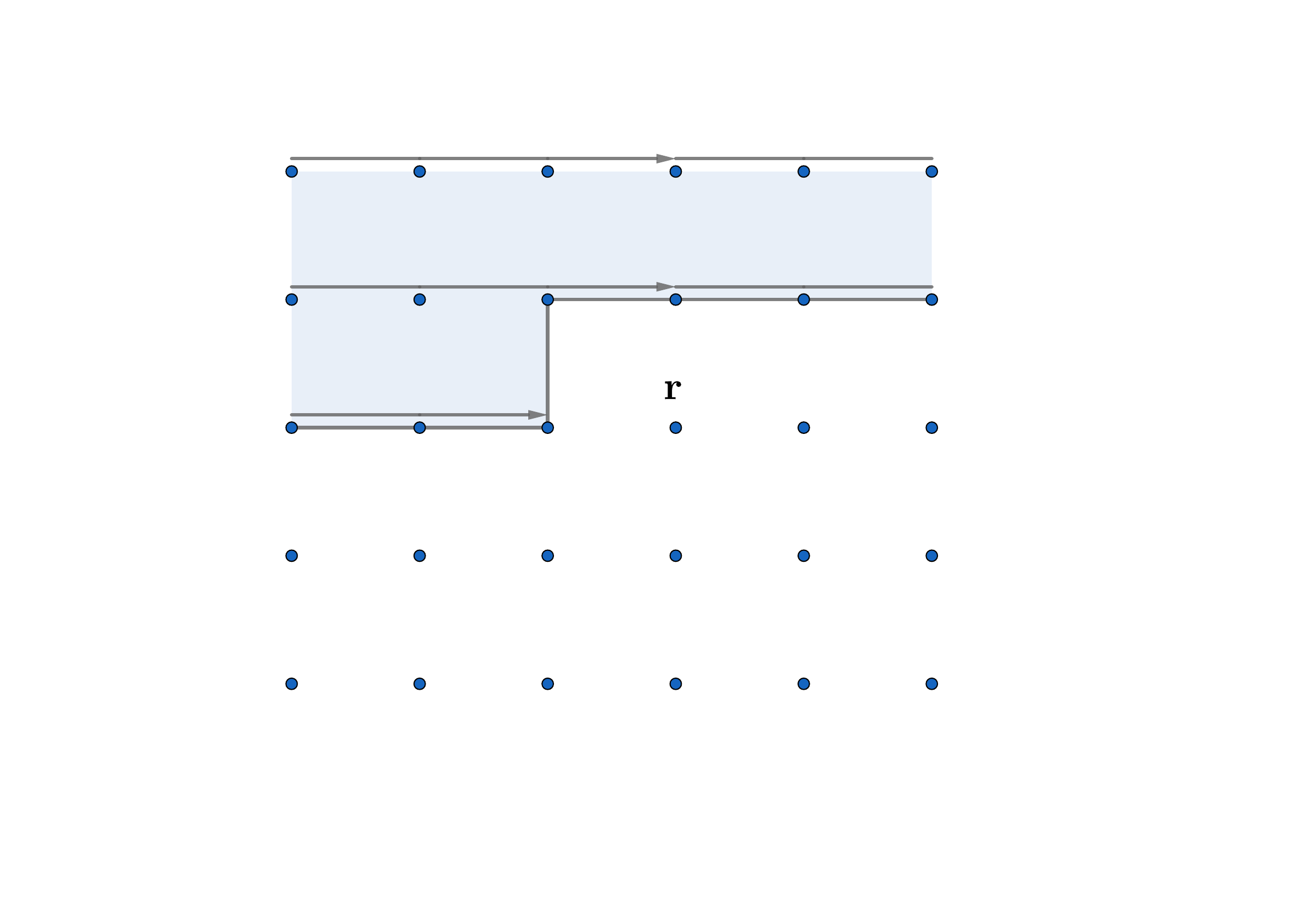}
   \caption{Physical spin-$\frac{1}{2}$ degrees of freedom reside at the blue sites of a 2D square lattice labeled by $\textbf{r}$. The light blue shaded region denotes the membrane operator made from the product of all the $\sigma^z(\textbf{r}')$ spin operators in such region, associated with the Jordan-Wigner fermion creation operator at site $\textbf{r}$ (see Eq.\eqref{basicJW}). The directed arrows illustrate our ``western typing'' convention for ordering the 2D lattice sites.}
   \label{jw2d}
  \end{figure}

Let us consider a two-dimensional square lattice with a spin-$\frac{1}{2}$ degree of freedom residing in each site denoted by ${\bf r}$, as depicted in Fig.\ref{jw2d}. These spin-$\frac{1}{2}$ degrees of freedom can also be viewed as hard-core bosons, according to the convention of Table \ref{Traduzione}. By choosing a convention for the ordering of sites, we can write the standard Jordan-Wigner fermion creation operators as follows (see Fig.\ref{jw2d}):

 \begin{equation}\label{basicJW}
    {f}^{\dag}(\textbf{r}) =  {b}^{\dag}(\textbf{r})   \prod_{1 \leq\textbf{r}' < \textbf{r}} \sigma^z(\textbf{r}').
  \end{equation}



We will order the sites using ``western typing" convention, as depicted in Fig.\ref{jw2d}. Since, $\sigma^z(\textbf{r})=\exp(i \pi n(\textbf{r}))$, it follows that for any pair of sites $\textbf{r},\textbf{r}'$ the boson hopping operators can be written as:

\begin{equation}\label{bdagb}
    {b}^{\dag}(\textbf{r}) {b}(\textbf{r}') =  {f}^{\dag}(\textbf{r})  e^{i \pi \sum_{ {\textbf{r}'} \leq \textbf{r}'' < {\textbf{r}}} {n}(\textbf{r}'')}  {f}(\textbf{r}').  
\end{equation}

 \vspace{0.10in}
  
\begin{figure}
  \centering
   \includegraphics[trim={12cm 8cm 0cm 6cm}, clip, width=0.7\textwidth]{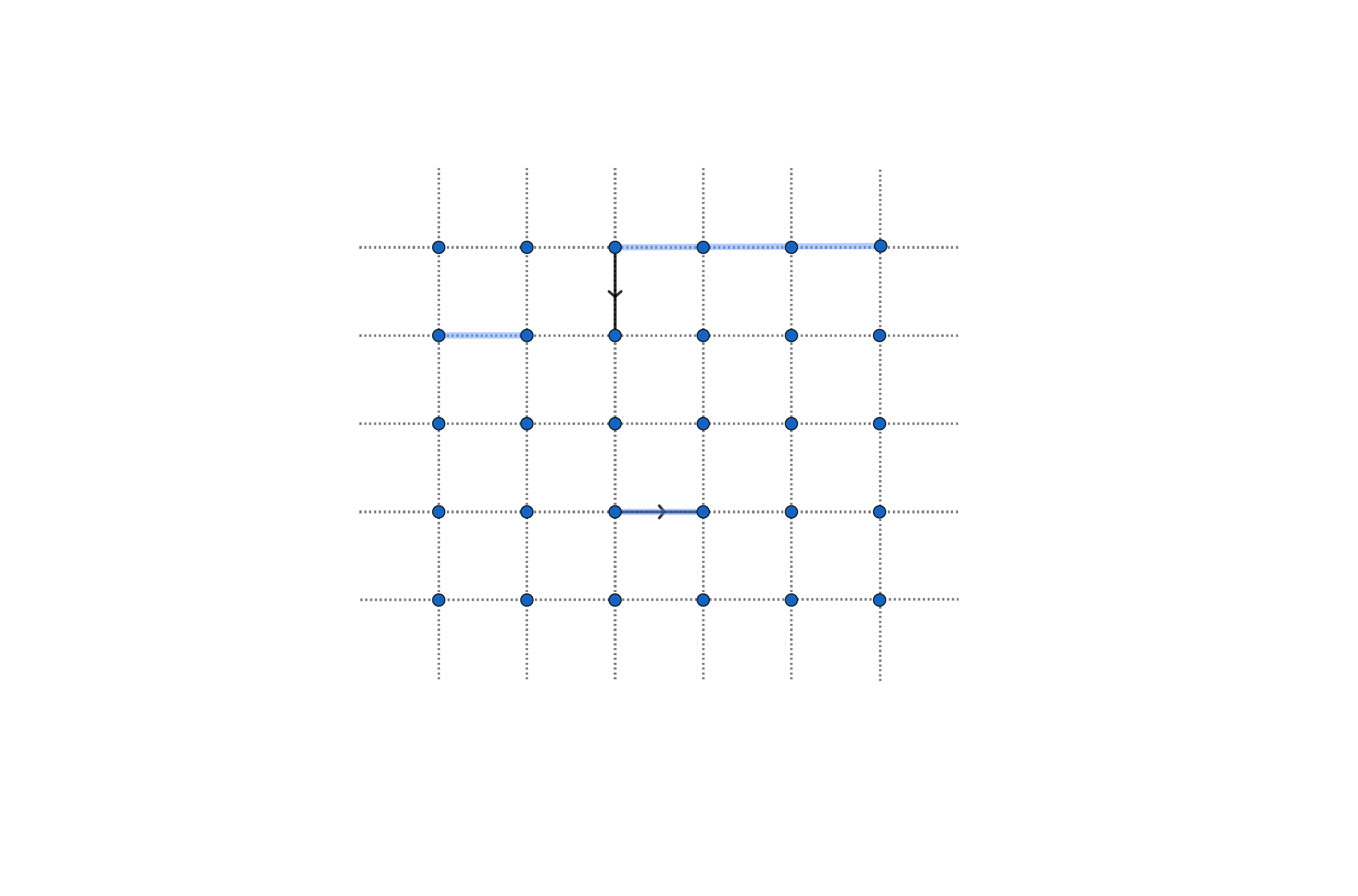}
   \caption{Solid directed arrows represent the local boson hopping operator between sites from site $\textbf{r}'$ towards site $\textbf{r}$ from Eq.\eqref{bdagb}. Blue lines represent the Jordan Wigner strings associated with the fermion representation of these same operators. We see that for our convention (see Fig.\ref{jw2d}), the horizontal boson hoppings remain local in the fermion representation, whereas the vertical hoppings have a non-local fermion representation.}
   \label{verticalhopping}
  \end{figure}

When $\textbf{r},\textbf{r}'$ are nearby, the above operator is clearly local in its physical bosonic representation, however is it is generally non-local in its dual fermion representation as illustrated in Fig.\ref{verticalhopping}. 


\begin{figure} 
   \includegraphics[trim={5cm 9cm 5cm 6cm}, clip, width=0.5\textwidth]{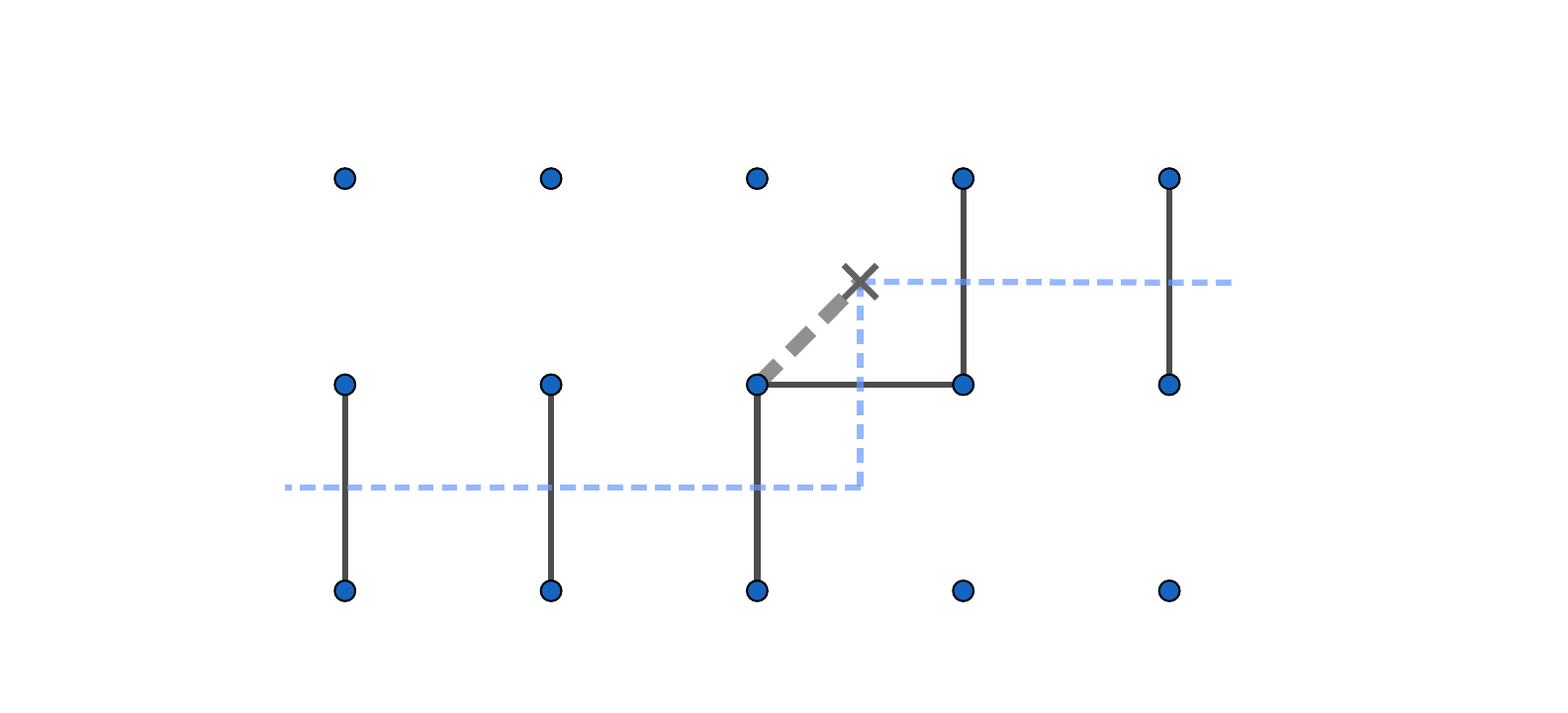}
   \caption{Flux attachment is performed by binding to each boson (located in the sites marked by blue dots) a thin solenoid depicted by the star which is located in the plaquette northeast from the boson site.  This thin solenoid carries $2\pi$ flux, whose vector potential is chosen to be concentrated in the two dotted lines connected to the star. The hopping operators (depicted by solid black lines) that intersect such dotted lines are multiplied by $-1$ when the solenoid is present, namely there is an extra $\pi$ phase for hopping across the dotted lines.}
   \label{2piflux}
  \end{figure}

Let us now demonstrate the equivalence of Eq.\eqref{bdagb} to $2\pi$-flux attachment. Consider spin-less fermions, ${f}^{\dag}(\textbf{r})$ located at the sites $\textbf{r}$ of the square lattice. We attach a thin solenoid to each of these fermions which we view as located in the center of the plaquette that is north-east to the site $\textbf{r}$ (see Fig.\ref{2piflux}). The  solenoid carries a $2 \pi$-flux and we choose a gauge that concentrates its vector potential, $\textbf{A}(\textbf{x})$, into two strings, depicted as dotted lines in Fig.\ref{2piflux}. This gauge is chosen so that the flux attachment exactly matches our specific choice of ``western typing'' ordering convention of the Jordan-Wigner transformation, and different ordering conventions lead to different gauge choices for the flux-attachment (see e.g. \cite{Fra89, Sem88, ES92, Azz93, PRB15,derzhko2001jordan}). Here $\textbf{x}$ can be viewed as a coordinate on the ambient 2D space in which the lattice is embedded. Each one of these strings is chosen so that the line integral of the vector potential across a path that intersects the strings is exactly $\pi$. Therefore, when another fermion hops across a bond that intersects one of these strings, its hopping amplitude will have an extra minus sign, relative to the hopping it has when the string is not present (namely each string acts as a ``branch-cut" that dresses the fermion hopping phase by $\pi$).

Therefore, the above convention fixes the vector potential $\textbf{A}(\textbf{x})$ to be a unique operator which is a function of all the fermion occupations operators, ${n}(\textbf{r})={f}^{\dag}(\textbf{r}){f}(\textbf{r}) $. Therefore, establishing the equivalence of the above flux attachment procedure to the Jordan-Wigner transformation reduces to demonstrating that the following operator identity holds:

\begin{equation}\label{equiv}
    \exp\left( i \pi \sum_{ {\textbf{r}'} \leq \textbf{r}'' < {\textbf{r}}} {n}(\textbf{r}'')\right) = \exp\left( i \int_{\textbf{r}'}^{\textbf{r}} \textbf{A}(\textbf{x}) \cdot d \textbf{x}\right).  
\end{equation}

\begin{figure}
  \centering
   \includegraphics[trim={1cm 7cm 4cm 6cm}, clip, width=0.4\textwidth]{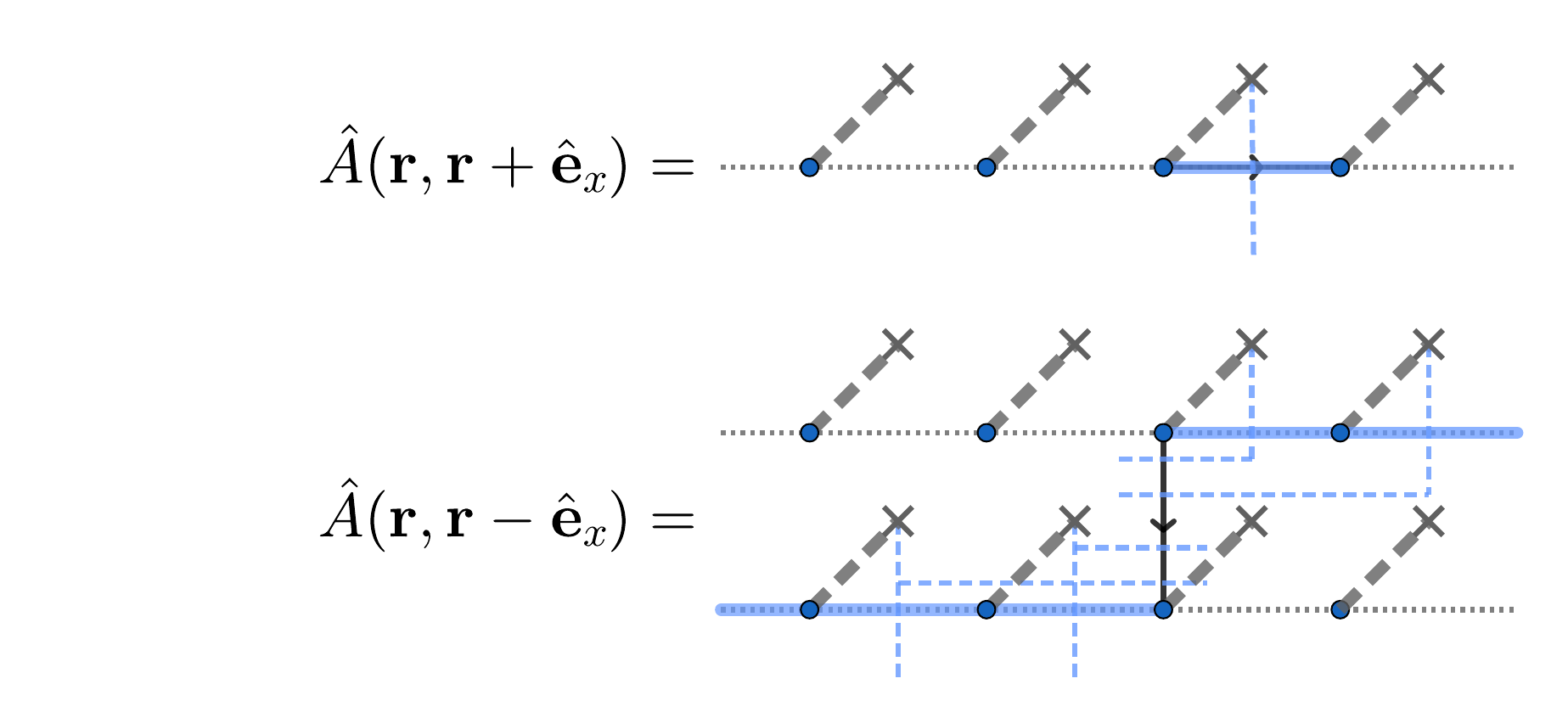}
   \caption{A local boson hopping operator (depicted by the directed black solid line), can be equivalently represented as a fermion hopping operator with its hoppings dressed by the vector potentials that capture the $2 \pi$-flux attachment, according to the rules depicted in Fig.\ref{2piflux} (see Eq.\eqref{clasgau}).}
   \label{gaugefield}
  \end{figure}

 To demonstrate the above relation, let us first consider the line integral of $\textbf{A}(\textbf{x})$ when $\textbf{r},\textbf{r}'$ are nearest neighbor sites. From Fig.\ref{gaugefield}, we can see that the following holds for the horizontal and vertical nearest neighbor hoppings:

\begin{equation}\label{clasgau}
      \begin{aligned}
         &\frac{1}{\pi}\int_{\textbf{r}}^{\textbf{r}+\textbf{e}_x} \textbf{A}(\textbf{x}) \cdot d \textbf{x}     =  {n}(\textbf{r}), \\
          &\frac{1}{\pi}\int_{\textbf{r}}^{\textbf{r}-\textbf{e}_y} \textbf{A}(\textbf{x}) \cdot d \textbf{x}  = \sum_{x \leq x'} {n}(x',y) + \sum_{x' <x} {n}(x',y-1),  \\
      \end{aligned}
  \end{equation}


  where $\textbf{r}=(x,y)$ are the coordinates of the lattice sites measured in units of lattice constant, and the integration path is chosen respectively to be the bonds $\{\textbf{r},\textbf{r} + \textbf{e}_x\}$ and $\{\textbf{r},\textbf{r} - \textbf{e}_y\}$ (see Fig.\ref{gaugefield}). The relations in Eq. \eqref{clasgau} are the same expected from Eq. \eqref{equiv}. 
  

  \vspace{0.10in}
  
\begin{figure}
 \centering
   \includegraphics[trim={9cm 9cm 0cm 6cm}, clip, width=0.65\textwidth]{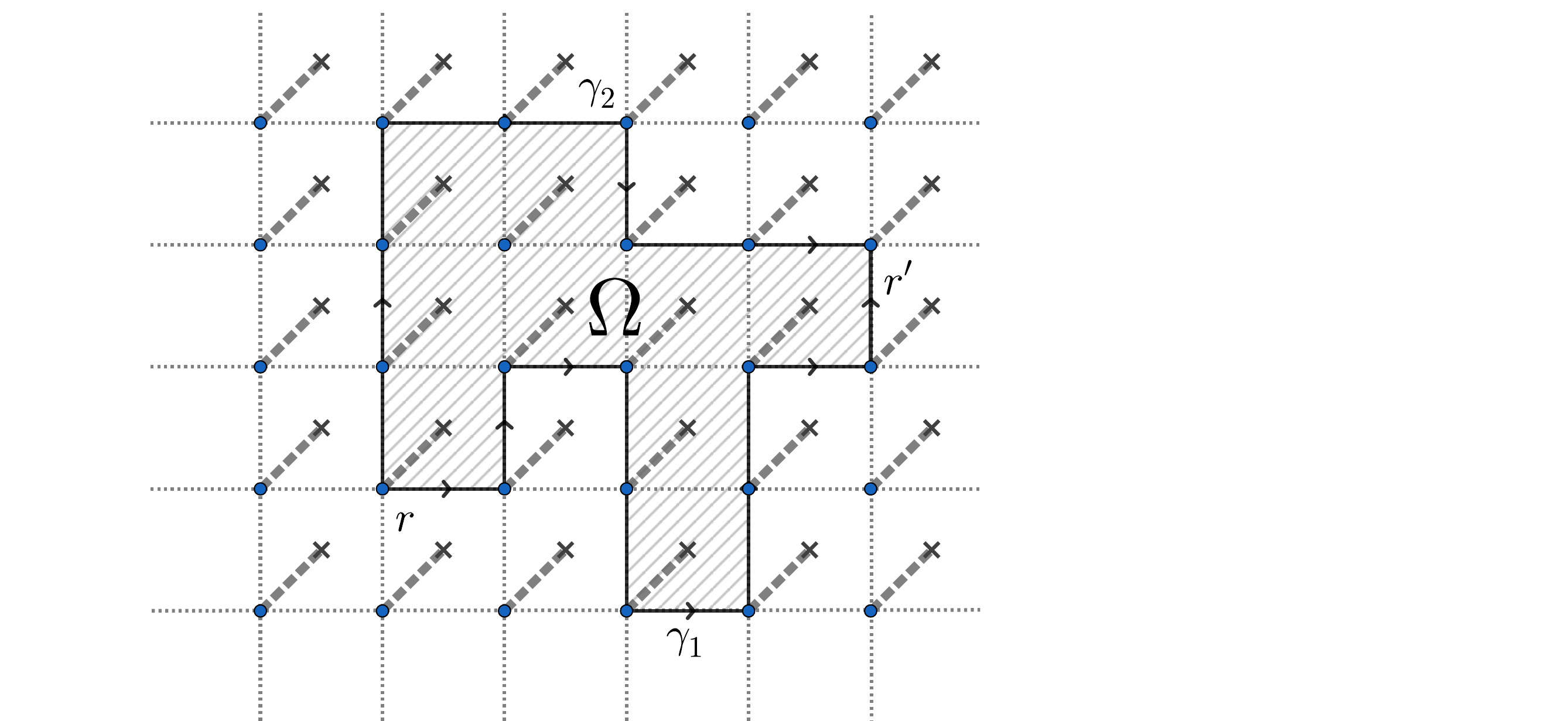}
  \caption{Illustration of the paths used to derive Eqs.\eqref{stokes} and \eqref{mod2pi}.}
  \label{imlem1}
 \end{figure}

 Let us now show that the line integral of $\textbf{A}(\textbf{x})$ in Eq.\eqref{equiv}  is independent of the specific path that connects the points $\textbf{r},\textbf{r}'$, modulo $2\pi$. Let us consider two paths $\gamma_1$ and $\gamma_2$ connecting $\textbf{r},\textbf{r}'$. These two paths define a closed path $\gamma$ which is the boundary of a region $\Omega$ (see Fig.\ref{imlem1}). From Stokes' theorem it follows that:

\begin{equation}
      \oint_{\gamma} \textbf{A}(\textbf{x}) \cdot d \textbf{x} = \iint_{\Omega} (\nabla \times \textbf{A}) (\textbf{x}) \cdot d \sigma = 2\pi \sum_{\textbf{r} \in \Omega}n(\textbf{r}).
      \label{stokes}
  \end{equation}

The sum over $\textbf{r} \in \Omega$ in the above expression is perfomed over those sites $ \textbf{r}$ for which the solenoid is strictly in the interior of $\Omega$ (see Fig.\ref{imlem1}). Therefore since the fermion number operators $n(\textbf{r})$ are integer valued, it follows from Eq. \eqref{stokes}, that:

\begin{equation}
        \int_{\gamma_1} \textbf{A}(\textbf{x}) \cdot d \textbf{x} = \int_{\gamma_2} \textbf{A}(\textbf{x}) \cdot d \textbf{x} \ \ {\rm mod}(2\pi).
        \label{mod2pi}
    \end{equation}

The above equation demonstrates that there is no ambiguity in the line integrals in the right hand side of Eq.\eqref{equiv}.  A detailed derivation that Eq. \eqref{equiv} holds for any pair $\textbf{r},\textbf{r}'$ is presented in Appendix \ref{2DfluxJWequiv}. 

Therefore, we see that there is a precise equivalence between the notion of the statistical transmutation of a hard-core boson and a ``composite fermion" carrying a solenoid of $2 \pi$ flux, and the statistical transmutation of spin-$\frac{1}{2}$ degrees of freedom onto Jordan-Wigner fermions in 2D lattices. The non-locality of the Jordan-Wigner transformation in 2D should not be viewed as ``bug" but rather as a ``feature" that secretly encodes the natural non-locality associated with flux attachment. This equivalence could also be useful to understand the precise lattice versions of transformations discussed within the web of dualities \cite{seiberg2016duality}.

\vspace{0.10in}

\section{Jordan Wigner/Composite fermions as extended partons in 2D quantum spin-ice}
\label{MFCF}
Quantum spin-ice in the 2D square lattice is a classic example of a lattice gauge theory, namely, a model with a set of local conservation laws \cite{Fra, Wen,RK88, Moessner_2004, PhysRevLett.108.247210, PhysRevLett.113.027204, Sode, PhysRevB.69.220403, PhysRevB.65.024504,  Banerjee_2013,tschirsich2019phase,ran2023fully,stornati2023crystalline,banerjee2021quantum}. For different values for these local conservation laws the Hilbert space can be reduced to that of the Quantum Six Vertex Model (Q6VM) or the celebrated Quantum Dimer Model (QDM) introduced by Rokhsar and Kivelson \cite{RK88}. In this chapter, we will develop a dual representation of these models in terms of Jordan-Wigner/composite fermions, and exploit the fact that these models of spins remain local in terms of their dual Jordan-Wigner/composite fermions. We will show that these Jordan-Wigner/composite fermions behave in certain sense like partons, such as Abrikosov-Scwhinger fermions \cite{WenA02, WenB02}, but with crucial qualitative differences arising from the fact that they carry not only lattice gauge charge, but also a lattice gauge dipole moment.



\subsection{2+1D Quantum spin-ice and its Jordan-Wigner Composite Fermion representation}
\label{QSIJW}
\begin{figure}
  \centering
   \includegraphics[trim={0cm 0cm 0cm 0cm}, clip, width=0.5\textwidth]{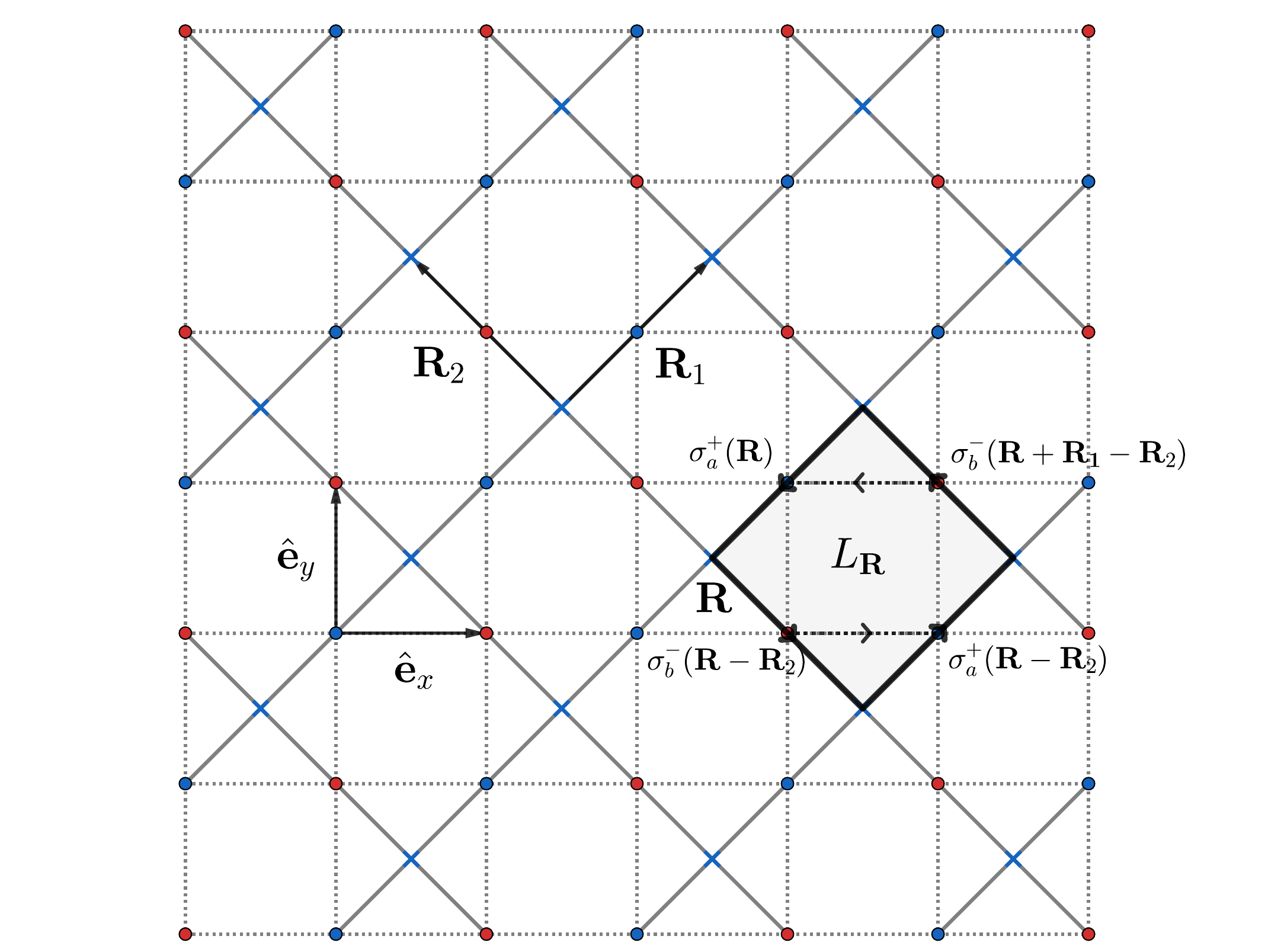}
   \caption{The original square lattice of spins is spanned by vectors $\hat{\textbf{e}}_x$ and $\hat{\textbf{e}}_y$, whose sites (blue and red dots) are denoted by $\textbf{r}$. The  ``spin-ice" lattice is the Bravais lattice spanned by vectors $\textbf{R}_1$ and $\textbf{R}_2$, and with a basis of two spin sites: the ``a'' sites (blue dots) and ``b'' sites (red dots). The plaquettes of the original square lattice are now separated into ``vertices'' located at $\textbf{R}$ and ``plaquettes'' (e.g. the square shaded in gray) of the  ``spin-ice lattice". The plaquette resonance operator of the RK model, $L_{\textbf{R}}$ from  Eq.\eqref{Lsquare}, is also illustrated.}
   \label{figulat12}
  \end{figure}

To describe quantum spin-ice models it is convenient to introduce a different lattice convention relative to that of the previous section. We first divide the plaquettes of the 2D square lattice into two sublattices, that we will now call ``vertices" and ``plaquettes", so that the spin-$\frac{1}{2}$ degrees of freedom are viewed as residing in the ``links" connecting such vertices (see Fig.\ref{figulat12}). These links therefore form another square lattice which is rotated $45^{\circ}$ relative to the orginal square lattice. The Bravais lattice of the spin-ice models is spanned by two vectors $\textbf{R}=n_1\textbf{R}_1+n_2\textbf{R}_2$, $n_{1,2} \in \mathbb{Z}$, which can be viewed as the position of vertices (see Fig.\ref{figulat12}). Therefore, the unit cell has a basis with two spin-$\frac{1}{2}$ degrees of freedom, which we will distinguish by subscripts $a,b$. For example, $\sigma_a^i(\textbf{R})$ will denote the $i$-th Pauli matrix associated with the site $(\textbf{R},a)$ (see Fig.\ref{figulat12}) \footnote{We will also continue to label the spin sites with lower-case letter $\textbf{r}$ when there is no need to specify its detailed Bravais lattice label, namely $\textbf{r}$ is also understood to be the physical coordinate of the spin site with Bravais lattice label $(\textbf{R},i)$ with $i=a,b$.}. From here on, we will assume that the original square lattice has an even number of spin sites both in the x- and y-directions, because this is needed in order to make spin-ice lattice periodic in a torus (see fig. \ref{figulat12}) .


\vspace{0.10in}
For every vertex, we define an ``ice charge operator" as the sum of the $z$-components of the spins in its four links:

\begin{equation}\label{chargemodel}
\small
  Q_{\text{ice}}(\textbf{R}) \doteq \sigma^z_a(\textbf{R})+ \sigma^z_b(\textbf{R}) + \sigma^z_a(\textbf{R} -\textbf{R}_1) + \sigma^z_b(\textbf{R} -\textbf{R}_2).
\end{equation}







The ice charge operators are the locally conserved quantities, and they are the generators of the following ``UV lattice gauge group" of unitary transformations:

\begin{equation}\label{gaugetrasnfo}
        G[\{\theta(\textbf{R})\}] = \exp\bigg(i \sum_{\textbf{R}} \theta(\textbf{R}) Q_{\text{ice}}(\textbf{R})\bigg).
    \end{equation}
    
where $\theta(\textbf{R})$ are arbitrary real numbers. The lattice gauge theory structure is imposed by demanding that the Hamiltonian, $H$, is invariant under the UV lattice gauge group, or equivalently, that it commutes with all the ice charge operators:
    
    

 \begin{equation}
        [{Q}_{\text{ice}}({\textbf{R}}), {H}] = 0, \,\,\,\forall \textbf{R}. 
\end{equation}



\begin{figure}[H]
  \centering
   \includegraphics[trim={2cm 8cm 5cm 4cm}, clip, width=0.5\textwidth]{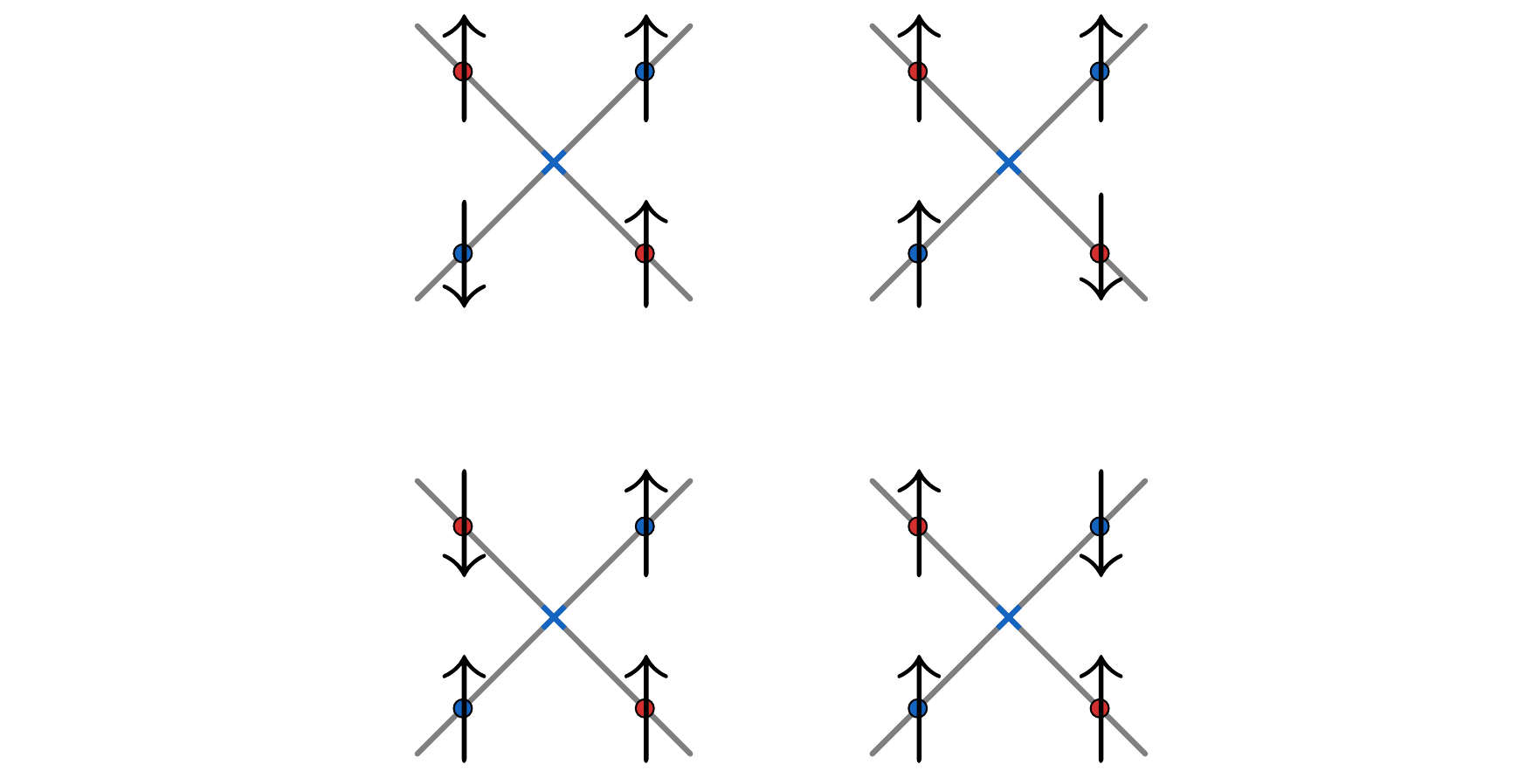}
   \caption{The four configurations of $\sigma^z$ on the links allowed by ${Q}_{\text{ice}}(\textbf{R}) =2$. This sector corresponds to the quantum dimer model, and the dimers are located at links  where spins are reversed relative to the state with all spins pointing up. Therefore, the dimers also mark the location of JW/composite-fermions.}
   \label{QDM}
  \end{figure}

\begin{figure}[H]
  \centering
   \includegraphics[trim={2cm 8cm 2cm 3cm}, clip, width=0.5\textwidth]{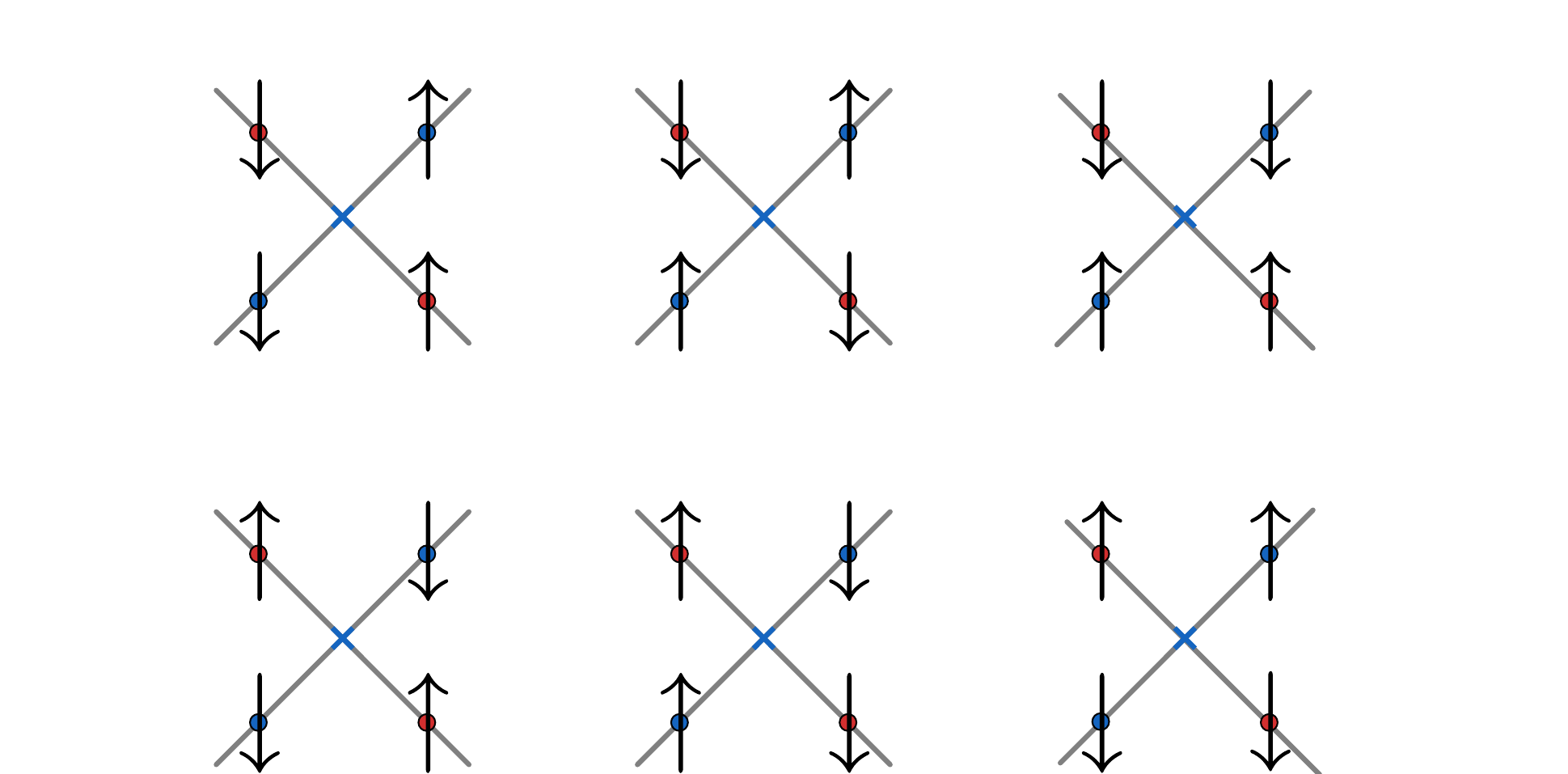}
   \caption{The six configurations of $\sigma^z$ on the links allowed by ${Q}_{\text{ice}}(\textbf{R}) =0$. This is the sector of to the six vertex model.}
   \label{6VM}
  \end{figure}

The subspace with ${Q}_{\text{ice}}({\textbf{R}})=2$ at every vertex is equivalent to that of the quantum dimer model (QDM), whereas the subspace with ${Q}_{\text{ice}}({\textbf{R}})=0$ is equivalent to the quantum six-vertex model (Q6VM) (see Figs. \ref{QDM} and \ref{6VM} for illustration of the allowed configurations). Gauge invariant operators include spin diagonal operators such as $\sigma^z_{a}(\textbf{R})$ (boson number), and products of spin/raising lowering operators (boson creation/ annihilation) over a sequence of links forming a closed loop, the smallest of which is the ``plaquette flipping" operator: 

\begin{equation}
    L_{\textbf{R}} = \sigma^+_a(\textbf{R}) \sigma^-_b(\textbf{R}-\textbf{R}_2) \sigma^+_a(\textbf{R}-\textbf{R}_2) \sigma^-_b(\textbf{R}+\textbf{R}_1-\textbf{R}_2).
    \label{Lsquare}
\end{equation}




The above operator can be viewed as centered around the plaquette that is neighboring to the right the vertex located at $\textbf{R}$ as shown in Fig.\ref{figulat12}. A classic gauge invariant Hamiltonian is the Rokhsar-Kivelson model:

\begin{equation}\label{RKhami}
        H = - t \sum_{\textbf{R}} L_\textbf{R} + L^{\dag}_\textbf{R} + v \sum_\textbf{R} L_\textbf{R}
L^{\dag}_\textbf{R} + L^{\dag}_\textbf{R} L_\textbf{R}.
\end{equation}

Additionally, when placed in a 2D torus each  gauge invariant subspace splits into ``winding" sectors, due to the existence of two conserved t'Hooft loop operators, one for each direction of the torus, defined as:    
     
\begin{equation}\label{loopdef}
    \ell_x \doteq   \sum_{\textbf{r} \in L_x} \sigma^z(\textbf{r}),  \
    \ell_y \doteq  \sum_{\textbf{r} \in L_y} \sigma^z(\textbf{r}).
\end{equation}

\begin{figure}
  \centering
   \includegraphics[trim={0cm 0cm 2cm 0cm}, clip, width=\linewidth]{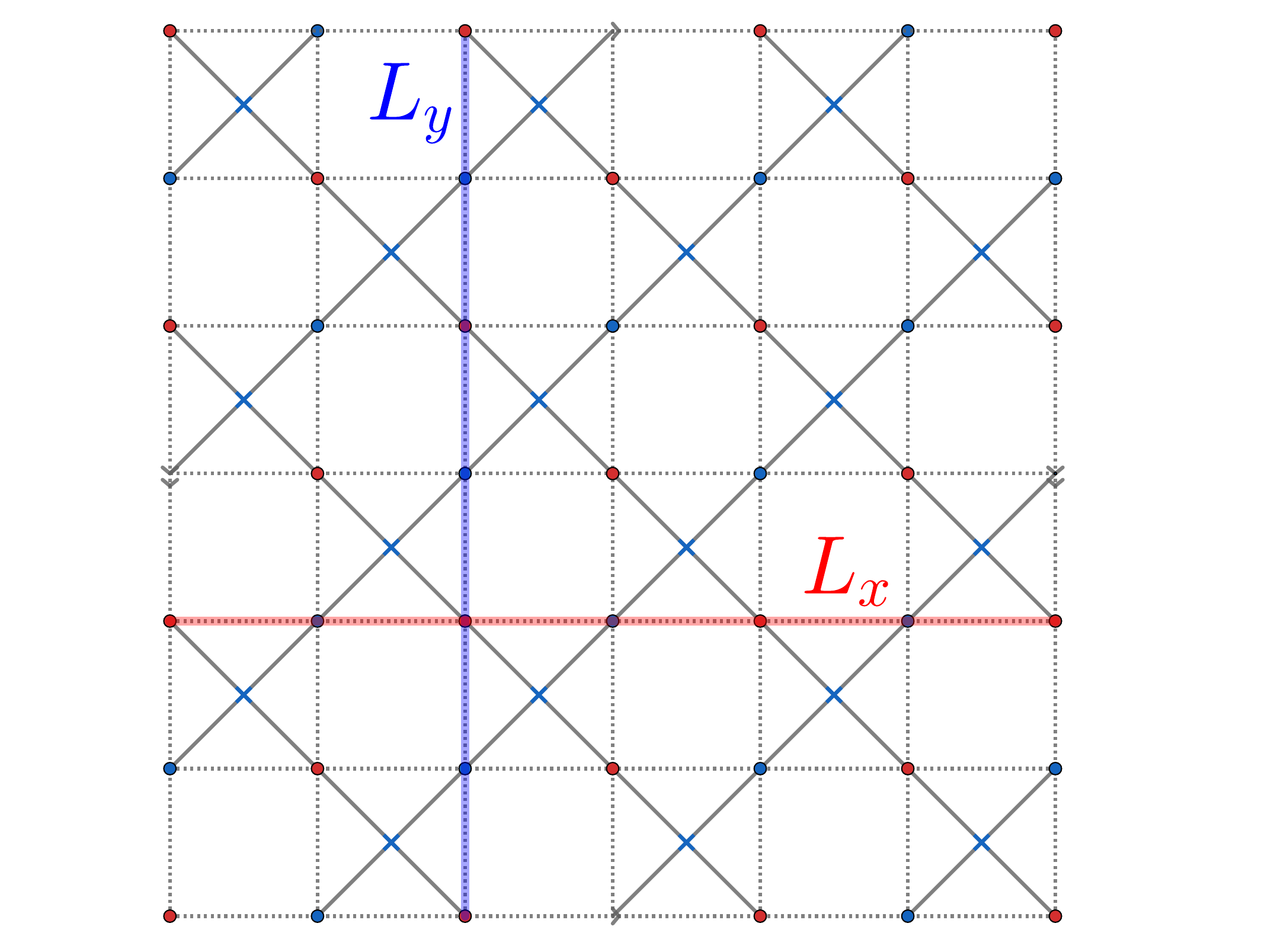}
    \caption{Non-contractible loops $L_{x,y}$ for the definition t'Hooft operators in Eq.\eqref{loopdef} for a lattice with periodic boundary conditions. Notice that in order to place the spin-ice lattice on a torus there needs to be an even number of spin along the $x$ and $y$ directions.} \label{tHooft}
  \end{figure}

Where $ \textbf{r} \in L_{x,y}$ denotes a sum over the sites in the non-contractible loops of the torus depicted in Fig.\Ref{tHooft}.

One of the remarkable properties of the lattice gauge structure of quantum spin-ice models, is that any local gauge invariant operator remains local in its dual Jordan-Wigner/composite-fermion representation. For example, the elementary plaquette flipping operator from Eq.\eqref{Lsquare}, after using the the Jordan-Wigner transformation described in Sec. \ref{equivJWCF}, can be written as: 

\begin{equation} \label{Lsquareferm}
    L_{\textbf{R}} = f_a(\textbf{R}) f^\dag_b(\textbf{R}+\textbf{R}_1) f_a(\textbf{R}+\textbf{R}_2) f^\dag_b(\textbf{R}).
\end{equation}
  
 
   Therefore we see that the RK model can be equivalently represented as a local model of interacting Jordan-Wigner/composite fermions. For larger gauge invariant loop operators (e.g. those enclosing two adjacent plaquettes), the dual fermion operators would also include the products of the fermion parities for the links inside the loop, but in general any local gauge invariant operator of spins maps onto a local fermion operator without any left-over trace of the long-range part of the Jordan-Wigner strings \footnote{This follows from the fact that plaquette operators, $L_{\bf R},L^\dagger_{\bf R}$, and $\sigma^z_{\bf r}$, form a complete algebraic basis from which any local gauge invariant operator can be obtained by addition and multiplication of these. Since this basis operators are mapped into local fermion operators via the JW transformation, it follows that any local gauge invariant operator remains local in its dual JW/composite-fermion representation.}.


The ice-charge operators are represented in terms of Jordan-Wigner/composite fermions as follows:

\begin{equation}\label{chargemodel2}
\small
  Q_{\text{ice}}(\textbf{R}) = 4- 2\,  n_{\text{ice}}(\textbf{R})= 4- 2\sum_{\textbf{r}\in \textbf{R}}f^\dag(\textbf{r})f(\textbf{r}),
\end{equation}

Where $ \textbf{r}\in \textbf{R} $ denotes the spin sites,  $\textbf{r}$, in the four links connected to vertex $\textbf{R}$, and $n_{\text{ice}}(\textbf{R})$ is the total number of fermions in such links. From the above we see that the subspaces obeying with different values of ice charge correspond to different lattice fillings of the Jordan-Wigner/composite-fermions. The QDM and Q6VM spaces have $\frac{1}{4}$ and $\frac{1}{2}$ filling of the fermion sites respectively. Some representative configurations illustrating these fillings are shown in Fig.\ref{unitcell}.

\begin{figure}
  \centering
  	\includegraphics[trim={0cm 0cm 0cm 0cm}, clip, width=0.45\textwidth]{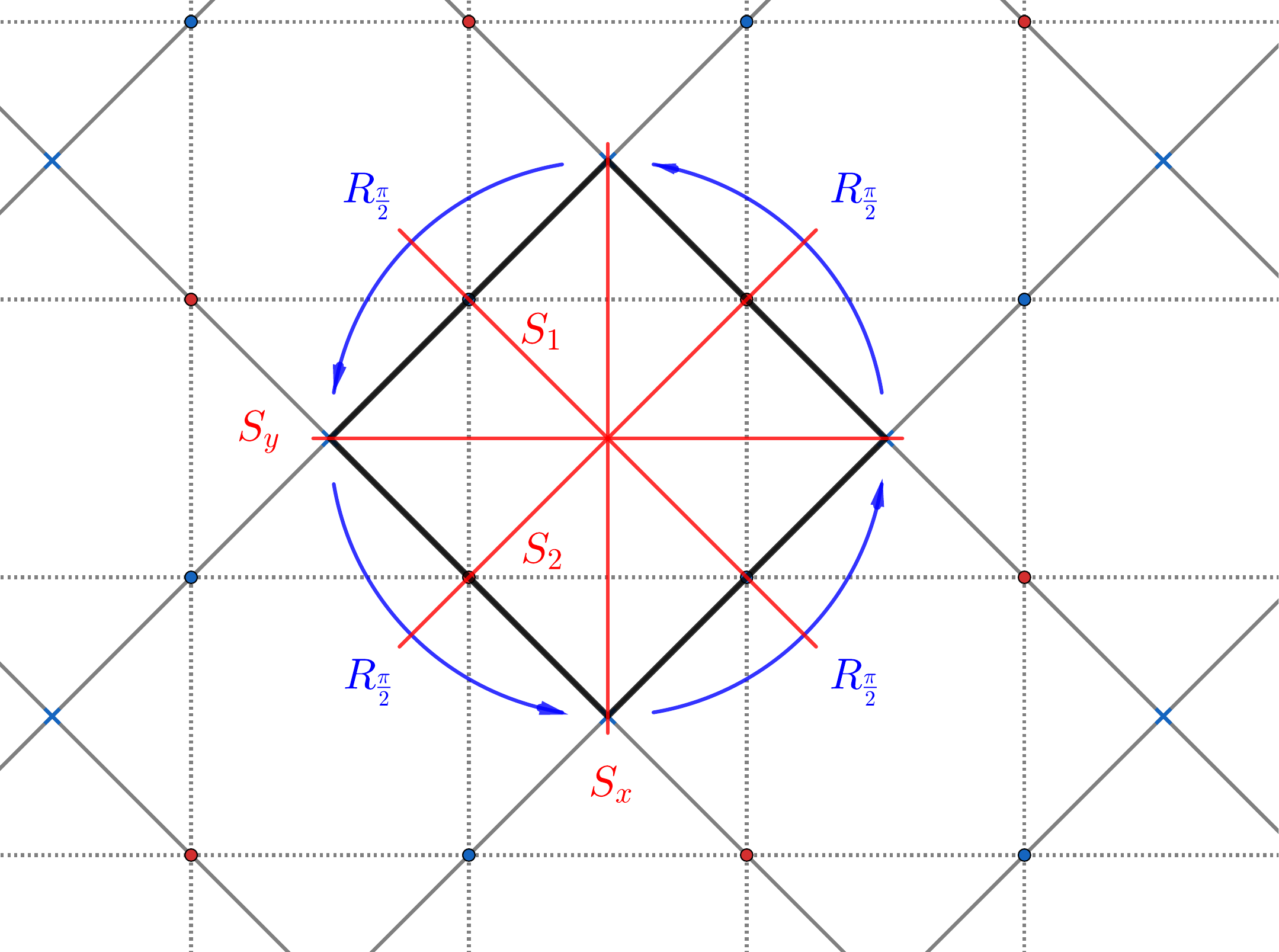}
    \caption{Illustration of the point group symmetry operations of the quantum spin-ice model centered on a spin-ice plaquette (corresponding to Dihedral Group $D_8$). The model also has a similar set of symmetry operations centered around the vertices, which we also consider for constructing states.}
    \label{D8}
  \end{figure}
\begin{figure*}
  \centering
 \includegraphics[trim={0cm 0cm 0cm 0cm}, clip, width=0.8\textwidth]{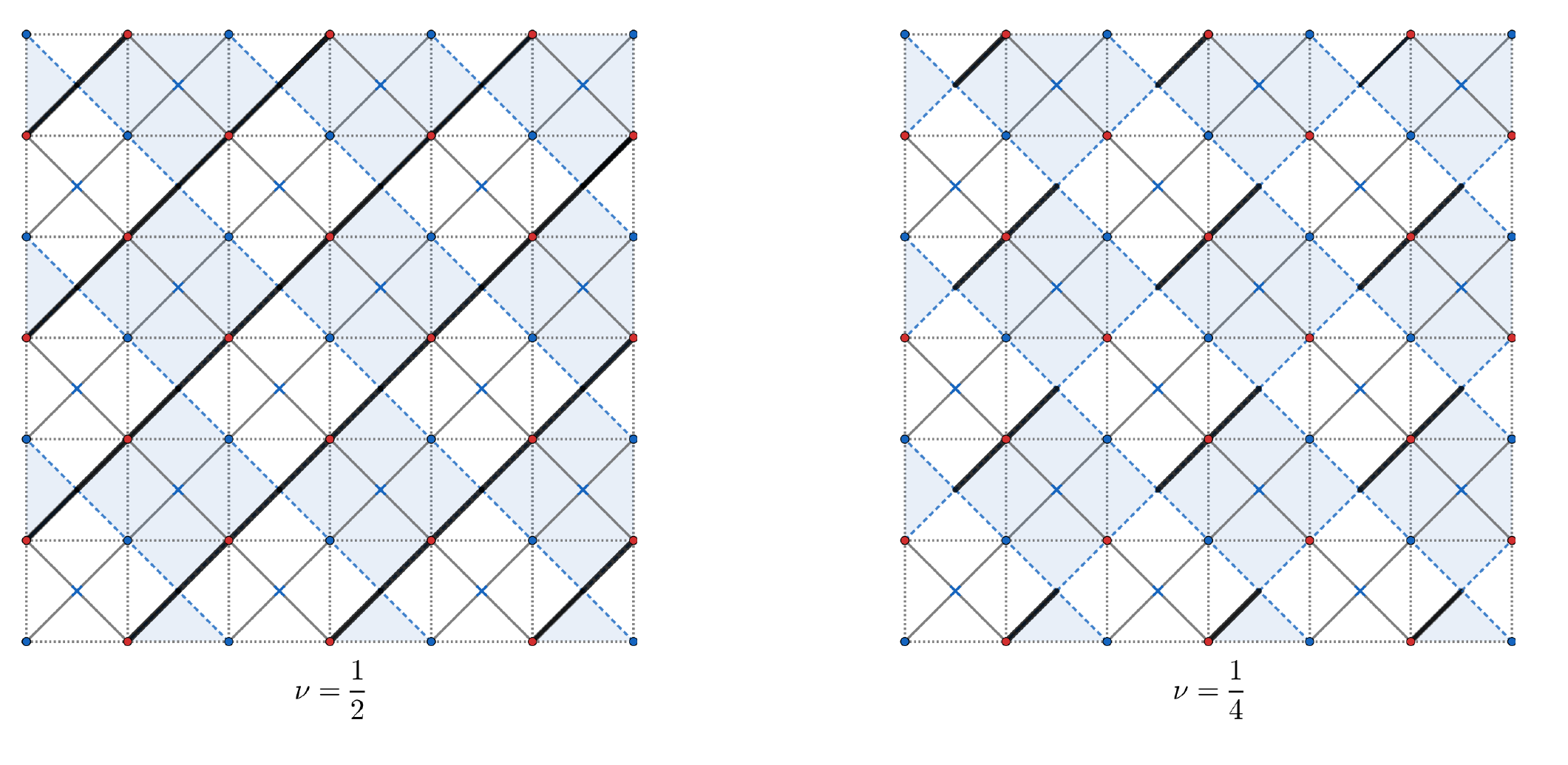}
   \caption{Left: depiction of a configuration with half of spins reversed (half-filling of JW/composite-fermions) relative to fully polarized state (denoted by solid bars) belonging to six-vertex subspace. Right: depiction of a configuration with one quarter of spins reversed (quarter-filling of JW/composite-fermions) relative to fully polarized state (denoted by solid bars) belonging to quantum dimer subspace.}
   \label{unitcell}
  \end{figure*}

In this work we will be interested in constructing spin-liquid states that are relevant not only for the RK model, but for the universality class that the RK Hamiltonian defines. This universality class is defined as the set of local spin Hamiltonians\footnote{The locality is defined with respect to the tensor product structure of the Hilbert space of underlying microscopic spin degrees of freedom.}, with the same spin-ice local conservation laws, and the same global symmetries of the RK model. Some of these global symmetries of the RK model are listed in Table \ref{symtable}, and the notation for some of its space symmetries is also depicted in figure \ref{D8}. The particle-hole symmetry can only be enforced for the filling associated with the subspace of the Q6VM (see Table \ref{symtable}).

\begin{table*}  
\centering
\begin{tabular}{c|c|c|c|c|c|c|c|c}
\toprule
Symmetries &Symbol       &        &Q6VM   &QDM &Linear &Antilinear  &Action on $b(\textbf{r})$     &Action on $Q_{\text{ice}}(\textbf{R})$\\
\hline
Time Reversal  &$\Theta$ &    &$\checkmark$            &$\checkmark$ & 

&$\checkmark$  &$b(\textbf{r})$   &$Q_{\text{ice}}(\textbf{R})$\\
\hline
\multirow{5}{*}{Spatial transformations}
& &$R_{\frac{\pi}{2}}$ & & & & & &\\
& &$S_x$ & & & & & &\\
&$U_d$ &$S_y$ &$\checkmark$ &$\checkmark$ &$\checkmark$ & &$b\big(U_d(\textbf{r})\big)$  &$Q_{\text{ice}}(U_d(\textbf{R}))$  \\
& &$S_1$ & & &  & & &\\
& &$S_2$ & & &  & & &\\ 
\hline
Particle-Hole &$X$ & &$\checkmark$            & &$\checkmark$ &     &$b^\dagger(\textbf{r})$ &$-Q_{\text{ice}}(\textbf{R})$\\
\bottomrule
\end{tabular}
 \caption{Table of symmetries of the Rokhsar-Kivelson Hamiltonian of quantum spin-ice.  $U_d(\textbf{r})$ and $U_d(\textbf{R})$ denote the image of the site $\textbf{r}$ and the vertex $\textbf{R}$ under the corresponding spatial transformation. See Fig.\ref{D8} for a definition of the spatial transformations and Fig.\ref{rot1} for a depiction of the action of $R_{\frac{\pi}{2}}$.}
 \label{symtable}
\end{table*}

\subsection{Review of Abrikosov-Schwinger parton states}
\label{B}
Before introducing the extended parton construction of states for Jordan-Wigner/composite-fermions in quantum spin-ice, we would like to review some of the key ideas of the more traditional construction of states for Abrikosov-Schwinger fermions, which we will sometimes refer to as “point-like” partons (for more detailed discussions see e.g. Refs.\cite{Aff, Dag,WenA02, WenB02}). The same previously discussed physical spin-$\frac{1}{2}$ degrees of freedom at the lattice site $\textbf{r}$ can be alternatively represented in terms of spinful Abrikosov-Schwinger fermions  $\psi^{\dag}_s(\textbf{r})$ ($s =\, \uparrow, \downarrow$):

\begin{equation}\label{parton}
    \sigma^i(\textbf{r}) = \sigma^i_{ss'} \psi^{\dag}_s(\textbf{r}) \psi_{s'}(\textbf{r}).
\end{equation}


where $\sigma^i_{ss'}$ is the $ss'$ element of the $i$-th Pauli Matrix. The above representation enlarges the physical Hilbert space from a two dimensional $\{\ket{\uparrow}, \ket{\downarrow}\}$ into a four dimensional $\{\ket{0},\ket{\uparrow}, \ket{\downarrow}, \ket{\uparrow \downarrow}\}$. In this case, the “UV lattice gauge group” is generated by the fermion number at each site:

\begin{equation}\label{charge}
    n(\textbf{r}) = \sum_{s} f^{\dag}_s(\textbf{r})f_s(\textbf{r}). 
\end{equation}

The above operator is the counterpart of the spin-ice charge for this  lattice gauge structure. Gauge invariant operators are defined as those commuting with $n(\textbf{r})$, and in this case they are  are the spin operators themselves, $\sigma^i(\textbf{r})$. The physical subspace is a gauge invariant subspace satisfying:

\begin{equation}\label{eq14}
    n(\textbf{r}) \ket{\psi} = \ket{\psi}.
\end{equation}

Therefore, in this parton construction physical states are restrited to have $\frac{1}{2}$ fermion filling of the lattice, which is already a crucial difference with respect to the Jordan-Wigner/composite-fermions.  

When restricted to the physical subspace, $n(\textbf{r}) = \mathbb{1}$, there is actually a $SU(2)$ group that leaves all gauge invariant, and which is larger than the $U(1)$ UV lattice gauge group generated by \eqref{eq14}. Such larger group of operations that leave the gauge invariant operators invariant, is called the \textit{parton Gauge group} (PGG). To construct spin-liquid states it is convenient to introduce an auxiliary mean-field Hamiltonian that parametrizes a Slater of fermions:

\begin{equation}\label{Hamiwen}
    H_{\rm MF} = \sum_{ss'}\sum_{\textbf{r},\textbf{r}'} t_{ss'}(\textbf{r},\textbf{r}') f^{\dag}_s(\textbf{r}) f_{s'}(\textbf{r}').
\end{equation}

The hopping elements $t_{ss'}(\textbf{r},\textbf{r}')$ in the mean-field Hamiltonian can be viewed as variational parameters of its Slater determinant ground state, that we will denote by $\ket{\Omega_0 [t_{ss'}(\textbf{r},\textbf{r}')]}$. The above mean field Hamiltonian conserves the total fermion number, and, therefore, it is invariant under a global $U(1)$ subgroup of the PGG. More generally, the group that leaves the mean-field Hamiltonian invariant is called invariant gauge group (IGG). The importance of the IGG is that it determines the expected true low energy emergent gauge group of the spin liquid state \cite{WenA02, WenB02} (assuming it does not suffer from instabilities such as gauge confinement). For the above mean-field Hamiltonian with a $U(1)$ IGG we expect then to have $U(1)$ spin liquid  \cite{WenA02, WenB02}\footnote{But had we chosen a BCS-like mean field state with a $\mathbb{Z}_2$ IGG, we would expect a $\mathbb{Z}_2$ spin liquid}. For concretenes in this work we will be focusing on spin liquids with low energy emergent $U(1)$ gauge groups.






The ground state, $\ket{\Omega_0 [t_{ss'}(\textbf{r},\textbf{r}')]}$, of the above mean field Hamiltonian generically is not invariant under the UV gauge group and violates the constraint of Eq.\eqref{eq14}. The correct physical mean-field state is obtained by projecting this state onto the physical gauge invariant subspace (Gutzwiller projection), as follows:

\begin{equation}\label{GutzJW}
    \ket{\Omega [t_{ss'}(\textbf{r},\textbf{r}')]} = \prod_{\textbf{r}} \bigg( \frac{1- (-1)^{n(\textbf{r})}}{2} \bigg) \ket{\Omega_0 [t_{ss'}(\textbf{r},\textbf{r}')]},
\end{equation}

The Gutzwiller projection is a nontrivial operation that generally makes difficult the calculation gauge invariant operators. It
is possible, however, to develop a precise understanding of
the symmetry properties of the Gutzwiller projected physical
state. To illustrate this, let us imagine that there is some global physical symmetry operation acting on the spins, denoted by $S$ (e.g. a lattice translation or a mirror symmetry). We say that two operations $S_1$ and $S_2$ defined by their action on the parton fermions represent the same physical symmetry, if they have the same action on all gauge invariant operators. However, if $S_1$ and $S_2$
differ by an element of the parton gauge group, their enforcement on $\ket{\Omega_0}$ can lead to two distinct
physical states $\ket{\Omega}$. In this case, then $S_1$ and $S_2$ are said to be two distinct projective symmetry group (PSG) implementations on the partons of the same underlying physical symmetry (for a recent discussion illustrating this, see e.g.\cite{Inti}).

\subsection{Extended parton states for quantum spin-ice}
\label{C}

We are now ready to present our extended parton construction of mean field states for composite fermions obtained from the JW transformation applied to the quantum spin-ice Hamiltonians. The idea is to parallel the construction for Abrikosov-Schwinger fermions, but for the UV gauge structure defined by the ice charge operators from Eq.\eqref{chargemodel}. We begin by introducing an auxiliary mean-field Hamiltonian of composite fermions:
 
\begin{equation}\label{mfhami}
       H_{\text{MF}} =  \sum_{ \textbf{r},  \textbf{r}' } t(\textbf{r},\textbf{r}')f^{\dag}(\textbf{r}) f(\textbf{r}').
   \end{equation}


 
here $f^{\dag}(\textbf{r})$ is the creation operator of the spinless Jordan-Wigner fermion at the spin site $\textbf{r}$. The hopping amplitudes, $t(\textbf{r},\textbf{r}')$, are again viewed as parametrizing the Slater determinant ground state of the mean field Hamiltonian, denoted by $\ket{\Phi_0 [t(\textbf{r},\textbf{r}')]}$. The physical spin orientation is encoded in the composite fermion occupation at each site, and therefore there is no enlargement of the full spin Hilbert space. Nevertheless, the composite fermion hopping bilinears in the above mean-field Hamiltonian generically do not commute with the generators of the UV lattice gauge transformations, and therefore its ground state, $\ket{\Phi_0}$ violates the ice rules. However, this is forbidden by Elitzur's theorem: local gauge symmetries cannot be spontaneously broken. As a consequence, the naive ground state of the above mean-field Hamiltonian is not a satisfactory approximation to the true gauge invariant ground states of quantum spin-ice models. However, this deficiency can be cured in an analogous way as in the case of Abrikosov-Schwinger partons, by projecting $\ket{\Phi_0}$ onto the Gauge invariant subspaces. Therefore, in analogy to the Gutzwiller projection, we introduce a projector into a Gauge invariant subspace, specified by the local ice charges $ Q_{\text{ice}}(\textbf{R})$ and the t'Hooft operators $(\ell_x,\ell_y)$ (for the case of a torus), given by:

\begin{equation}\label{proiector}
\begin{aligned}
    &P(\{Q_{\text{ice}}(\textbf{R}),\ell_x, \ell_y\})\doteq P(\ell_x)  P(\ell_y) \prod_{\textbf{R}} P(Q_{\text{ice}}(\textbf{R})).\\
&\ket{\Phi[t(\textbf{r},\textbf{r}')]} = P(\{Q_{\text{ice}}(\textbf{R}),\ell_x, \ell_y\})\ket{\Phi_0[t(\textbf{r},\textbf{r}')]}.
    \end{aligned}
\end{equation}


The above projected state is also parametrized by the hoppings, $t(\textbf{r},\textbf{r}')$, that could be in principle optimized as variational parameters to minimize the energy of RK-like Hamiltonians.

\begin{table*}
\centering
\begin{tabular}{c|c|c}
\toprule
 &Abrikosov-Schwinger Fermions & Jordan-Wigner Composite-Fermions  \\
 \hline
Local Hilbert Space Enlargement  &YES &NO\\
\hline
Internal degrees of freedom &Spin $\frac{1}{2}$ & spinless\\
\hline
UV Gauge Transformations generators &$n(\textbf{r}) = \sum_{s} f^{\dag}_s(\textbf{r})f_s(\textbf{r})$  & $Q_{\text{ice}}(\textbf{R}) = 4-2 \sum_{\textbf{r}\in \textbf{R}}f^\dag(\textbf{r})f(\textbf{r})$\\
\hline
Physical lattice fillings &$n(\textbf{r})=1$ & Any
(e.g. $\langle f^\dag(\textbf{r})f(\textbf{r})\rangle=\frac{1}{4}$ for QDM)\\
\bottomrule
\end{tabular}
\caption{Comparison between traditional point-like Abrikosov-Schwinger fermion partons and the extended Jordan-Wigner/composite-fermion parton constructions for 2D quantum spin-ice. Here $s \in \{\uparrow,\downarrow\}$, ${\bf r}$ denotes spin sites, ${\bf r}\in {\bf R}$ denotes the four spins adjacent to a quantum spin-ice vertex located at ${\bf R}$. See Eq.\eqref{chargemodel2} and Fig.\ref{figulat12} for definitions and depictions.}
\label{tab_dmm_1}
\end{table*}



\subsubsection{Symmetry implementation on JW composite fermions: general considerations}\label{C1}


Let us now consider the implementation of symmetries on these mean field states of Jordan-Wigner/composite-fermions. As in the case of Abrikosov-Schwinger fermions, the key idea is that the task of enforcing symmetries in the physical projected states is traded by the easier task of enforcing symmetries in the un-projected mean-field Hamiltonians. However, one needs to develop a set of consistency criteria for these implementations because there are multiple ways in which one given symmetry can be implemented in the un-projected state, leading to the rich structure of projective symmetry group implementations \cite{WenA02, WenB02}. 

At first glance it might appear as if there was no freedom on how to implement symmetries on  the Jordan-Wigner fermions, because any prescription on how physical symmetries act on the underlying spin-$\frac{1}{2}$ degrees of freedom would fix a unique symmetry action of the Jordan-Wigner/composite-fermion operators. We will refer to this underlying symmetry implementation as the ``bare'' symmetry action. However, this bare symmetry implementation cannot be suitably enforced in the mean field Hamiltonians from Eq.\eqref{mfhami}. This is because the specific choice for implementing the Jordan-Wigner ordering of the 2D lattice (e.g. the western typing convention of Sec. \ref{equivJWCF}) does not manifestly preserve the symmetries of the lattice, and thus, for example, the bare action of the bare implementation of a $\pi/2$ lattice rotation would map the  fermion bilinear mean field Hamiltonian from Eq.\eqref{mfhami} onto a complex operator which is no longer fermion bilinear Hamiltonian and does not appear local in its dual fermion representation.  Our goal in this subsection will be therefore to develop a precise but more flexible notion of symmetry implementations on the Jordan-Wigner/composite-fermions that is amenable to enforcement on mean-field Hamiltonians.

Some of these difficulties of bare symmetry actions are not peculiar to the 2D Jordan-Wigner transformation but are also reminiscent of those appearing in the 1D Jordan-Wigner transformation, e.g. in the anomalous implementation of lattice translations, which we will now discuss in order to motivate the 2D construction. For example, consider a standard 1D finite lattice with periodic boundary conditions and a standard translational symmetry implemented on the microscopic spin operators located at site $r$ as follows:

\begin{equation}
     T \sigma^i({r}) T^{\dag} = \sigma^i(r+1)
  \end{equation}


However, when this ``bare" symmetry is implemented on the JW fermions it does not act like a standard fermionic lattice translation, which we denote by $\tau$, defined as:

\begin{equation}
      \tau f^{\dag}(r) \tau^{\dag} = f^{\dag}(r+1)\neq T f^{\dag}(r)T^{\dag} 
\end{equation}

The above arises because the JW string becomes translated by $T$ and therefore it does not follow the initial JW convention (it does not start at spin ``1" any more)\footnote{For a recent discussion of the connection between lattice translational symmetries, anomalies and dualities associated with the 1D JW transformation see Ref. \cite{seiberg2023majorana}.}. However while $T$ and $\tau$ are different operations when acting on the single fermion operator, they would act identically on fermion bilinear operators supported in the interior of the 1D chain: 

\begin{equation}
      \tau f^{\dag}(r)  f(r) \tau^{\dag} =  T f^{\dag}(r)f(r) T^{\dag}=f^{\dag}(r+1) f(r+1)
\end{equation}

Therefore we can say that when the symmetry operations $T$ and  $\tau$ are restricted to act on parity even operators, they are essentially the same symmetry\footnote{Up to corrections associated with boundary terms, but in this work we will focus on implementations of symmetry in the bulk.}. The parity restriction in 1D plays an analogous role to the spin gauge structure in 2D, in the sense that local spin operators that are invariant under the UV lattice gauge symmetry, remain local in their dual fermion representation after the JW map. In other words, after a quantum spin-ice model is mapped onto fermions via the JW map, it appears to be a bona fide local fermionic model, similar to how a parity even spin Hamiltonian looks like an ordinary fermionic model after 1D JW map.\\

Therefore we define a generalized notion of equivalence among symmetries of the 2D quantum spin-ice model when these are implemented on JW transformation, as follows:
\\

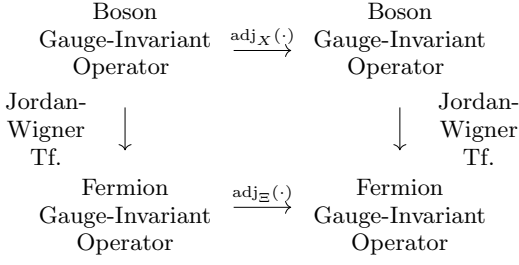
\begin{figure}
\begin{tikzcd}
\parbox{8em}{ \centering
    Boson Gauge-Invariant\\ Operator \vspace{0.10in}
} \arrow[swap]{d}{\parbox{6em}{\centering
    Jordan-Wigner\\ Tf.
}}  \arrow{r}{\text{adj}_X(\cdot)}  & \parbox{8em}{\centering
    Boson Gauge-Invariant\\ Operator \vspace{0.10in}
}    \arrow{d}{\parbox{6em}{\centering
    Jordan-Wigner\\ Tf.
}}  
\\
 \parbox{8em}{\centering 
 \vspace{0.10in}
    Fermion Gauge-Invariant\\ Operator
}  \arrow{r}{\text{adj}_{\Xi}(\cdot)}  &  \parbox{8em}{\centering
  \vspace{0.10in}  Fermion Gauge-Invariant\\ Operator
 }
\end{tikzcd}
\caption{Illustration of the notion of equivalence of  symmetry actions of operators. Two operators $X$ and $\Xi$ are equivalent implementations of a symmetry, if their action is identical on all the operators that are invariant under the spin-ice UV lattice gauge transformations (defined in Eq.\eqref{gaugetrasnfo}). This notion allows us to trade the possibly complicated ``bare'' action of the microscopic symmetry, $X$, on the JW/composite-fermions, by a simpler but equivalent symmetry implementation, $\Xi$, which maps JW fermion bilinears onto JW fermion bilinears. This is a natural extension of the notion of symmetry equivalence in standard parton constructions Abrikosov-Schwinger fermions (see e.g. Refs.\cite{WenA02, WenB02}).}
    \label{commut}
  \end{figure}

\begin{mdframed}
{\it For a quantum spin-ice model, we say that two operators $S_1$ and $S_2$ that implement a symmetry action are equivalent, when they have the same action on all local operators that are invariant under the UV lattice gauge transformations defined in Eq.\eqref{gaugetrasnfo}} (see Fig.\ref{commut}).
\end{mdframed}

\vspace{0.10in}

\par The usefulness of this notion of equivalent symmetries is that instead of enforcing the non-trivial ``bare" action of a symmetry, $S_1$, we can enforce instead a simpler but equivalent symmetry implementation, $S_2$, which maps fermion bilinear Hamiltonians onto fermion bilinear Hamiltonians. By enforcing $S_2$ on the fermion bilinear mean-field Hamiltonian from Eq.\eqref{mfhami}, then the expectation value of any gauge invariant operator computed from its corresponding Gutzwiller projected state from Eq.\eqref{proiector}, will obey the same symmetry constraints as if we had enforced the bare symmetry action $S_1$. In particular, if $G$ is a lattice UV gauge transformation from Eq.\eqref{gaugetrasnfo}, then $S$ and $G S$ are equivalent implementations of a symmetry. Interestingly, as we will see, enforcing symmetries that differ by such a pure UV lattice gauge transformation, $G$, on the mean-field Hamiltonian of Eq.\eqref{mfhami}, can lead to physically distinct states after the generalized Gutzwiller projection of Eq.\eqref{proiector}. This situation is analogous to that of PSG implementations of symmetry on the Abrikosov-Schwinger partons (see discussion following Eq.\eqref{GutzJW}). However several interesting qualitative differences will appear between these two cases, and this is partly why we call our construction an extended projective symmetry group implementation (see Table \ref{tab_dmm_1}).


\subsubsection{A specific implementation of symmetries of 2D quantum spin-ice on JW composite fermions.}\label{C1}

We will now construct a concrete example of extended projective symmetry implementation for the symmetries of the quantum spin-ice model (see Table \ref{symtable}). Our objective is to illustrate the general ideas by constructing interesting and perhaps even energetically competitive spin liquid states (although we will not compute explicitly their energy). It is clear, in analogy to ordinary parton constructions \cite{WenA02, WenB02}, that there is a large landscape of possible extended projective symmetry implementations beyond the ones we will illstrate concretely. We leave to future work the development of a more global understanding and classification of the large and colorful landscape of extended projective symmetry group implementations.

\begin{figure}
  \centering
   \includegraphics[trim={1cm 0cm 0cm 0cm}, clip, width=0.50\textwidth]{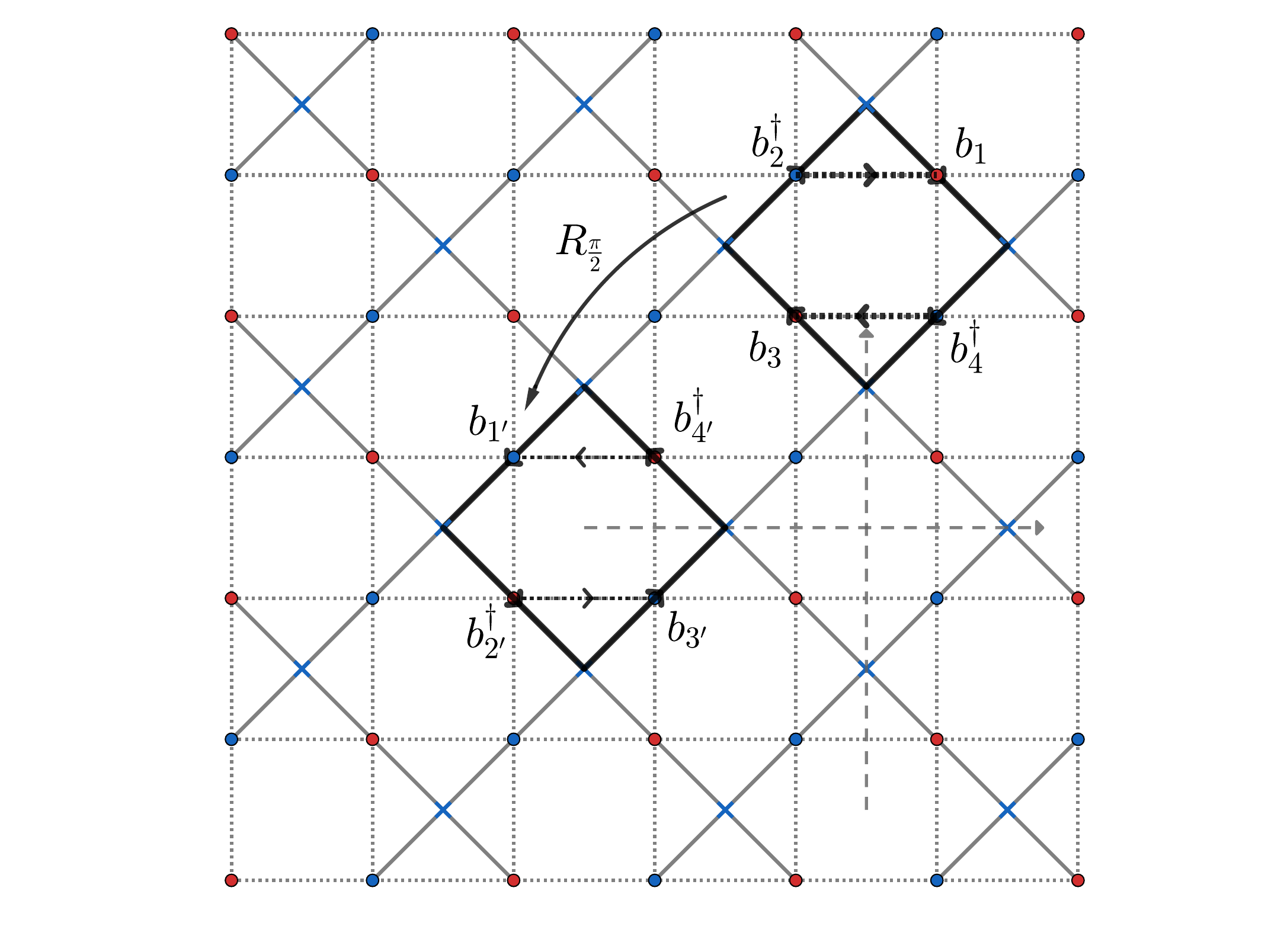}
   \caption{Action of a 90° rotation (denoted by $R_{\frac{\pi}{2}}$) centered on the plaquette where the dashed dotted lines intersect, acting on the plaquette resonace operators marked by thick squares (according to convention in fig. \ref{figulat12}).}
   \label{rot1}
  \end{figure}

Let us begin by considering a $\pi/2$ spatial rotation centered on a plaquette (see Fig.\ref{rot1}), denoted by $R_{\frac{\pi}{2}}$. We define its action on the microscopic degrees of freedom from an implementation that is natural when viewed as spinless bosons, namely as follows:

 \begin{equation}\label{Ppi2micro}
  \begin{aligned}
      R_{\frac{\pi}{2}} b^{\dag}({\bf r})  R_{\frac{\pi}{2}}^\dagger 
      &= b^{\dag}(R_{\frac{\pi}{2}}{\bf r}).
      \end{aligned}
  \end{equation}

Here $b^{\dag}({\bf r})$ is the hard-core boson equivalent of the spin lowering operator (see Table \ref{Traduzione}), and $R_{\frac{\pi}{2}}{\bf r}$ is the image of site ${\bf r}$ under the rotation. The action of $R_{\frac{\pi}{2}}$ on the gauge invariant plaquette operator from Eq.\eqref{Lsquare} is thus simply:
  \begin{equation}\label{Ppi2}
  \begin{aligned}
      R_{\frac{\pi}{2}} (b_2^{\dag} b_1 b^{\dag}_4 b_3) R_{\frac{\pi}{2}}^\dagger 
      &= b^{\dag}_{4'} b_{1'}  b_{2'}^{\dag} b_{3'},
      \end{aligned}
  \end{equation}
  where $1,2,3,4$ denote the sites in the plaquette from figure \ref{rot1} and $1',2',3',4'$ their images after the $\frac{\pi}{2}$ rotation. As discussed in Eq.\eqref{Lsquareferm}, this same plaquette operator can be alternatively written as product of JW/composite-fermion operators.
  However, while the action of $R_{\frac{\pi}{2}}$ is simple on this four fermion operator, it is complex and cumbersome on JW/composite-fermion operators themselves, as it involves a $\pi/2$ rotation of the JW-strings. More importantly $R_{\frac{\pi}{2}}$ does not map fermion bilinear operators onto fermion bilinears, because, for example, it maps a fermion horizontal hopping into a fermion vertical hopping dressed by JW-strings (see Fig.\ref{verticalhopping}). Therefore, we would like to find an alternative but gauge equivalent implementation of $R_{\frac{\pi}{2}}$ to overcomes this difficulty. 
  
  To do so, we define a collection of auxiliary operators associated with each of the microscopic symmetries listed in Tables \ref{symtable} and \ref{symtablejw}, whose action is defined by replacing the boson operator, $b^{\dag}({\bf r})$ with the fermion operator  $b^{\dag}({\bf r})$ in Table \ref{symtable}. For example, for the microscopic symmetry $R_{\frac{\pi}{2}}$, we associate the auxiliary fermion operator $P_{\frac{\pi}{2}}$, whose action is obtained from Eq.\eqref{Ppi2micro} by replacing $b^{\dag}({\bf r})\rightarrow f^{\dag}({\bf r})$, leading to:
\begin{equation}\label{rotfer}
    P_{\frac{\pi}{2}}  f^{\dag} (\textbf{r}) P_{\frac{\pi}{2}}^\dagger = f^{\dag}(R_{\frac{\pi}{2}}\textbf{r}).
\end{equation}
Thus the idea is that these auxiliary fermion operators are intuitive and natural symmetry implementations on fermions, but they are not necessarily equivalent implementations of the microscopic symmetries on gauge invariant operators, as we now explain. This auxiliary fermion rotation acts on the same plaquette operator from Eq.\eqref{Ppi2}, which can be equivalently represented with fermions using Eq.\eqref{Lsquareferm}, as follows:

\begin{equation}
P_{\frac{\pi}{2}}  (f_2^{\dag} f_1 f^{\dag}_4 f_3) P_{\frac{\pi}{2}}^\dagger = -f^{\dag}_{4'} f_{1'}  f_{2'}^{\dag} f_{3'} 
\end{equation}

\begin{figure*}
  \centering
   \includegraphics[trim={0cm 0cm 0cm 3cm}, clip, width=0.95\textwidth]{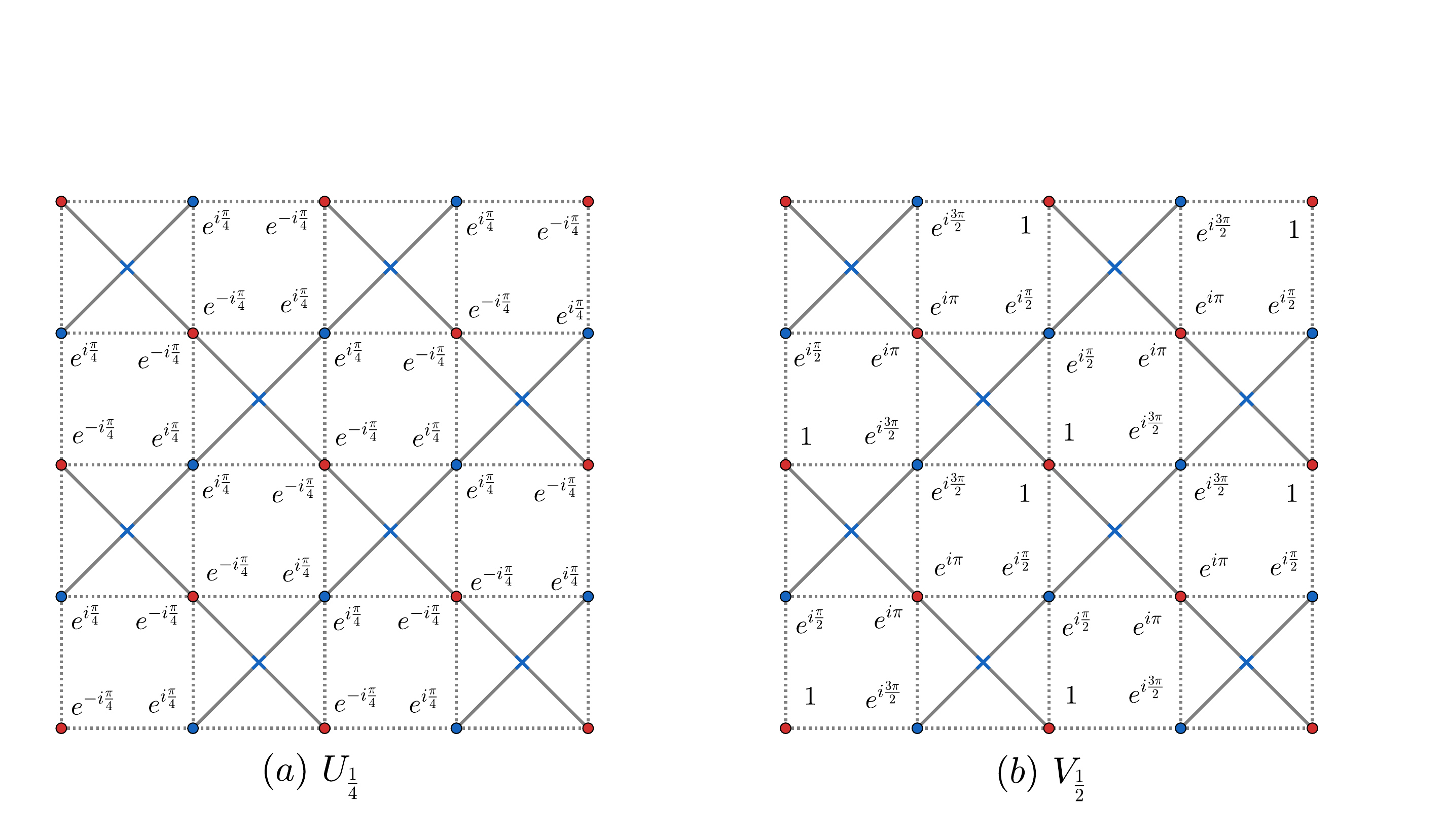}
   \caption{Phases gained by creation operators $f^\dag(\textbf{r})$ or  $b^\dag(\textbf{r})$ under the action of the site dependent $U(1)$ transformations: (a) $U_{\frac{1}{4}}$ (from Eq.\eqref{U1/4}), (b) $V_{\frac{1}{2}}$ (from Eq.\eqref{V1/2}).}
    \label{u14}
  \end{figure*}

  Therefore the fermion rotation, $P_{\frac{\pi}{2}}$, is not an equivalent implementation of the the underlying physical symmetry, $R_{\frac{\pi}{2}}$, because it additionally multiplies the gauge invariant plaquette operator by a global minus sign. The extra minus sign can be removed by dressing $P_{\frac{\pi}{2}}$ with a staggered $U(1)$ transformation that we call $U_{\frac{1}{4}}$, which rotates the phase of bosons with opposite signs in the $a$ and $b$ sublattices (see Fig.\ref{u14} $(a)$) as follows:
  

  \begin{equation*}
  \begin{aligned}
           U_{\frac{1}{4}}  b^{\dag}_{a} (\textbf{R}) U^{\dag}_{\frac{1}{4}} &= e^{i \frac{\pi}{4}} b^{\dag}_{a}(\textbf{R}),\\
           U_{\frac{1}{4}} b^{\dag}_{b}(\textbf{R}) U^{\dag}_{\frac{1}{4}} &= e^{-i \frac{\pi}{4}} b^{\dag}_{b}(\textbf{R}).\\
  \end{aligned}
  \end{equation*}
   
   where we are using the Bravais labels of the sites of the model (see Sec. \ref{QSIJW} for the convention). Notice that the action of $U_{\frac{1}{4}}$ on boson and JW fermion operators is the same. Therefore, its action on the plaquette operator is:

\begin{equation}\label{U1/4}
U_{\frac{1}{4}}  (f_2^{\dag} f_1 f^{\dag}_4 f_3) U_{\frac{1}{4}}^{\dag} = -f_2^{\dag} f_1 f^{\dag}_4 f_3.
\end{equation}

    \begin{figure}
  \centering
\includegraphics[trim={4cm 0cm 0cm 0cm}, clip, width=0.55\textwidth]{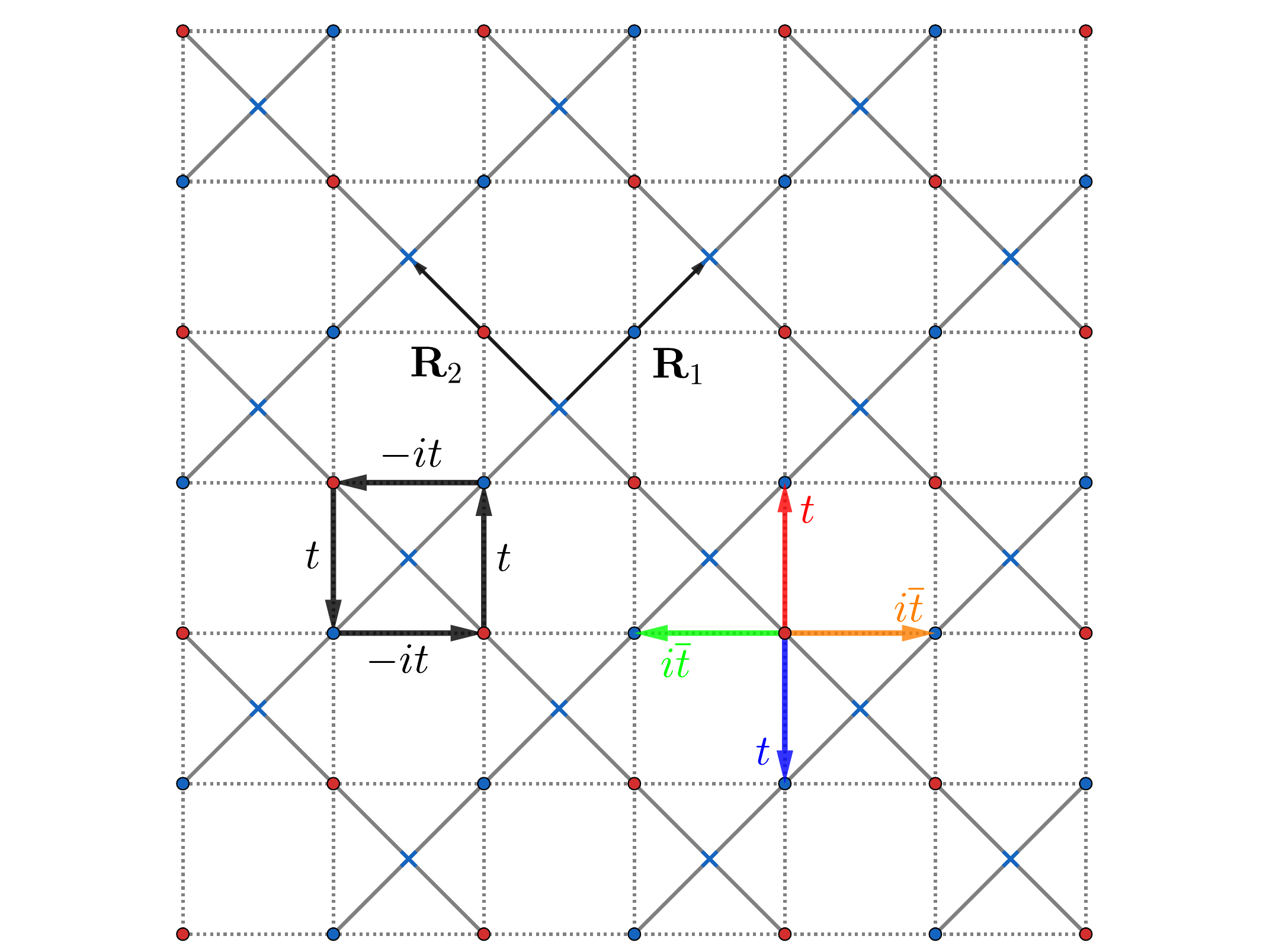}
   \caption{Right: Depiction of the allowed nearest neighbour JW/composite-fermion hoppings associated with $\Theta_x$ and $\Theta_y$ extended projective symmetry groups from Table \ref{implsym}. Here the amplitude $t$ is real for $\Theta_x$ and purely imaginary for $\Theta_y$. Left: The phase of fermion hopping around a vertex or plaquette of the original square lattice (dotted line) is $\pi$. This $\pi$ phase is behind the Dirac spectrum of JW/composite-fermions for these state (see Fig.\ref{fsfil}).}
   \label{mfhop1}
  \end{figure}
  Therefore the action of $R_{\frac{\pi}{2}}$  and $U_{\frac{1}{4}}P_{\frac{\pi}{2}}$ on plaquette operators is identical. Moreover, their action is also identical on $\sigma^z(\textbf{r})$ (or equivalently the on-site fermion number operator). Since these operators together with the plaquette operators form a complete algebraic basis for all local gauge invariant operators, it follows that $U_{\frac{1}{4}}P_{\frac{\pi}{2}}$ is an equivalent implementation of the underlying physical symmetry $R_{\frac{\pi}{2}}$ on gauge invariant operators:

\begin{equation}\label{equivbosonfermi}
    R_{\frac{\pi}{2}} \equiv U_{\frac{1}{4}}P_{\frac{\pi}{2}}.
\end{equation}

Table \ref{symtablejw} presents a list of the microscopic symmetries of the quantum spin-ice model and a corresponding equivalent symmetry operation acting on the JW-fermions. We see that in addition to the rotations, the natural fermionic implemention of diagonal mirrors $S_1$ and $S_2$ (see Fig.\ref{D8}) also need to be dressed by $U_{\frac{1}{4}}$ in order to make them equivalent to the underlying microscopic symmetries. We will also enforce Bravais lattice translational symmetries, which are understood to act identically on bosons and fermions (up to boundary terms) and thus are not listed explicitly in Table \ref{symtablejw}. Details of the derivations for these additional symmetries can be found in Appendix \ref{PSI}.

\begin{table*}
\centering
\begin{tabular}{c|c|c|c}
\toprule
Symmetry &Bare microscopic boson       &Auxiliary fermion transformation        &Equivalent JW fermion symmetry  \\
\hline
Time Reversal  &$\Theta$ &$\Theta$    &$ \Theta$          \\
\hline
\multirow{5}{*}{Spatial}
&$R_{\frac{\pi}{2}}$ &$P_{\frac{\pi}{2}}$ &$ U_{\frac{1}{4}} P_{\frac{\pi}{2}}$ \\
&$S_x$ &$\Sigma_x$ &$ \Sigma_x$ \\
&$S_y$ &$\Sigma_y$ &$ \Sigma_y$ \\
&$S_1$ &$\Sigma_1$ &$ U_{\frac{1}{4}} \Sigma_1$ \\
&$S_2$ &$\Sigma_2$ &$ U_{\frac{1}{4}} \Sigma_2$ \\ 
\hline
Particle-Hole &$X$ &$\Xi$            &$\Xi$\\
\bottomrule
\end{tabular}
 \caption{Summary of RK Hamiltonian symmetries and their implementation on bosons (spins) and JW fermions. The operations listed under the column ``bare microscopic boson'' are the underlying bare microscopic symmetries implemented on the  boson creation operators, for example the 90° rotation $R_{\frac{\pi}{2}}$ acts as defined in Eq.\eqref{Ppi2micro}. For each of these we introduce an  ``auxiliary fermion transformation'' which acts in a  simple way as expected for spinless fermions, such as the fermion rotation $P_{\frac{\pi}{2}}$ defined in Eq.\eqref{rotfer}. However, this auxiliary fermion transformation is not always equivalent to the ``bare microscopic boson'' symmetry (see Fig.\ref{commut} for notion of equivalence), and might need to be dressed by an extra site dependent $U(1)$ gauge transformation to make it equivalent, as listed under ``equivalent JW fermion symmetry'' (see Eq.\eqref{equivbosonfermi} as an example for the $R_{\frac{\pi}{2}}$ rotation and Fig.\ref{u14} for a definition of  $U_\frac{1}{4}$). The above ``equivalent JW fermion symmetries'' define only one possible extended projective symmetry group implementation on the JW/composite-fermions. Two other examples, that are the focus of this work, are described Table \ref{implsym}. In all the examples we implement the translations by Bravais vectors ${\bf R}_1,{\bf R}_2$  in the standard trivial non-projective way for bosons and fermions without dressing the auxiliary fermion transformations by gauge transformations.}
 \label{symtablejw}
\end{table*}

\vspace{0.10in}

This set of equivalent symmetries listed under the Jordan-Wigner fermion column of Table \ref{symtablejw} maps fermion bilinears onto fermion bilinears. Therefore, any such equivalent symmetry implementation, denoted by $S$, can be used to enforce the symmetry on the fermion mean-field Hamiltonian $H_{\text{MF}}$ of Eq.\eqref{mfhami}, by determining the hoppings that are satisfy by the following relation:

\begin{equation}
      S   H_{\text{MF}} S^{-1} =  H_{\text{MF}}.
\end{equation}

 Interestingly one can show that after enforcing all the equivalent symmetries from Table \ref{symtablejw} and Bravais lattice translations, there are no allowed nearest neighbor fermion hoppings in the lattice. For typical RK models, one expects that short distance correlations determine a sizable portion of the energy-density of the state and one would therefore expect that this projective symmetry implementation from Table \ref{symtablejw} is not a very energetically favorable choice for reasonably simple microscopic Hamiltonians. However, as mentioned before, the Fermionic symmetries listed in Table \ref{symtablejw} are only one choice among a large set of possibilities.

\begin{figure*}
  \centering
   \includegraphics[trim={0cm 6cm 0cm 0cm}, clip, width=0.95\textwidth]{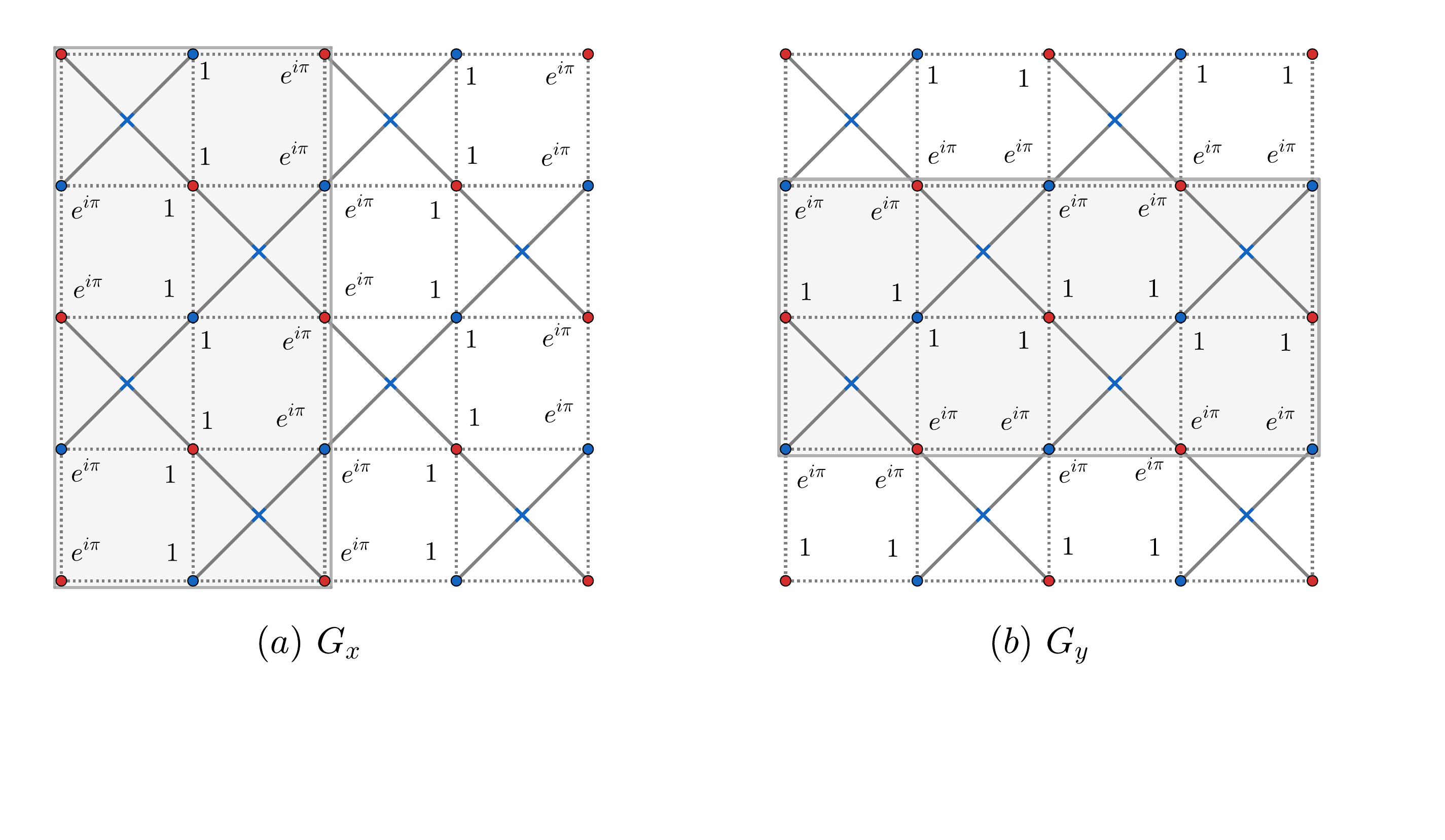}
   \caption{Phases gained by creation operators $f^\dag(\textbf{r})$ or  $b^\dag(\textbf{r})$ under the action of the UV gauge transformations: (a) $G_x$, (b) $G_y$. The transformations are obtained by choosing $\theta({\bf r})=\frac{\pi}{2}$ in Eq.\eqref{gaugetrasnfo} over the vertices contained in the gray regions and zero in the remainder.}
    \label{u141}
  \end{figure*}

It is therefore interesting to consider the following question: can we construct an alternative equivalent symmetry implementations that impose all the symmetries of the RK spin-ice model but which allow for the nearest neighbor hoppings to be non-zero? We have found two modified symmetry implementations that allow for non-zero nearest neighbor hoppings, which we shall denote as $\Theta_x$ and $\Theta_y$ implementations, and on which we focus for the remainder of the paper. These projective symmetry implementations are obtained by dressing the implementations of Table \ref{symtablejw} with the operations listed in Table \ref{implsym}, which are obtained after composition with the following UV lattice gauge group elements $G_x$ and $G_y$:
\begin{equation}\label{GxGy}
\begin{aligned}
              G_x b^{\dag}(\textbf{r}) G^\dagger_x &=  (-1)^{x} b^{\dag}(\textbf{r}).\\
              G_y b^{\dag}(\textbf{r}) G^\dagger_y &=  (-1)^{y} b^{\dag}(\textbf{r}).\\
\end{aligned}
\end{equation}
   Here we write the sites as $\textbf{r}=(x,y)$, where $x,y$ are understood to be integers. These two transformations can be viewed as generated by the UV lattice gauge transformations from Eq. \eqref{gaugetrasnfo}, by choosing $\theta(\textbf{r})$ so that it takes the values depicted respectively in  Fig.\ref{u141} $(a)$ and \ref{u141} $(b)$.

\begin{table}
\centering
\begin{tabular}{c|c|c}
\toprule
Symmetry &$\Theta_x$ (extended PSG) &$\Theta_y$ (extended PSG)   \\
\hline
Time Reversal  &$G_x \Theta$    &$G_y \Theta$      \\
\hline
\multirow{5}{*}{Spatial}
  &$U_{\frac{1}{4}} P_{\frac{\pi}{2}}$ &$U_{\frac{1}{4}} P_{\frac{\pi}{2}}$\\
&$G_x \Sigma_x$ &$G_y \Sigma_x$ \\
 &$G_x \Sigma_y$ &$G_y \Sigma_y$ \\
 &$G_y U_{\frac{1}{4}} \Sigma_1$ &$G_x U_{\frac{1}{4}} \Sigma_1$ \\
 &$G_y U_{\frac{1}{4}} \Sigma_2$ &$G_x U_{\frac{1}{4}} \Sigma_2$ \\ 
\hline
Particle-Hole &$G_y \Xi$ &$ G_x\Xi$  \\
\bottomrule
\end{tabular}
 \caption{The two distinct extended projective symmetry group implementations on the JW/composite-fermions that are the focus of this work. These symmetries are all equivalent to the microscopic symmetries listed in Tables \ref{symtable} and \ref{symtablejw} (see Fig-\ref{commut} for summary of notion of equivalence).}
 \label{implsym}
\end{table}

 

\subsubsection{Connection to pseudo-scalar spin liquids.}\label{pseudo}

So far we have used an implementation of microscopic symmetries which is more natural when we view the microscopic degrees of freedom as hard-core bosons, but which is not necessarily natural when we view them as spin-$\frac{1}{2}$. However, thanks to the large set of microscopic symmetries of the RK model of spin-ice, we are implicitly also enforcing symmetries whose action is the natural one when we view the microscopic degrees of freedom as spins.

For example the time-reveral operator $\Theta$, defined in Table \ref{symtable}, acts as complex conjugation in the standard choice of Pauli matrices where only $\sigma^y$ is imaginary, and $\sigma^{x,z}$ are real. Therefore it does not square to $-1$. The more standard time-reversal operator of spin-$\frac{1}{2}$, would act on the spin at site ${\bf r}$, as $\mathcal{T}=i\sigma^y({\bf r}) \Theta$. However, the operator $i\sigma^y({\bf r})$ is equivalent to composition $U(\pi) \sigma^x({\bf r})$, where $U(\pi)$ is a $\pi$ spin rotation around the z-axis, which acts on the fermions as:  $U(\pi)f^\dag(\textbf{r})U^\dagger(\pi)=-f^\dag(\textbf{r})$. Therefore $i\sigma^y({\bf r})$ is equivalent to a composition of the particle-hole $X$, implemented $\sigma^x({\bf r})$, and a boson global U(1) symmetry operation, implemented by $U(\pi)$, which we are already enforcing, namely, we have: 
\begin{equation}\label{spinTR}
    \mathcal{T}=U(\pi) X \Theta.
\end{equation}
Therefore, we are also implicitly enforcing such standard time-reversal action, $\mathcal{T}$, on spin-$\frac{1}{2}$, and one can similarly understand other natural spin symmetries of the RK model, as products of the natural boson symmetries that we are already enforcing. 

To determine the explicit action of $\mathcal{T}$ on JW/composite-fermions, let us first describe the action of $\Theta$. On spin raising/lowering operators this acts as a trivial anti-unitary operator (complex conjugation):
\begin{equation*}
    \Theta \sigma^+({\bf r})\Theta^{-1}=\sigma^+({\bf r}).
\end{equation*} 

Therefore this operator acts similarly on JW/composite-fermions:
\begin{equation}\label{parton}
    \Theta f^\dag(\textbf{r}) \Theta^{-1} = f^\dag(\textbf{r}).
\end{equation}

Let us now describe the implementation of the particle-hole conjugation of hard-core bosons, denoted by $X=\prod_{\bf r} \sigma^x({\bf r})$. From the action of this operator on spin operators, $X \sigma^+({\bf r})X^\dagger=\sigma^-({\bf r})$, one obtains the action on the JW/composite-fermions:
\begin{equation}\label{XFermionPH}
    X f^\dag(\textbf{r}) X^\dagger = (-1)^{L_s(\textbf{r})}f(\textbf{r}).
\end{equation}
Where $L_s(\textbf{r})$ is the length of the JW string. The factor $(-1)^{L_s(\textbf{r})}$ can be viewed as pure UV gauge transformation, and therefore $X$ is gauge equivalent to the natural JW/composite-fermion particle-hole conjugation, denoted by $\Xi$ (see Table \ref{symtablejw}), and defined as:
\begin{equation}\label{FermionPH}
    \Xi f^\dag(\textbf{r}) \Xi^\dagger = f(\textbf{r}).
\end{equation}
The spin time-reversal symmetry, $\mathcal{T}$, reverses the direction of all the spin components, and in particular the z-direction: $\mathcal{T} \sigma^z(\textbf{r}) \mathcal{T}^{-1} = - \sigma^z(\textbf{r})$. Therefore, since $\sigma^z(\textbf{r})$ encodes the information of the JW/composite-fermion, it is clear that $\mathcal{T}$ maps a fermion particle into a hole and viceversa. Therefore, we see that $\mathcal{T}$ is therefore a type of anti-unitary particle hole conjugation on the JW/composite-fermion operator, which explicitly reads as:
\begin{equation}\label{parton}
    \mathcal{T} f^\dag(\textbf{r}) \mathcal{T}^{-1} = (-1)^{L_s(\textbf{r})+1}f(\textbf{r}).
\end{equation}
Where $L_s(\textbf{r})$ is the length of the JW string, and the factor $(-1)^{L_s(\textbf{r})+1}$ is a pure UV gauge transformation identical to $G_x$ defined in Eq.\eqref{GxGy}\footnote{Assuming the lattice  has an even number of sites in the x-direction, which is natural for quantum spin-ice in a torus (see Fig.\ref{tHooft}).}. Because of the above we see the JW/composite-fermion behaves under $\mathcal{T}$ as a pseudoscalar spinon, in the sense defined in Ref. \cite{Inti}. 

We have introduced other space symmetries in their natural boson representation in Tables \ref{symtable} and \ref{symtablejw} that also would act as particle-hole conjugations on the  JW/composite-fermions when implemented as standard spin-$\frac{1}{2}$ symmetries. For example, for the space mirror operations $S_x,S_y,S_1,S_2$ $\sigma^z(\textbf{r})$ transform as a scalar, e.g.: $S_y \sigma^z(\textbf{r}) S_y^{-1}=\sigma^z(S_y\textbf{r})$. However, its spin  spin version would include an additional boson particle-hole conjugation, leading to the standard action of spins which are pseudo-vectors under mirrors, and which would reverse $\sigma^z(\textbf{r})$ because it is parallel to these mirror planes. Therefore, these mirrors act as unitary particle-hole conjugations on the JW/composite-fermions, and the spin liquid states that we will be discussing in this paper can be viewed as pseudo-scalar spin liquids with respect to spin implementations of time-reversal and space mirror symmetries in the sense defined in Ref.\cite{Inti}.\\

\subsubsection{Dirac and Fermi surface mean-field states for the six-vertex model and quantum dimer models}\label{C2}

    The procedure described in the previous section allows us to fix the nearest neighbour hopping amplitudes in Eq.\eqref{mfhami}. The resulting pattern of hoppings is illustrated in fig. \ref{mfhop1}, and the corresponding mean-field hamiltonian reads as:

\begin{widetext}
      \begin{equation}\label{hami12345}
  \begin{aligned}
      H_{\text{MF}} = &\sum_{\textbf{R}} i t^* \big(f^{\dag}_{a}( \textbf{R} - \textbf{R}_1+ \textbf{R}_2) f_b (\textbf{R}) +   f^{\dag}_{a} (\textbf{R}) f_b (\textbf{R}) \big) +t \big( f^{\dag}_{a} (\textbf{R}   -  \textbf{R}_1) f_b (\textbf{R}) +   f^{\dag}_{a} (\textbf{R} + \textbf{R}_2) f_b (\textbf{R}) \big)+ h.c.
      \end{aligned}
  \end{equation}
  \end{widetext}

Which in crystal momentum basis can be re-expressed as:

   \begin{equation*}
       H_{\text{MF}} = \sum_{\textbf{q} \in \text{BZ}}  \begin{pmatrix}
       f^{\dag}_{a} (\textbf{q}) & f^{\dag}_{b} (\textbf{q})
       \end{pmatrix}  \begin{pmatrix}
       0 &h_{ab}(\textbf{q})\\
       h_{ab}^*(\textbf{q}) &0
       \end{pmatrix} \begin{pmatrix}
       f_{a} (\textbf{q}) \\ f_{b} (\textbf{q})
       \end{pmatrix},
   \end{equation*}
   
   where we are using the crystal momentum basis $f^{\dag}_{a} (\textbf{R})= N_{\Lambda}^{-1/2} \sum_{\textbf{q} \in \text{BZ}} e^{-i \textbf{q} \cdot \textbf{R}} f^{\dag}_{a} (\textbf{q})$, and the matrix entry is:

   \begin{equation*}
       h_{ab}(\textbf{q}) = 2 e^{  \frac{i}{2} (q_1-q_2) } \bigg[ i t^*  \,    \cos\big( \frac{q_1-q_2}{2} \big)    + t \,    \cos\big( \frac{q_1+q_2}{2} \big) \bigg].  
   \end{equation*}

Where $q_i={\bf q}\cdot{\bf R_i}$, $i=1,2$. The associated band energy dispersions is:

   \begin{equation}\label{dispersionrelation}
        \epsilon_{\pm}(\textbf{q}) = \pm 2 |t| \sqrt{\cos\big(\frac{q_1-q_2}{2}\big)^2 + \cos\big(\frac{q_1+q_2}{2}\big)^2}.
    \end{equation}

\begin{figure*}
 \centering
   \includegraphics[trim={0cm 0cm 0cm 0cm}, clip, width=\textwidth]{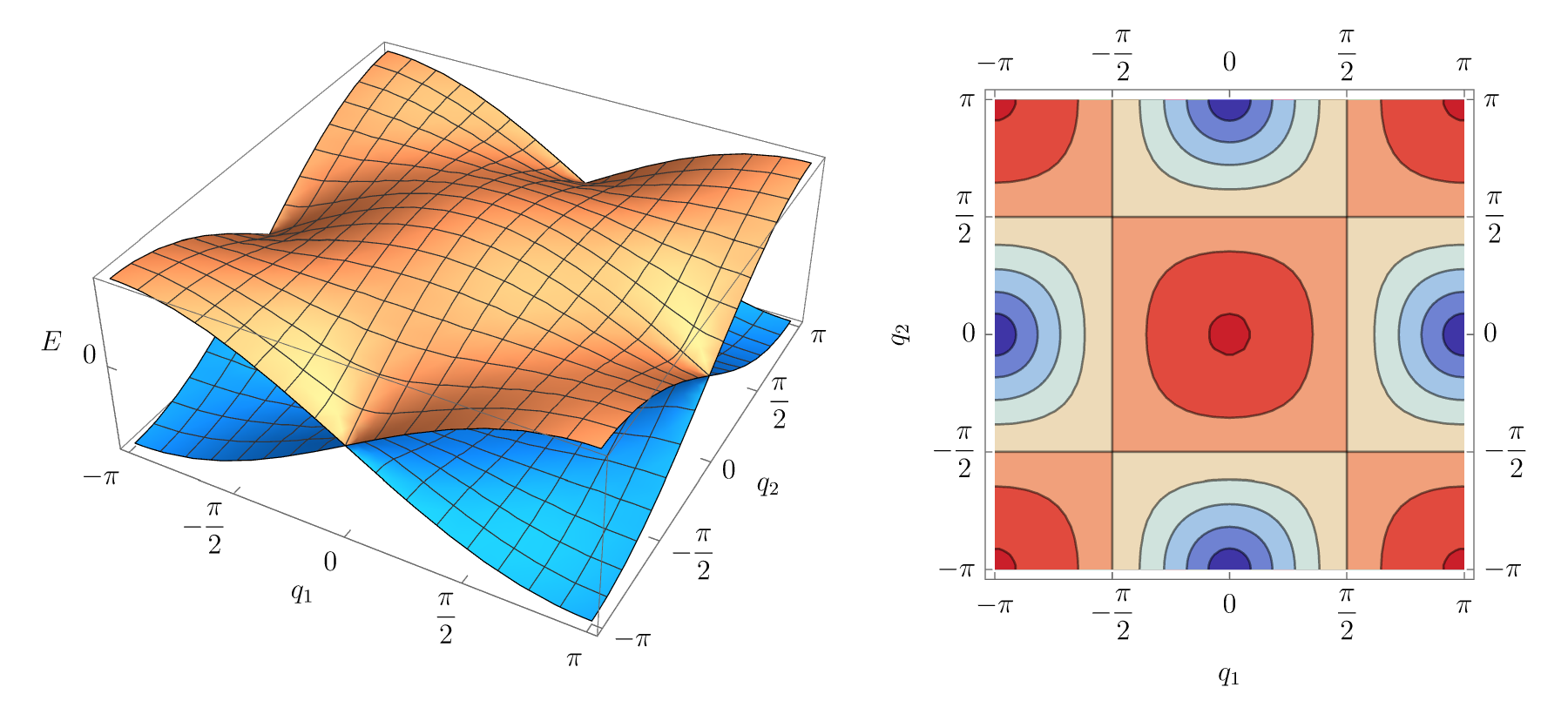}
   \caption{Dispersions of the JW/composite-fermions associated with the extended projective symmetry groups $\Theta_x$, $\Theta_y$ (see Table \ref{implsym} for their definitions, Fig.\ref{mfhop1} for the hoppings, and Eq.\eqref{dispersionrelation} for dispersion). There are two Dirac cones at $(0,\pi)$ and $(\pi,0)$ and a Fermi surface at $\frac{1}{4}$-filling which is perfectly nested, and correspons to the straight lines separating blue from red regions in the equal energy contours shown in the right panel. The crystal momenta is defined as $q_{1,2}={\bf q}\cdot{\bf R}_{1,2}$ and therefore measure along the ${\bf R}_{1,2}$ directions (see Fig.\ref{mfhop1}).}
   \label{fsfil}
  \end{figure*}

These bands are illustrated in Fig.\ref{fsfil}. The two extended projective symmetry implementations $\Theta_x$ and $\Theta_y$ (see table \ref{implsym}) impose different constraints on the hopping amplitude $t$ be either purely real or purely imaginary:

\begin{equation*}
       \begin{cases}      t=t^*\,\,\,\,\,\,\,\,\,\text{for the} \,\,\Theta_x\\
          t=-t^*\,\,\,\,\text{for the}\,\, \Theta_y
       \end{cases}
   \end{equation*}


Details on how the above follows from implementing the symmetries from Table \ref{implsym} are shown in Appendix \ref{PSI}.\\

Despite their similarity, the extended projective symmetry group implementations $\Theta_x$ and $\Theta_y$ (see Table \ref{implsym}) are inequivalent. This can be seen by considering the action of a particular unitary transformation, denoted by $V_{\frac{1}{2}}$, which acts on the fermion operator, $f^\dag(\textbf{r})$, as a local site dependent U(1) transformation multiplying it by the specific phases shown in Fig.\ref{u14}(b).  It turns out that $V_{\frac{1}{2}}$ maps the $\Theta_x$ mean-field Hamiltonian onto the $\Theta_y$ mean-field Hamiltonian, as can be seen from its action of the following fermion bilinears (the $1,2,3,4$ sub-indices below are the sites shown in Fig.\ref{mfhop1}):

 \begin{equation}\label{V1/2}
 \begin{aligned}
        t f_3^{\dag} f_1 &\to it f_3^{\dag} f_1\\
        -it f_1^{\dag} f_2 &\to t f_1^{\dag} f_2 \\
         t f_2^{\dag} f_4 &\to it f_2^{\dag} f_4\\
         -it f_4^{\dag} f_3 &\to  t f_4^{\dag} f_3 
        \end{aligned}
   \end{equation}

  On the other hand, the operators $L_{\textbf{R}}$ and $L_{\textbf{R}}^{\dag}$ that enter in the microscopic RK Hamiltonian (see Eq.\eqref{Lsquare}) can be shown to be odd under the action $V_{\frac{1}{2}}$. Since $L_{\textbf{R}}$ is a invariant under the UV Lattice Gauge group, it follows that $V_{\frac{1}{2}}$ is not a pure gauge transformation but a transformation with non-trivial action within the gauge invariant subspaces, and therefore the $\Theta_x$ and $\Theta_y$ mean-field Hamiltonians are not gauge equivalent, but rather realizing to two physically distinct  generalized projective symmetry group implementations. This implies that only one them will have lower energy as a trial ground state for a specific  microscopic RK Hamiltonian. Since the plaquette resonance term $L_{\textbf{R}}$ is odd under $V_{\frac{1}{2}}$, the one that is more energetically favorable will be determined by the sign of the plaquette resonance term in the microscopic Hamiltonian\footnote{Notice that $(V_{\frac{1}{2}})^2$ would map both the $\Theta_x$ and $\Theta_y$ mean-field Hamiltonians into minus themselves. However, $(V_{\frac{1}{2}})^2$ leaves all the gauge invariant operators unchanged, and it is therefore an element of UV gauge group. Therefore we see that changing the global sign of $t$ in either the $\Theta_x$ or $\Theta_y$ mean-field Hamiltonians leads to the same physical state.}.


%
  
  
  As described in Sec.\ref{QSIJW}, for the cases of the $\mathcal{H}_{\text{Q6VM}}$ and $\mathcal{H}_{\text{QDM}}$, the system is respectively at half-filling and quarter filling, therefore, as depicted in Fig.\ref{fsfil}, these systems have a mean field dispersion featuring two massless Dirac cones and a Fermi surface, respectively.    
   The Dirac points are located at ${\bf q}_0=(q_1,q_2) = (\pi,0)$ and ${\bf q}_0=(q_1,q_2) = (0,\pi)$. By writing ${\bf q}={\bf q}_0+{\bf p}$ and expanding the mean-field Hamiltonian to linear order in the momentum ${\bf p}$, we obtain the following effective Dirac Hamiltonian for the $\Theta_x$ extended PSG ($t \in \mathbb{R}$) is:

 \begin{equation}\label{linhamire}
      \begin{aligned}
          h({\bf q}_0+{\bf p}) \simeq v \begin{cases}
                p_x \tau^x + p_y \tau^y \,\,\,\,\,\text{for}\,\, {\bf q}_0 = (\pi, 0) \\
               p_x \tau^x - p_y \tau^y \,\,\,\,\,\text{for}\,\,   {\bf q}_0 = (0, \pi) \\
          \end{cases}
      \end{aligned}
  \end{equation}

where $\tau^{x,y}$ are Pauli matrices in the $a/b$ sublattice space, $v=\sqrt{2}t|{\bf R}_1|$, and $p_x=({\bf p}\cdot\hat{{\bf R}}_1-{\bf p}\cdot\hat{{\bf R}}_2)/\sqrt{2}|{\bf R}_1|, p_y=({\bf p}\cdot\hat{{\bf R}}_1+{\bf p}\cdot\hat{{\bf R}}_2)/\sqrt{2}|{\bf R}_1|$. On the other hand, for the $\Theta_y$ extended PSG ($t \in  i \mathbb{R}$) the linearized Hamiltonian is:

\begin{equation}\label{linhamire2}
      \begin{aligned}
          h({\bf q}_0+{\bf p}) \simeq v \begin{cases}
               p_x \tau^y + p_y \tau^x \,\,\,\,\,\text{for}\,\, {\bf q}_0 = (\pi, 0) \\
               p_x \tau^y - p_y \tau^x \,\,\,\,\,\text{for}\,\,   {\bf q}_0 = (0, \pi) \\
          \end{cases}
      \end{aligned}
  \end{equation}



 where $v=-i \sqrt{2}t |{\bf R}_1|$. 

 \vspace{0.05in}

  \begin{figure}
  \centering
   \includegraphics[trim={2cm 0cm 2cm 0cm}, clip, width=0.45\textwidth]{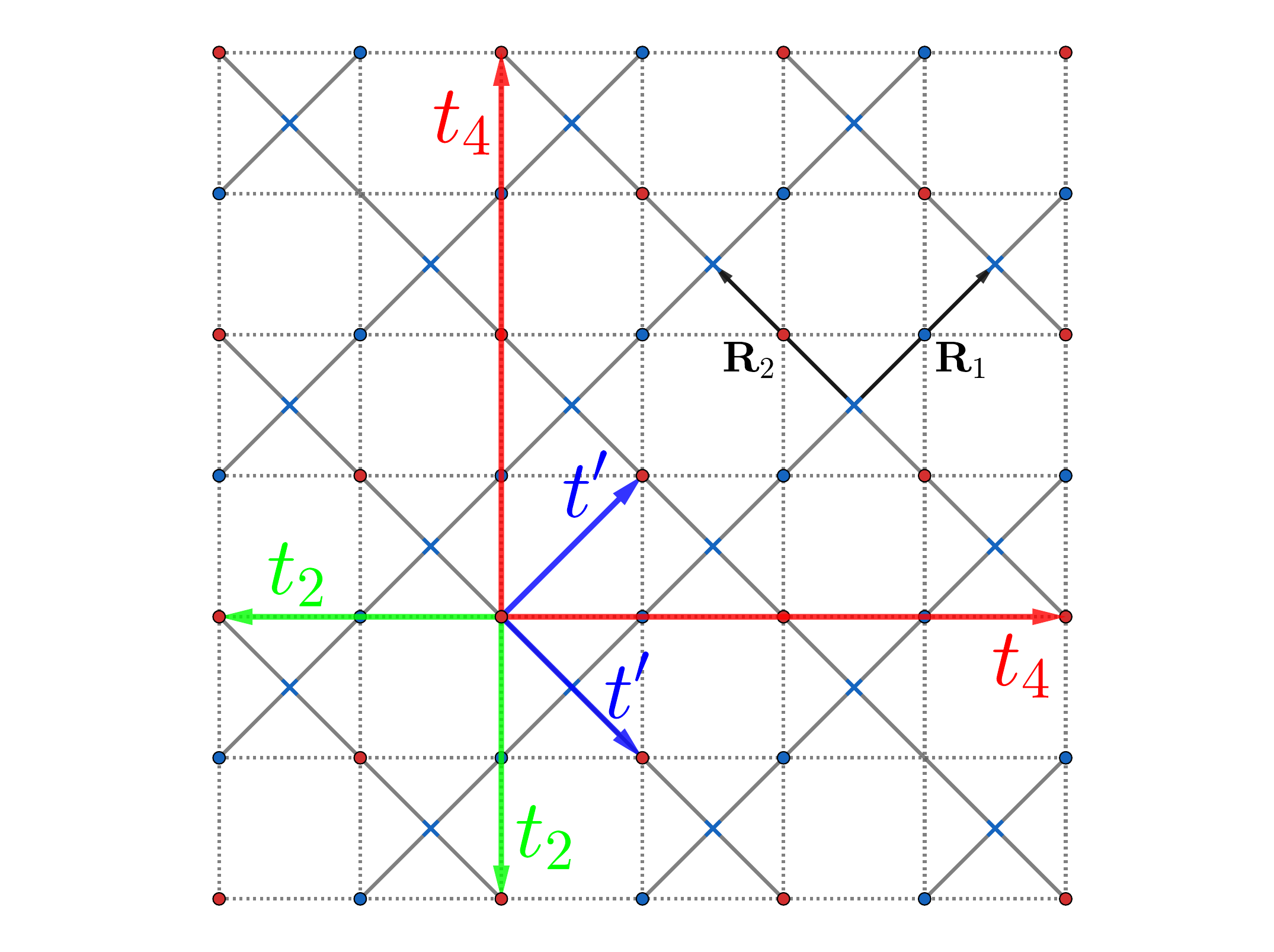}
   \caption{Further neighbour hoppings considered to deform the Fermi surface at $\frac{1}{4}$-filling and remove its perfect nesting. $t'$ is forbidden by the $\Theta_x$ and $\Theta_y$ extended projective symmetry groups. $t_2$ and $t_4$ are both allowed, but $t_2$ alone does not remove the perfect nesting (see Eq.\eqref{hoppingt2}). Such resilience of the nesting is indicative of the fragility of the JW/composite-fermi liquid state at $\frac{1}{4}$-filling, and thus it might be a useful parent to understand the competing confined broken symmetry states of the RK-like quantum dimer models (e.g. columnar and resonant plaquette states).} 
   \label{fh}
  \end{figure}
 
\begin{figure}
  \centering
   \includegraphics[trim={0cm 0cm 0cm 0cm}, clip, width=0.50\textwidth]{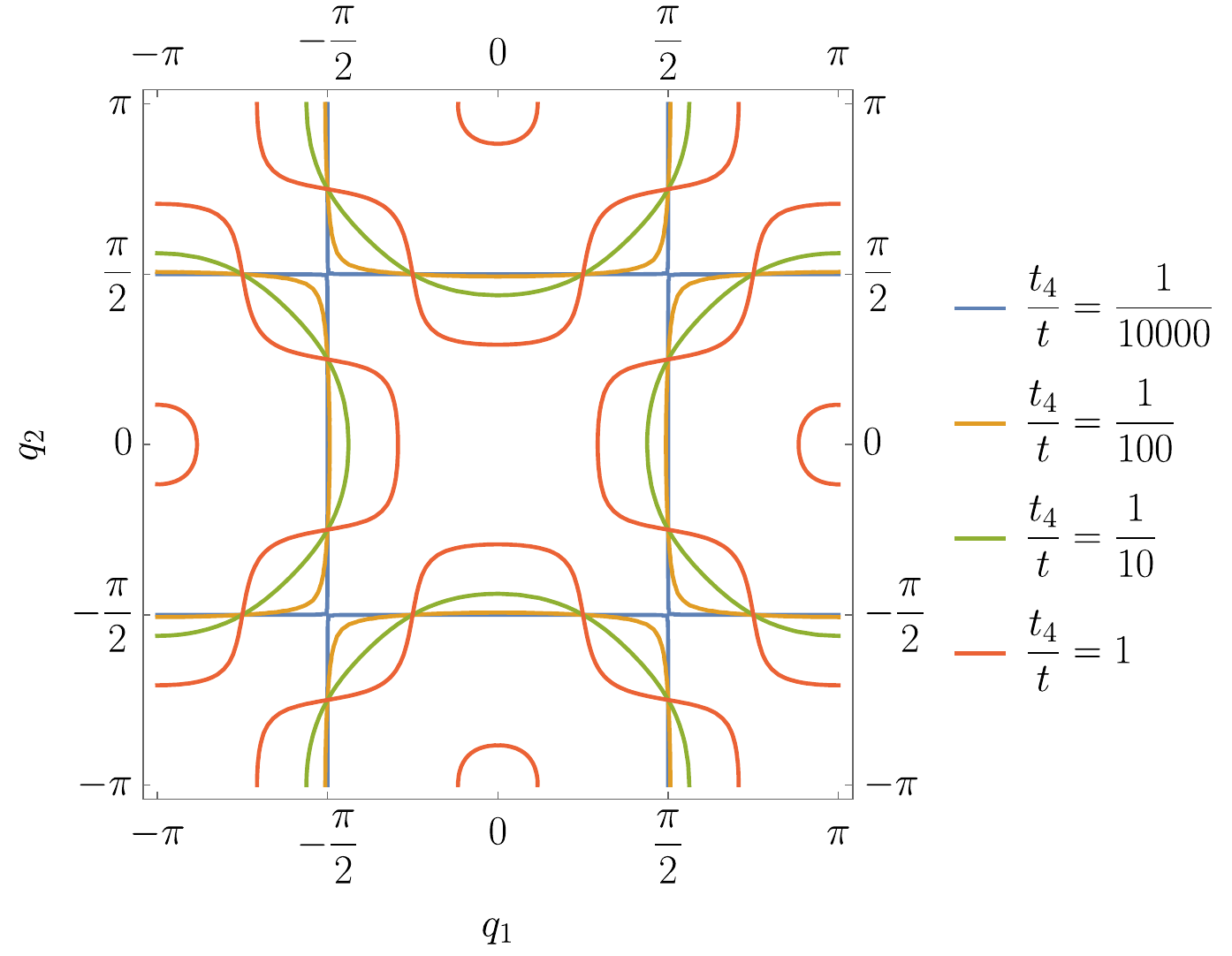}
   \caption{Fermi Surfaces at quarter-filling for various values of $t_4/t$, illustrating the lifting the perfect nesting of the JW/composite-fermi surface associated with the QDM model,  by adding this further-neighbor hopping. }
   \label{lift}
  \end{figure}

 Now for the case of the subspace of the QDM model which corresponds to a quarter filling of the bands by the JW/composite fermions, there is a Ferrmi surface that consists of straight lines that are perfectly nested by $(\pi,0)$ and $(0,\pi)$ vectors (see Figs. \ref{fsfil},\ref{lift}). This indicates that such putative composite Fermi liquid state would be highly unstable towards forming a state which spontaneously breaks the lattice translational symmetry and gaps the Fermi surface. This perfect nesting occurs only for the strict nearest neighbor mean-field Hamiltonian, and therefore can be removed by adding longer range hoppings which are allowed by the extended projective symmetry implementations under consideration ($\Theta_{x},\Theta_{y}$ from Table \ref{implsym}). 
 To illustrate this, we consider the further neighbor hoppings depicted in Fig.\ref{fh}. One can show that the second neighbor hopping, denoted by $t'$ and depicted by blue arrows in Fig.\ref{fh}, vanishes for the $\Theta_{x},\Theta_{y}$ symmetry implementations. The further neighbor hopping denoted by $t_2$ and depicted by green arrows in Fig.\ref{fh} is allowed by $\Theta_{x},\Theta_{y}$ symmetry implementations, and it leads to the following sublattice-diagonal entries to the mean-field Hamiltonian:

 \begin{equation}\label{hoppingt2}
    {\Delta h_2}({\bf q})_{ab} \sim  t_2 [\cos( q_1 + q_2) + \cos( q_1 - q_2)] \delta_{ab}.
   \end{equation}


 
However, since $\cos(q + \pi/2) + \cos(q - \pi/2)=0$, the above correction vanishes exactly along the lines that define the nested Fermi surface (see Figs.\ref{fsfil}  and \ref{lift}), and therefore does not remove the perfect nesting. Nevertheless there are symmetry-allowed hoppings that lifts the nesting. One of them is denoted by $t_4$ and depicted in Fig.\ref{fh} by the red arrows. This hopping adds the following sublattice-diagonal entries to the mean-field Hamiltonian:
   \begin{equation*}
      \Delta h_4({\bf q})_{ab} \sim t_4 [\cos(2 q_1 + 2 q_2) + \cos(2 q_1 - 2 q_2)] \delta_{ab}.
   \end{equation*}
   Figure \ref{lift} illustrates how the perfect nesting is destroyed as $t_4$ increases relative to $t$, leading to a composite Fermi liquid state with two Fermi surfaces centered around $(\pi,0)$ and $(0,\pi)$. The above illustrates that a composite fermi liquid state at $\frac{1}{4}$-filling could be in principle a true stable spin liquid ground state. However the strong resilience of the nesting to near neighbor hopping corrections is indicative that the Fermi surface has a strong tendencies to be  gapped out and destroyed via instabilities of composite-fermion particle-hole  pair condensation with finite crystal momentum, leading to ordinary confined states with spontaneously broken lattice translational symmetries, such as the columnar, staggered and resonant plaquette phases that are believed to be typically realized for RK Hamiltonians of quantum dimer models.


   \vspace{0.10in}
    

We would like to close this subsection by noting that our mean-field states associated with the $\Theta_{x},\Theta_{y}$ extended projective symmetry groups have a resemblance to the classic $\pi$-flux state of standard Abrikosov-Schwinger fermions introduced in Refs. \cite{PhysRevB.37.3774,PhysRevB.37.3664}. In fact from  Fig.\ref{mfhop1}, we see that the fermions are hopping around every plaquette of the original square lattice (which are now subdivided into vertices and plaquettes of the ``spin-ice'' lattice)  accumulating a phase $\pi$ over the closed loop. There are, however, several crucial physical differences with the classic $\pi$-flux state of Abrikosov-Schwinger fermions. First, the classic $\pi$-flux state is a spin singlet in which each spin specie of Abrikosov-Schwinger fermions 
 has the same hoppings in the square lattice, whereas in our construction the JW/composite-fermions are spin-less with only one fermion specie hopping around the plaquette, in a state that is not a spin singlet\footnote{The RK Hamiltonian of quantum spin-ice is highly anisotropic in spin space and far from having SU(2) symmetry.}. More fundamentally, an onsite U(1) transformation such as the $V_{\frac{1}{2}}$ transformation defined in Fig.\ref{u14}, would be an element of the UV parton gauge of Abrikosov-Schwinger fermions and therefore, the analogue of the $\Theta_{x},\Theta_{y}$ projective symmetry groups would be two physically equivalent states for the classic $\pi$-flux state of Abrikosov-Schwinger fermions. Generally speaking symmetries are much more constraining for the JW/composite-fermions relative to Abrikosov-Schwinger fermions, as they fix the phases of hopping and different phase might lead to physically distinct states.

Nevertheless, the fact that our mean-field hamiltonians of JW/composite-fermions can be viewed as states with $\pi$-flux in each plaquette of the original square lattice is useful for understanding the properties of the mean-field states. For example, for a $\pi$-flux mean-field state there exists an intra-unit cell magnetic translation that is not part of the spin-ice Bravais lattice, which can be taken to be a translation by $(\textbf{R}_1+\textbf{R}_2)/2$ (see Fig.\ref{mfhop1}). This magnetic translation would anti-commute with the ordinary elementary translations along either of the two basis vectors of the Bravais lattice $\textbf{R}_1,\textbf{R}_2$, because the parallelogram spanned by  $\textbf{R}_1$ and $(\textbf{R}_1+\textbf{R}_2)/2$ encloses $\pi$-flux, and similarly for the parallelogram spanned by  $\textbf{R}_2$ and $(\textbf{R}_1+\textbf{R}_2)/2$. As a consequence this magnetic translation boosts the standard crystal momentum by $(q_1,q_2)\rightarrow (q_1+\pi,q_2+\pi)$, and this explains why the mean-field fermion dispersions that we have found display this translational symmetry in momentum space (see Figs.\ref{fsfil},\ref{fh}). However, while this magnetic translation by $(\textbf{R}_1+\textbf{R}_2)/2$ is a symmetry of the unprojected mean-field Hamiltonian, this cannot be a symmetry of the microscopic RK-model of quantum spin-ice, because a translation by $(\textbf{R}_1+\textbf{R}_2)/2$ would map spin-ice vertices onto spin-ice plaquettes, which are clearly distinct in the RK model, and in any typical model with the same spin-ice rules, since the ice rules themselves are incompatible with a symmetry that would exchange vertices and plaquettes (except for trivial models without quantum fluctuations). However this symmetry of the bare-unprojected mean-field state would not be present for the full physical trial state obtained after the spin-ice Gutzwiller projection, because the Gutzwiller projection operator from Eq.\eqref{proiector} is not invariant unders such translation by $(\textbf{R}_1+\textbf{R}_2)/2$, since it is defined by projecting onto the spin-ice rules associated with the vertices. As we will see the effective Hamiltonian capturing the gauge field fluctuations that we will discuss in the next section, in fact does not have any associated translational symmetry by $(\textbf{R}_1+\textbf{R}_2)/2$, and thus this symmetry of the bare mean-field state will be lifted by gauge fluctuations.

\section{Gauge field fluctuations and effective low energy continuum field theory}
\label{gauge}

The Gutzwiller projection is a non-trivial operation that substantially changes the character of the un-projected mean-field state. Computing analytically the properties of the projected state is a however a highly non-trivial task. In a sense, this projection can be viewed as giving rise to the appearance of strong gauge field fluctuations around the mean-field state \cite{Wen, WenA02, WenB02}, and, accounting for such fluctuations is necessary to capture, even qualitatively, the correct behavior of the phase of matter in question at low energies. \\

The mean-field description introduced in the previous section still conceals the emergence of low-energy dynamical gauge fields which can be viewed as arising from fluctuations of the hopping amplitudes of the mean-field state in Eq. \eqref{mfhami}. While a desciption of these gauge field fluctuations is often performed by enforcing constraints and performing saddle point expansions in a path integral representation (see e.g. Ref.\cite{lee2005u,hermele20072}), we will device here a more phenomenological approach to infer the field content and emergent gauge structure of the low energy field theory that describes the phase of matter for the itinerant liquids of JW/composite-Fermions associated with the mean field states constructed in the previous section. \\

    We will include only the fluctuations of the phases of the complex hoppings $t(\textbf{r},\textbf{r}')$ of the mean field Hamiltonian (see Eq.\eqref{mfhami}) but not the fluctuations their amplitudes, because we assume that the latter can be viewed as being gapped and thus not important at low energies. To capture the fluctuations of such phases, we introduce additional bosonic degrees of freedom associated with the non-zero hopping, $t(\textbf{r},\textbf{r}')$, of the mean field state from Eq.\eqref{mfhami}, that connect a pair of fermion lattice sites ${\bf r}, \textbf{r}'$. We denote the deviation of the phase from its mean field value by $A(\bf{r},\bf{r}')$, and we promote the mean field Hamiltonian to a new Hamiltonian including this phase fluctuation variables as follows:
 
 \begin{equation}\label{psub}
    \begin{aligned}
    H[t] &\mapsto H[t,A],\\
           t(\textbf{r},\textbf{r}') f^{\dag}(\textbf{r}) f(\textbf{r}') &\to  t(\textbf{r},\textbf{r}') f^{\dag}(\textbf{r})  e^{i A(\textbf{r},\textbf{r}')}  f(\textbf{r}').
    \end{aligned}
    \end{equation}

The scalar phase $A(\textbf{r},\textbf{r}')$ can be interpreted as a lattice version of  $\int_{\textbf{r}'}^{\textbf{r}} \textbf{A} \cdot d \textbf{x}$. Notice that hermiticity demands that $A(\textbf{r},\textbf{r}')=-A(\textbf{r},\textbf{r}')$ and $t(\textbf{r},\textbf{r}')=t^*(\textbf{r}',\textbf{r})$. The above Hamiltonian describes the coupling of the matter fields to the gauge fields, and therefore we need to provide another Hamiltonian for the ``pure" gauge field sector. This Hamiltonian can be otained by demanding invariance under a generalized version of lattice UV gauge structure and simple symmetry considerations. In the next section \ref{3A}, we will review this construction first for the case of usual Abrikosov-Schwinger partons and subsequently in section \ref{3A} we will apply it to the case of the extended parton constructions for quantum spin-ice.

\subsection{Review of Gauge field fluctuations for $U(1)$ spin liquids from standard parton constructions}\label{3A}

In this section we will derive the effective field theory governing a U(1) spin liquid associated with the standard Abrikosov-Schwinger parton mean field states (see section \ref{B}). 
The conclusion in this section will be simple and well established before, namely, when the spin-liquid state associated with a given mean-field parton state is stable, the low-energy de-confined gauge structure will be given by the invariant gauge group (IGG) \cite{WenA02, WenB02}. We will illustrate this for a mean-field parton state with a global U(1) particle-conservation symmetry, and thus a U(1) IGG, leading, therefore, to a low-energy U(1) gauge group minimally coupled to the parton fermions (i.e. an standard U(1) spin-liquid). We wish, however, here to rederive these results in what is hopefully a more conceptually intuitive construction, so that we can use it as a template of reasoning for deriving the new results of the emergent low energy gauge structure of our extended parton constructions of JW/composite-fermion states in the next section. \\

We begin by promoting the phases of the hoppings into dynamical degrees of freedom and the mean field Hamiltonian from Eq.\eqref{Hamiwen}, into the following Hamiltonian capturing the field matter coupling:

\begin{equation}\label{hamipe}
      H[t, A] \doteq \sum_{s,s'}\sum_{ \textbf{r},\textbf{r}'} t_{ss'}(\textbf{r},\textbf{r}')  e^{i A(\textbf{r},\textbf{r}')}f^{\dag}_s(\textbf{r}) f_{s'}(\textbf{r}'), 
\end{equation}

Here $A(\bf{r},\bf{r}')$ is viewed as a dynamical compact  periodic phase taking values $A(\textbf{r},\textbf{r}') \in [0,2 \pi)$. We would like now to define an extension of the local UV parton gauge symmetry, but which acts not only on the fermions but also on the dynamical gauge fields 
$A(\bf{r},\bf{r}')$. As discussed in section \ref{B}, the local U(1) transformations of the parton gauge group are generated by the local fermion occupations $n(\textbf{r})$, which transform the fermion bilinears as (see Eq.\eqref{Hamiwen}):

\begin{equation}\label{gaugeASF}
    f^{\dag}_s(\textbf{r})f_s(\textbf{r}') \xrightarrow[]{\text{Gauge}}  e^{-i [\theta(\textbf{r})- \theta(\textbf{r}')]}f^{\dag}_s(\textbf{r})f_s(\textbf{r}')
\end{equation}

Therefore, in order to leave the Hamiltonian from Eq. \eqref{hamipe} invariant, we demand that these transformations act on the dynamical phase gauge degrees of freedom as follows:
\begin{equation}\label{gaugetrasnfoA}
      A(\textbf{r},\textbf{r}') \xrightarrow[]{\text{Gauge}} A(\textbf{r},\textbf{r}') + \theta(\textbf{r})- \theta(\textbf{r}')  
\end{equation}

For simplicity, from now on we will assume that the hoppings only connect nearest neighbour sites $\textbf{r}$ and $\textbf{r}'=\textbf{r} + \textbf{e}_i$ (with $i=x,y$) and we will label the bond connecting them by $(\textbf{r}, i)$, and the gauge fields by $A({\bf r},i)$. To implement the transformation from Eq.\eqref{gaugetrasnfoA} quantum-mechanically, we introduce a canonically conjugate variable to the vector potentials denoted by $E({\bf r},i)$, and take these variables to satisfy the following commutation relations:

\begin{equation}\label{CCR}
\begin{cases}
   [A(\textbf{x},i), E(\textbf{y},j)  ] &= -i \delta_{\textbf{x},\textbf{y}}\delta_{ij} \\
    [E(\textbf{x},i), E(\textbf{y},j)  ] &= 0\\
    [A(\textbf{x},i), A(\textbf{y},j)  ] &= 0\\
\end{cases}
\end{equation}

Since $A(\textbf{x},i)$ is an angle, $E(\textbf{x},i)$ is an angular momentum with integer-valued spectrum. It is then easy to show, that the generalized UV gauge transformations acting on matter and dynamical phase gauge fields are generated by exponentials of the following operators:

 \begin{equation}\label{nochargediscrete}
 \begin{aligned}
 G(\textbf{r}) &= n(\textbf{r}) - \nabla \cdot E(\textbf{r}).
      \end{aligned}
  \end{equation}

  where:

  \begin{equation}\label{divE}
      \nabla \cdot E(\textbf{r}) =
      E(\textbf{r},x) + E(\textbf{r},y) - E(\textbf{r} - \textbf{e}_x,x) - E(\textbf{r} - \textbf{e}_y,y) 
  \end{equation}
 
\begin{figure}
  \centering
   \includegraphics[trim={5cm 10cm 0cm 6.5cm}, clip, width=0.70\textwidth]{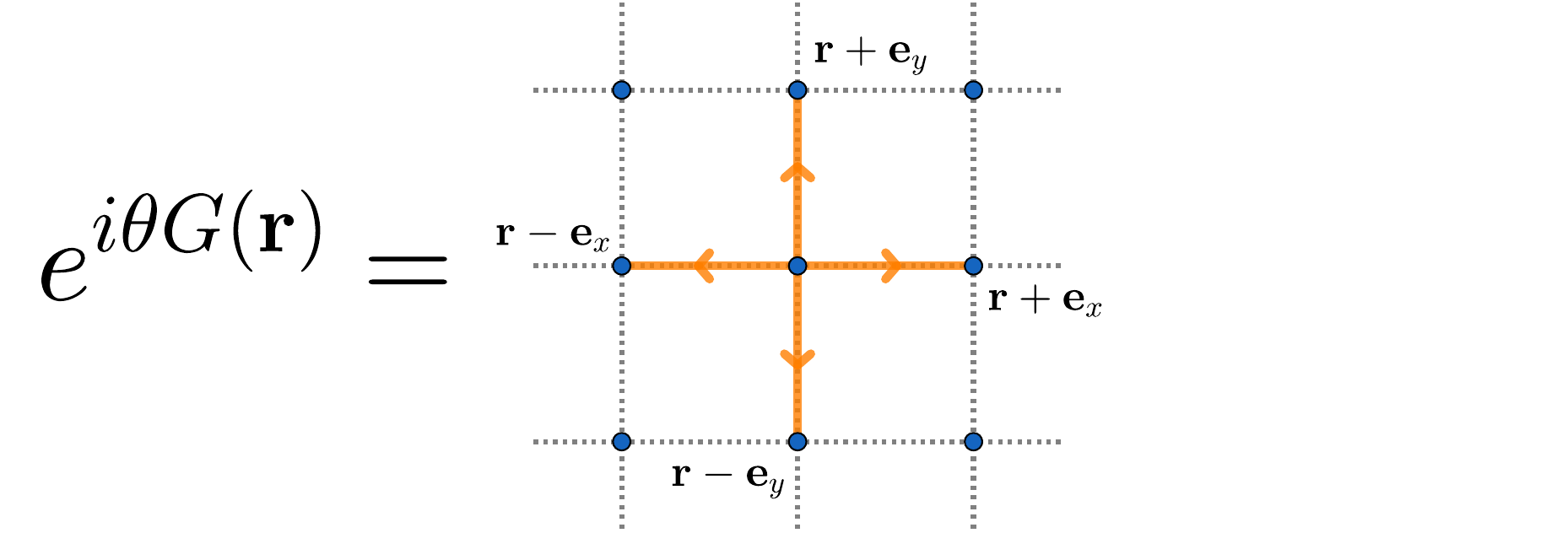}   \caption{Depiction of generator of generalized gauge transformations (see Eq.\eqref{nochargediscrete}) acting on the matter (residing on blue sites) and vector potentials (residing on links), relevant for the emergent lattice U(1) gauge theory of standard Abrikosov-Schwinger partons.}
   \label{gauss1}
  \end{figure}

We will demand that the combined effective Hamiltonian of matter and gauge fields is invariant under the above local gauge group, and we will interpret then the values of $G(\textbf{r})$ as a constraint that can be consistently imposed on the states in order to represent the subspace of physical interest. The subspace of physical interest will be that for which $G(\textbf{r})=0$ for all $\textbf{r}$, and therefore this constraint can be viewed as a lattice version of Gauss's law (see Fig.\ref{gauss1} for a depiction).
\\

Let us now determine the simplest operators that are made only from gauge fields which commute with every $G(\textbf{r})$. It is easy to verify that one of them is the magnetic field operator associated with the curl of $A$ around a plaquette: 

\begin{equation}\label{B}
    B(\textbf{r}) \doteq A(\textbf{r},x) + A(\textbf{r}+\textbf{e}_x,y) - A(\textbf{r}+\textbf{e}_y,x) - A(\textbf{r},y) 
\end{equation}

Here we view the plaquette of interest as being northeast from lattice site $\textbf{r}$, and thus we are using this as a label of the plaquette as well. The canonically conjugated variable to $B(\textbf{r})$ can be shown to be the lattice curl of $E$:

    \begin{equation*}
    \nabla \times E( \textbf{r}) \doteq E(\textbf{r},x) + E(\textbf{r}+\textbf{e}_x,y) - E(\textbf{r}+\textbf{e}_y,x) - E(\textbf{r},y) 
\end{equation*}

\begin{figure}
  \centering
   \includegraphics[trim={35cm 11cm 7cm 7cm}, clip, width=0.3\textwidth]{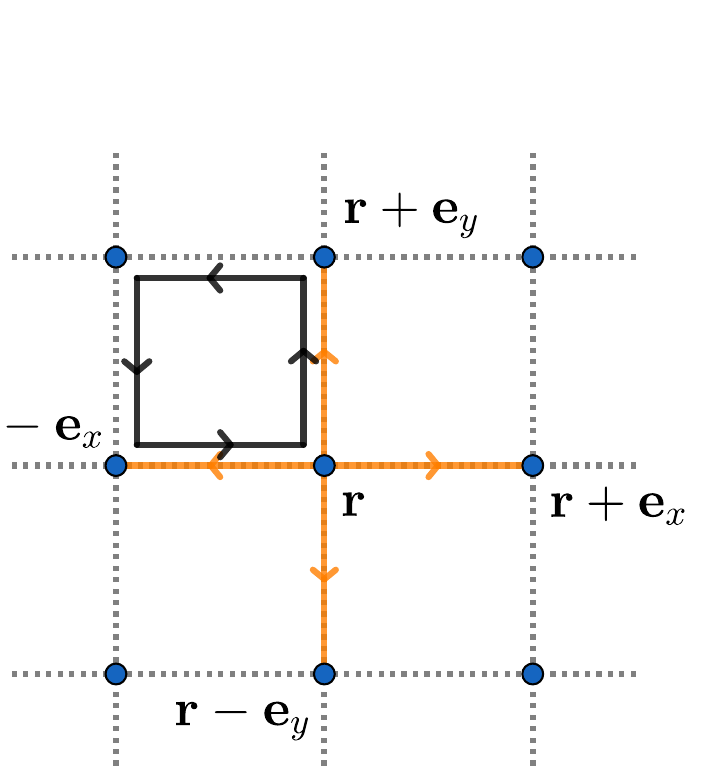}
\caption{The solid black lines depict the sum of vector potentials that enter the definition of the magnetic flux operator $B(\textbf{r}-\textbf{e}_x)$ (from Eq.\eqref{B}). The orange lines depict the sums of electric fields that enter the divergence operator (from Eq.\eqref{divE}). The fact that these two operators commute can be visualized by noting that the number of segments in which parallel black and orange arrows overlap equals the number of segments in which anti-parallel arrows overlap.}
   \label{loops}
  \end{figure}
  

Notice that at any site $\textbf{r}$, $\nabla \times E(\textbf{r})$ and $(\nabla \cdot E)(\textbf{r})$ are two indipendent degrees of freedom. Following an analogous reasoning to the one we did to define the action of Gauge transformations on gauge fields, we extend the symmetries in \ref{symtablejw} onto gauge fields by requiring that the interaction hamiltonian \eqref{gaugeASF} remains invariant. Importantly, the action of symmetries on gauge fields is independent of the specific extended projective symmetry group implementation for the fermionic matter, because the ``projective" factors are already fully taken into account in the fermion transformation rules and the choice of mean-field hopping amplitudes. Moreover, under space transformations, the vector potential $A(\textbf{r}, i)$ transforms as a vector directed along the bond  $(\textbf{r}, i)$. Its transformation under time reversal, $\Theta$, can be fixed by demanding that the exponent in Eq.\eqref{hamipe} is left invariant:

 \begin{equation}\label{Eqtheta}
 e^{i A(\textbf{r},i)} = \Theta  e^{i A(\textbf{r},i)} \Theta^{-1} = e^{-i \Theta A(\textbf{r},i) \Theta^{-1}},
 \end{equation}

 thus, $\Theta A(\textbf{r},i) \Theta^{-1}=- A(\textbf{r},i)$. \\







So far we have kept track of the compactification of the $A$ field. When the low energy phase is deconfined, it is appropriate to simplify the description by neglecting the compactification and view the fields $A$ as taking values on the real axis.  With this simplification and after enforcing the symmetries it is easy to see that the simplest Hamiltonian that is bilinear in the local gauge-invariant fields, $E$ and $B$, is the standard Maxwell Hamiltonian in the lattice, given by:

  \begin{equation}\label{lqed}
      H_{\text{Gauge}} =  \frac{\epsilon}{2}  \sum_{\textbf{r}}   ( E^2(\textbf{r},x) +  E^2(\textbf{r},y) )
      + \frac{1}{2\mu}  \sum_{\textbf{r}} B^2(\textbf{r}).
 \end{equation}

Where $\epsilon$ and $\mu$ are constants. The above Hamiltonian can be diagonalized in terms of ``normal modes" of the pure-gauge in the absence of coupling to fermionic matter. Since we have two independent scalar degrees of freedom per unit cell, associated with $A(\textbf{r},x)$ and $A(\textbf{r},y)$, but one non-dynamical constraint per unit cell (since $\nabla \cdot E (\textbf{r})$ commutes with $H$), there is only one truly dynamical Harmonic oscillator degree of freedom per unit cell, associated with the magnetic field $B(\textbf{r})$. Its equations of motion can be determined easily from the Hamiltonian using the commutators from eq.\eqref{CCR}, and read as follows:

 \begin{equation}\label{eqm}
      \begin{aligned}
        \frac{d B(\textbf{r})}{dt} &=  -  \nabla \times E (\textbf{r}) \\
        \frac{d }{dt} \nabla \times E (\textbf{r}) &= \frac{4}{\mu \epsilon}  B(\textbf{r}) -  \frac{1}{\mu \epsilon} \sum_{\bm{\xi} = \pm \textbf{e}_x \atop \bm{\xi} =\pm \textbf{e}_y} B(\textbf{r}+\bm{\xi}). 
        \end{aligned}
  \end{equation}  
 
\begin{figure*}
  \centering
   \includegraphics[trim={0cm 0cm 0cm 0cm}, clip, width=\textwidth]{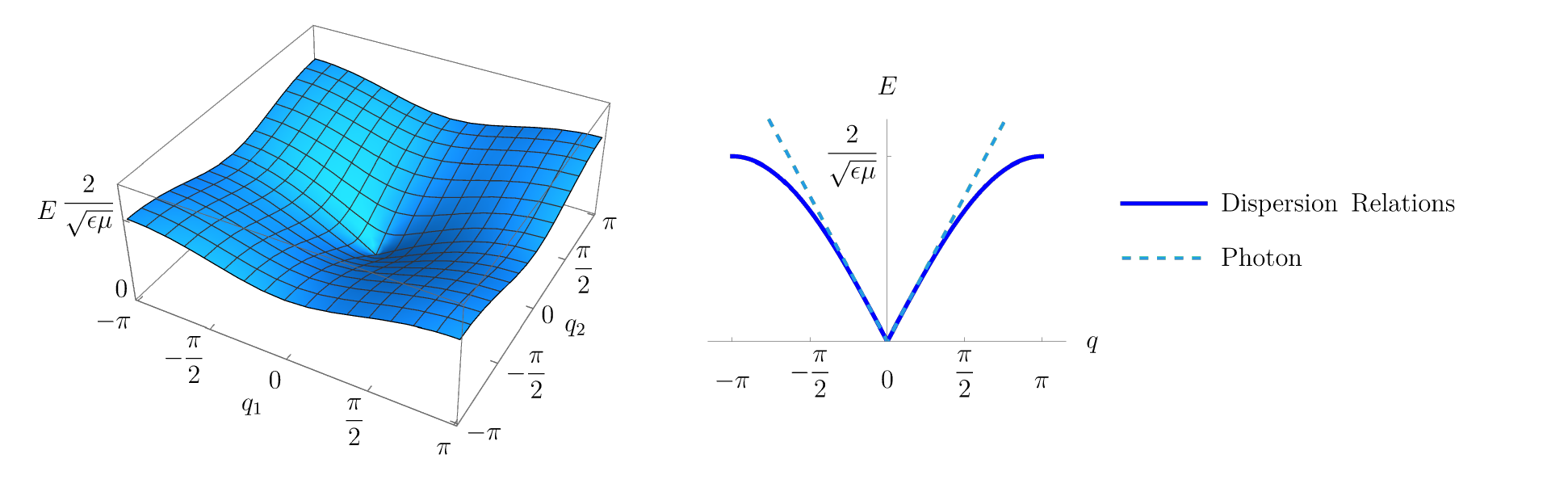}
   \caption{Left: Dispersion relation of the standard emergent photon of a U(1) spin liquid of Abrikosov-Schwinger fermions, from Eq.\eqref{dis}. Right: Cut of the dispersion relations along $q_y=0$, with the dashed line illustrating the linearized photon dispersion near $(q_x,q_y)=(0,0)$.}
   \label{photon1}
  \end{figure*}

The above can be solved by expanding fields in crystal momentum basis (Fourier transform), to obtain:
  
  \begin{equation}\label{LatticeMaxwell}
      \frac{d^2 B(\textbf{q})}{dt^2} +  \frac{1}{\mu \epsilon}[4 - 2\cos(q_x)-2\cos(q_y)]
      B(\textbf{q}) = 0,
  \end{equation}
  
and therefore the dispersion of the normal modes is (illustrated in Fig.\ref{photon1}):

\begin{equation}\label{dis}
    \omega^2(\textbf{q}) = \frac{1}{\epsilon \mu}[4 - 2\cos(q_x)-2\cos(q_y)].
\end{equation}

 The above dispersion features a linearly dispersing photon-like mode centered at momentum $(q_x,q_y)=(0,0)$ with a speed of (see Fig.\ref{photon1}):
 \begin{equation*}
     v = \frac{1}{\sqrt{\mu \epsilon}}.
 \end{equation*}
 
  This photon would be minimally coupled to the fermionic matter through Eq.\eqref{psub}. We see, therefore, that our phenomelogical procedure is able to describe the low energy field content expected at low energies for a $U(1)$ spin liquid associated with the standard Abrikosov-Schwinger parton construction \cite{WenA02, WenB02}. Let us pause to consider what protects the gaplessness of this photon mode?. Once deconfinement is presumed, so that it is valid to replace vector potentials by continuum real-valued variables, the lattice Faraday law from Eq.\eqref{eqm}, can be re-interpreted as a continuity equation:
 \begin{equation}\label{Faraday}
       \frac{\partial B}{\partial t}(\textbf{r},t) + \nabla \cdot \varepsilon = 0,
 \end{equation}
 where $\varepsilon$ is a dual electric field. It is a rotated version of the previously defined electric field, so that its lattice divergence is centered on the plaquettes, and is defined as:
 \begin{equation*}
    \varepsilon_i =  \epsilon_{ij} E_j ,
 \end{equation*}
 where $\epsilon_{ij}$ is the 2D Levi-civita symbol. The photon can be viewed as a Goldstone mode of a spontaneously broken global U(1) 1-form symmetry associated with the conservation of magnetic flux, as it is usually discussed in boson-vortex dualities in 2+1D \cite{peskin1978mandelstam,dasgupta1981phase,fisher1989correspondence}. In the absence of gapless fermionic matter and due compact nature of the gauge fields, the above photon would ultimately become gapped at low energies due to Polyakov confinement \cite{polyakov1975compact}, because the global conservation law of magnetic flux would be explicitly broken by fluctuations associated with local magnetic flux creation and destruction events. 
 

\subsection{Gauge field fluctuations for $U(1)$ spin liquids from extended parton constructions in 2D quantum spin-ice}\label{3B}
 
\begin{figure}
  \centering
   \includegraphics[trim={2cm 1cm 0cm 0cm}, clip, width=0.50\textwidth]{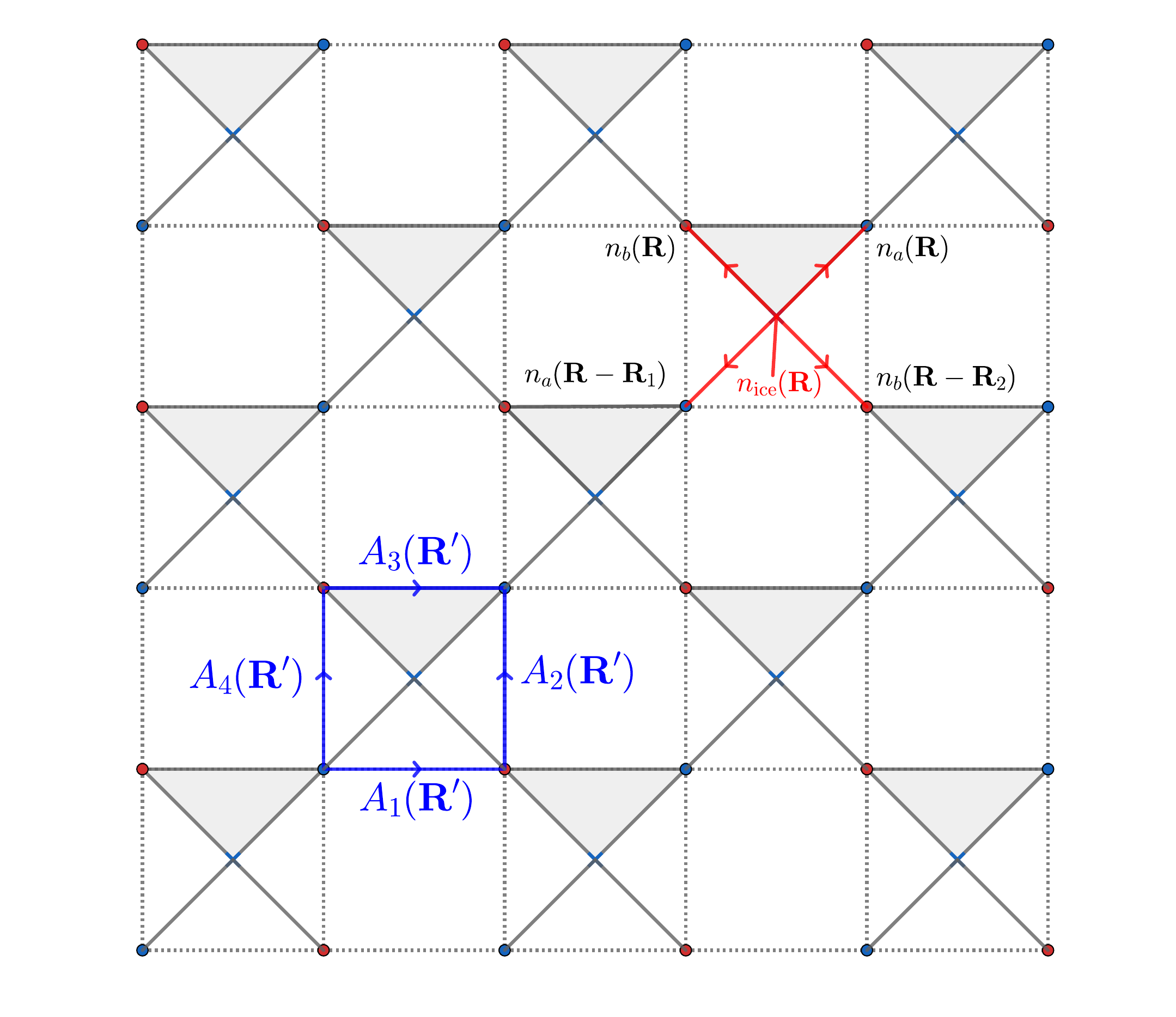}
   \caption{Bottom left: convention for labeling gauge fields. These reside at the links (blue lines) that connect the JW/composite fermion sites (located at the solid dots).  Top right: depiction of the operators entering in the generator of generalized spin-ice gauge transformations, $n_{\text{ice}}(\textbf{R})$ from Eq.\eqref{nice},  which is ceneterd at the spin-ice vertices.}
   \label{modqed}
  \end{figure}

Let us now generalize the previous construction to try to elucidate the low energy emergent gauge structure associated with the extended parton composite fermi liquid states of quantum spin-ice models discussed in Sec.\ref{C}. Just as we did for the Abrikosov-Schwinger fermions, we begin by writing the mean field Hamiltonian of the composite fermions and introduce a real-valued variable that captures the fluctuations of the phase of the hoping amplitude connecting a pair of fermion sites $(\bf{r},\bf{r}')$ and denote it by $A(\bf{r},\bf{r}')$. For concreteness we will focus on the fluctuations of the mean-field states described in Sec.\,\ref{C2} which had non-zero hoppings only for $(\textbf{r},\textbf{r}')$ being nearest neighbour sites, so that the reasulting mean field Hamiltonian, analogously to \eqref{hamipe}, reads as:
\begin{widetext}
       \begin{equation}\label{hami}
  \begin{aligned}
      H(t, A) = \sum_{\textbf{R}} &i t^* e^{-i A_1(\textbf{R}+ \textbf{R}_2)}f^{\dag}_{a}( \textbf{R} - \textbf{R}_1+ \textbf{R}_2)f_{b} (\textbf{R})
      +it^* e^{i A_3(\textbf{R})}   f^{\dag}_{a} (\textbf{R})f_{b} (\textbf{R})  \\
      +& t  e^{-i A_4(\textbf{R})} f^{\dag}_{a} (\textbf{R}   -  {\textbf{R}_1}) f_{b} (\textbf{R})  +t  e^{i A_2(\textbf{R}+ \textbf{R}_2)} f^{\dag}_{a} (\textbf{R} + {\textbf{R}_2}) f_{b} (\textbf{R})+ h.c.
      \end{aligned}
  \end{equation}
  \end{widetext}
where the convention for the labelling of Gauge fields is depicted in Fig.\ref{modqed}. 
As before we promote the above phases into angular quantum-rotor bosonic degrees of freedom, with an associated canonically conjugate degree of freedom denoted by $E(\textbf{r},\textbf{r}')$, with the same commutation relations described in Eqs.\eqref{CCR}. However, the first crucial difference that appears for the extended partons is that the UV $U(1)$ gauge transformations are not acting as in Eq.\eqref{gaugeASF} for the Abrikosov-Schwinger fermions. Instead the UV $U(1)$ gauge transformation are generated by the spin-ice charge operators from Eqs.\eqref{chargemodel},\eqref{chargemodel2}, or equivalently by the total number of fermions in the links connected to vertex ${\bf R}$, denoted by $n_{\text{ice}}(\textbf{R})$, and which in Bravais lattice notation reads as (see Fig.\ref{modqed}):
  \begin{equation}\label{nice}
      n_{\text{ice}}(\textbf{R}) = n_a(\textbf{R}) + n_b(\textbf{R}) + n_a(\textbf{R}-\textbf{R}_2) + n_b(\textbf{R}-\textbf{R}_1). 
  \end{equation}
  
  Therefor, the generator of the generalized lattice gauge transformations analogous to the one from Eq.\eqref{nochargediscrete}, that also acts on the dynamical phase degrees of freedom, is a sum of the corresponding four generators from Eq.\eqref{nochargediscrete}, and is given by:
\begin{equation}\label{Gice}
    \ G_{\text{ice}}(\textbf{R}) \doteq n_{\text{ice}}(\textbf{R})-\sum_{{\bf r}\in \textbf{R}}(\nabla \cdot E) (\textbf{r}),
\end{equation}
where $\textbf{r}\in \textbf{R}$ denotes the four spin sites that contribute to the ice-rule associated to the vertex \textbf{R}, as depicted in Fig.\ref{modqed}, and the lattice divergence $(\nabla \cdot E)(\textbf{r})$ is defined in the same way as in Eq.\eqref{divE}.\\

\begin{figure*}
  \centering
   \includegraphics[trim={0cm 0cm 0cm 0cm}, clip, width=\textwidth]{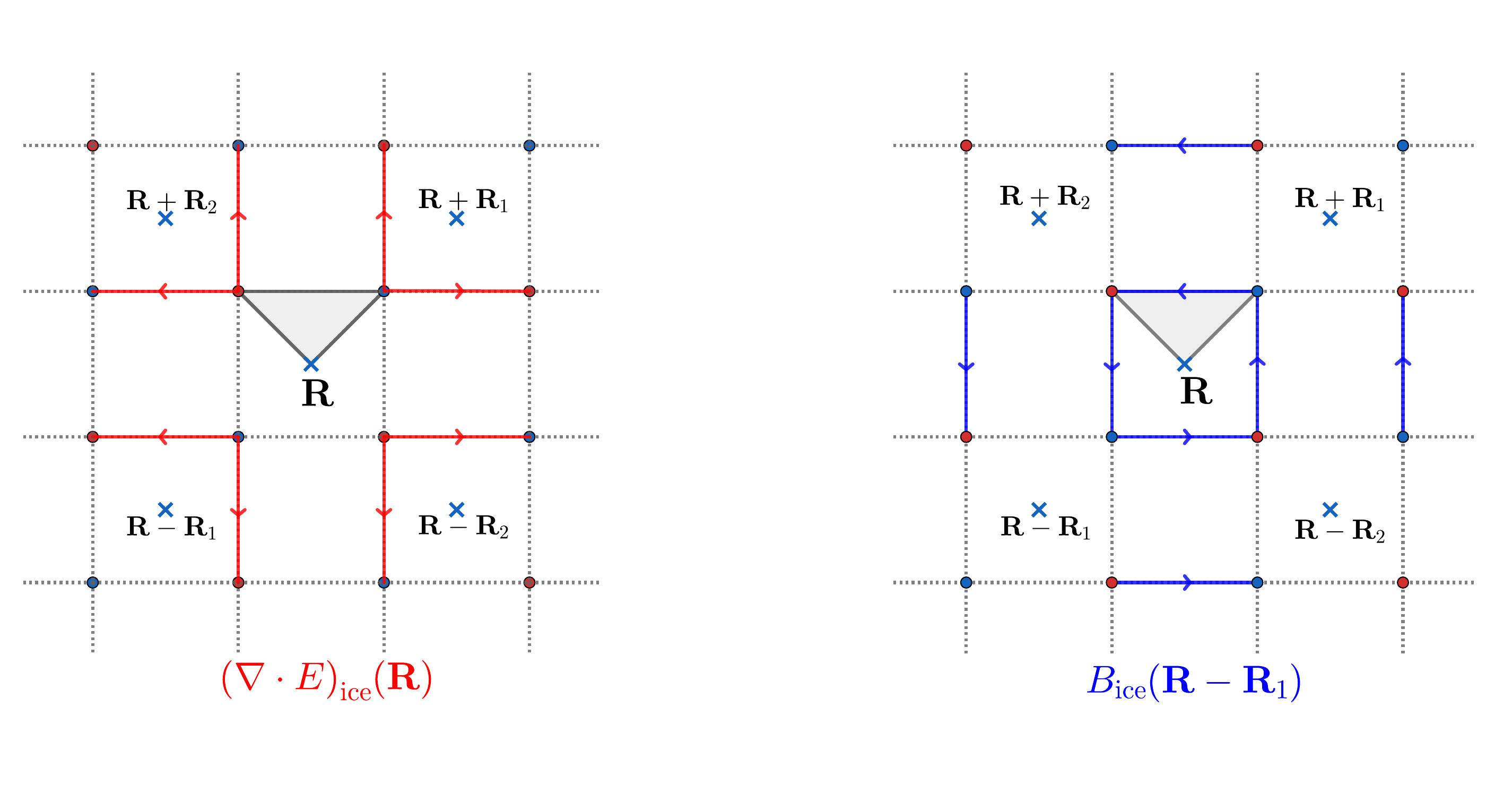}
   \caption{Left: depiction of generalized spin-ice electric field divergence operator,  $(\nabla \cdot E)_{\text{ice}}=\sum_{{\bf r}\in \textbf{R}}(\nabla \cdot E)(\textbf{r})$ from Eqs.\eqref{Gice},\eqref{divmodqed}. The red arrows depict the convention for adding electric fields. 
   Right: depiction of spin-ice magnetic field operator $B_{\text{ice}}(\textbf{R}-\textbf{R}_1)$ from Eq.\eqref{defBice}. The blue arrows depict the convention for adding vector potentials. The blue crosses mark the location of the spin-ice vertices.}
   \label{divmod}
  \end{figure*}

As before we demand that $G_{\text{ice}}$ commutes with every term in the Hamiltonian and interpret the physical Hilbert space as the one satisfying the constraint $G_{\text{ice}}(\textbf{R})=0$ for every $\textbf{R}$, which can be re-written as a Gauss law of the form:
\begin{equation}\label{icemod}
    (\nabla \cdot E)_{\text{ice}}(\textbf{R}) = n_{\text{ice}}(\textbf{R}),
\end{equation}
where $ (\nabla \cdot E)_{\text{ice}}$ is given by (see Fig.\ref{divmod}):
\begin{equation}\label{divmodqed}
    (\nabla \cdot E)_{\text{ice}}(\textbf{R}) = \sum_{{\bf r}\in \textbf{R}}(\nabla \cdot E)(\textbf{r}) .
\end{equation}


We can also write a canonically conjugate partner to the above gauge constraint operator, given by:

\begin{equation}\label{divAice}
    (\nabla \cdot A)_{\text{ice}}(\textbf{R}) = \sum_{{\bf r}\in \textbf{R}}(\nabla \cdot A)(\textbf{r}) .
\end{equation}
  
Let us now construct the analogue of the Maxwell Hamiltonian from Eq.\eqref{lqed}. To do so, we need to find all the linearly independent gauge field operators that commute with the gauge field part of the  constraint operator $G_{\text{ice}}(\textbf{R})$ from Eq.\eqref{Gice}, namely with $(\nabla \cdot E)_{\text{ice}}(\textbf{R})$ and its canonical partner $(\nabla \cdot A)_{\text{ice}}(\textbf{R})$. Since the Bravais unit cell contains four scalar vector potential degrees of freedom (see Fig.\ref{modqed}), but there is one Gauss law constraint per unit cell, we expect three independent harmonic oscillator modes per cell and therefore three dynamical gauge field bands. To find a basis for such modes, we notice that since $(\nabla \cdot E)_{\text{ice}}(\textbf{R})$ is a sum of divergences from the previous section on Abrikosov-Schiwnger fermions (see Eq.\eqref{divE}), the gauge invariant operators we discussed in the previous section, would also be gauge invariant in the new spin-ice construction. These include the magnetic operators $B(\textbf{r})$ from Eq.\eqref{B}, but now there are two such operators per spin-ice Bravais unit cell, one associated with the spin-ice vertex and one with the spin-ice plaquette, which we denote respectively by $B_V(\textbf{R})$, $B_P(\textbf{R})$, and together with their canonically conjugate partners are given by are explicitly given by (see Fig.\ref{modqed}): 

\begin{equation}\label{BVBP}
    \begin{aligned}
        B_V(\textbf{R}) =& A_1(\textbf{R})+A_2(\textbf{R})-A_3(\textbf{R}) - A_4(\textbf{R}), \\
        (\nabla \times E)_V(\textbf{R}) =& E_1(\textbf{R})+E_2(\textbf{R})-E_3(\textbf{R}) - E_4(\textbf{R}), \\
        B_P(\textbf{R}) =& A_3(\textbf{R}-\textbf{R}_2) +A_4(\textbf{R}+\textbf{R}_1 -\textbf{R}_2) \cdots \\ &- A_1(\textbf{R}+\textbf{R}_1) - A_2(\textbf{R}),\\
         (\nabla \times E)_P(\textbf{R}) =& E_3(\textbf{R}-\textbf{R}_2) +E_4(\textbf{R}+\textbf{R}_1 -\textbf{R}_2) \cdots \\ &- E_1(\textbf{R}+\textbf{R}_1) - E_2(\textbf{R}),
    \end{aligned}
\end{equation}

\begin{figure*}
  \centering
   \includegraphics[trim={0cm 8cm 0cm 4cm}, clip, width=1\textwidth]{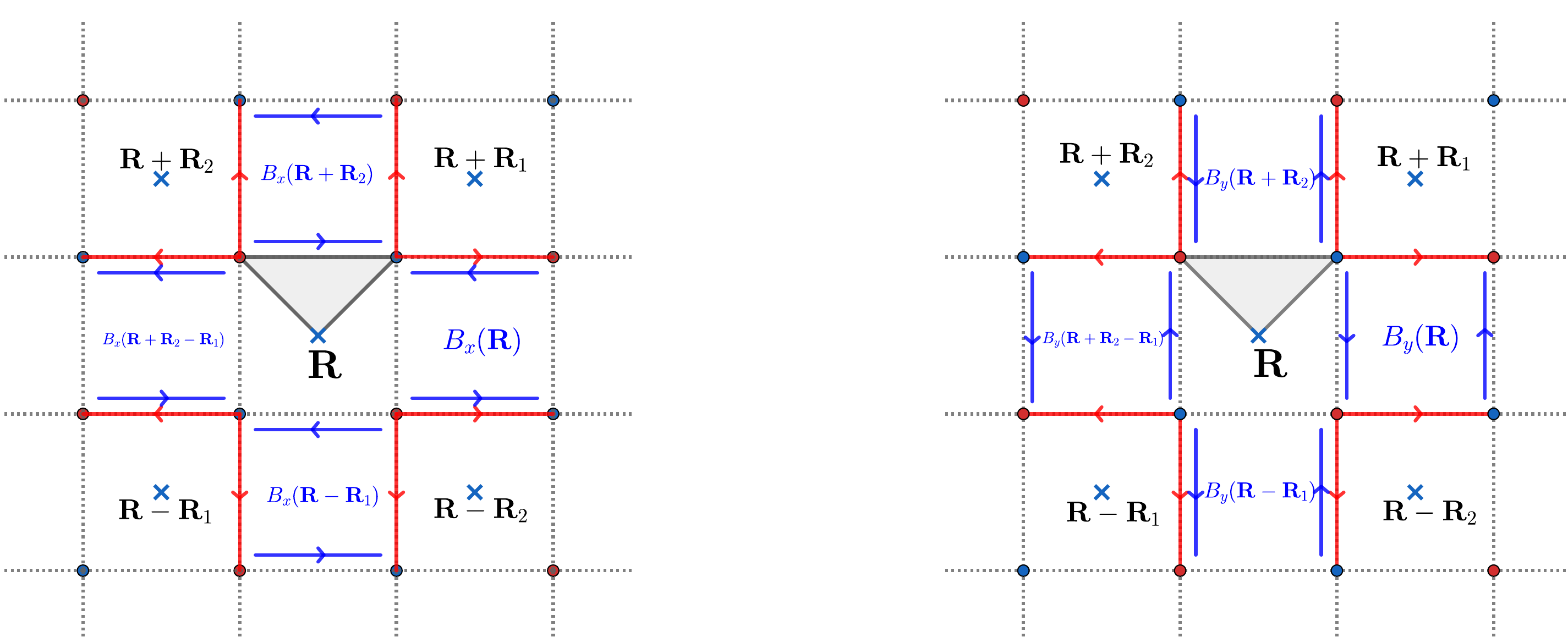}
\caption{Left: depiction of $B_x$ operators (from Eq.\eqref{gaugefildi}), as pairs of blue arrows. Notice that whenever the blue arrows (vector potentials) of a $B_x$ operator overlap with the red arrows (electric fields) of a $(\nabla \cdot E)_{\text{ice}}$ operator, there is always an equal number of parallel and antiparallel arrows, illustrating that  these operators commute. Right: analgous depictions for the $B_y$ operators  (from Eq.\eqref{gaugefildi}). The blue crosses mark the location of the spin-ice vertices.}
   \label{bx}
  \end{figure*}

where $B_V(\textbf{R})$ can viewed as a lattice curl centered around vertex $\textbf{R}$ and $B_P(\textbf{R})$ as a curl centered around the plaquette which is neighboring to the right the vertex $\textbf{R}$ (see Fig.\ref{bx}). However, there are certain additional operators containing only gauge fields, that commute with every $(\nabla \cdot E)_{\text{ice}}(\textbf{R})$, but which would not be gauge invariant under the convention of previous section, namely they would not commute with all the divergences of electric fields defined in Eq.\eqref{divE}. These operators and their canonically conjugate partners (see Fig.\ref{bx}) can be taken to be:

\begin{equation}\label{gaugefildi}
    \begin{aligned}
        B_x(\textbf{R}) &= A_3(\textbf{R}-\textbf{R}_2) - A_1(\textbf{R}+\textbf{R}_1), \\
       E_x(\textbf{R}) &= E_3(\textbf{R}-\textbf{R}_2) - E_1(\textbf{R}+\textbf{R}_1), \\
       B_y(\textbf{R}) &=   A_4(\textbf{R}+\textbf{R}_1 -\textbf{R}_2) - A_2(\textbf{R}), \\
        E_y(\textbf{R}) &=   E_4(\textbf{R}+\textbf{R}_1 -\textbf{R}_2) - E_2(\textbf{R}),
    \end{aligned}
\end{equation}
  
   where the $B_x(\textbf{R}),B_y(\textbf{R})$ fields can viewed as centered around the plaquette of the spin-ice model which is neighboring to the right the vertex $\textbf{R}$ (see Fig.\ref{bx}). Notice that $B_P(\textbf{R})=B_x(\textbf{R})+B_y(\textbf{R})$. Therefore, the set of linearly independent dynamical fields could in principle be chosen to be $B_x(\textbf{R}),B_y(\textbf{R}),B_V(\textbf{R})$. There is however a much better choice of local gauge invariant fields that will highly simplify the dynamics and the final physical picture. The idea is that instead of $B_V(\textbf{R})$, we would like to construct a local magnetic field strength that fits more naturally within the spin-ice gauge structure, which we will denote by $B_{\text{ice}} (\textbf{R})$. This quantity and its canonical partner can be chosen as follows: 

  \begin{widetext}
       \begin{equation}\label{defBice}
   \begin{aligned}   
  B_{\text{ice}} (\textbf{R}) &=  B_x(\textbf{R}) + B_x(\textbf{R}+\textbf{R}_1+\textbf{R}_2)  + B_y(\textbf{R}+\textbf{R}_2) + B_y(\textbf{R}+\textbf{R}_1) + 2 B_V(\textbf{R}+\textbf{R}_1), \\
  (\nabla \times E)_{\text{ice}} (\textbf{R}) &= E_x(\textbf{R}) + E_x(\textbf{R}+\textbf{R}_1+\textbf{R}_2) + E_y(\textbf{R}+\textbf{R}_2) + E_y(\textbf{R}+\textbf{R}_1) + 2 (\nabla \times E)_V(\textbf{R}+\textbf{R}_1).
   \end{aligned}
  \end{equation}
  
  \end{widetext}
  
 Figure \ref{divmod} illustrates the terms that enter into $B_{\text{ice}}(\textbf{R})$ making more clear why it has a natural interpretation of a spin-ice lattice curl. Notice that $B_{\text{ice}}(\textbf{R})$ is naturally viewed as centered around the vertex $\textbf{R}+\textbf{R}_1$ (see Fig.\ref{divmod}), but it will be convenient to keep its position label as $\textbf{R}$ as we will see later on. 
 The three fields $B_x(\textbf{R}),B_y(\textbf{R}),B_{\text{ice}}(\textbf{R})$ and their canonical conjugate partners $E_x(\textbf{R}),E_y(\textbf{R}),(\nabla \times E)_{\text{ice}} (\textbf{R})$, commute with the 
 gauge constraint field, $(\nabla \cdot E)_{\text{ice}}(\textbf{R})$, and its canonical partner, $(\nabla \cdot A)_{\text{ice}}(\textbf{R})$, and thus form a basis for the three independent modes of physical gauge fluctuations. The advantage of this basis over the $B_x(\textbf{R}),B_y(\textbf{R}),B_V(\textbf{R})$ basis, is that these fields form a sets of decoupled canonical coordinates, namely their mutual commutators vanish:

\begin{equation*}
\begin{aligned}
        [B_x(\textbf{R}),E_y(\textbf{R}')]&=[B_x(\textbf{R}), (\nabla \times E)_{\text{ice}}(\textbf{R}')]=0,\\
            [B_y(\textbf{R}),E_x(\textbf{R}')]&=[B_y(\textbf{R}), (\nabla \times E)_{\text{ice}}(\textbf{R}')]=0,\\
                [B_{\text{ice}}(\textbf{R}),E_x(\textbf{R}')]&=[B_{\text{ice}}(\textbf{R}), E_y(\textbf{R}')]=0.\\
\end{aligned}
\end{equation*}

The action of the microscopic lattice space symmetries on these fields is the same as in the case of Abrikosov-Schwinger fermions, and the additional pure gauge group transformation that enter into the extended projective symmetry group implementation on the fermions do not affect the gauge fields, therefore the fields $A_i(\textbf{R})$ transform as ordinary vectors according to the directions specified by the sites they connect, which is depicted in Fig.\ref{modqed}. From this the transformations of dynamical fields under space symmetries follow easily. The action of time reversal ($\Theta$ in Table \ref{symtable}) can be also inferred analogously to Eq.\eqref{Eqtheta}, and one concludes that: 
\begin{equation}\label{ThetaA}
\begin{aligned}
    \Theta  A_i(\textbf{R}) \Theta^{-1}&= -A_i(\textbf{R}) \\
    \Theta  E_i(\textbf{R}) \Theta^{-1}&= E_i(\textbf{R})
\end{aligned}
\end{equation} 
where $i=1,2,3,4$ are the components depicted in Fig.\ref{modqed}, and the transformations of $E_i(\textbf{R})$ can be inferred from its canonical commutator with $A_i(\textbf{R})$. Let us now consider the action of the microscopic particle-hole conjugation of hard-core bosons, denoted by $X$ (see Table \ref{symtable}). From its action on JW/composite-fermions (see Eq.\eqref{XFermionPH}) we obtain that the phases dressing the mean-field Hamiltonian should transform as: 
\begin{equation}\label{divAice}
\begin{aligned}
    X  e^{i A(\textbf{r},\textbf{r}')} X^{\dagger}&= e^{i A(\textbf{r}',\textbf{r})}=e^{-i A(\textbf{r},\textbf{r}')}
\end{aligned}
\end{equation} 
where we used that $A(\textbf{r}',\textbf{r})=-A(\textbf{r},\textbf{r}')$ (hermiticity). Therefore the fields transform as:
\begin{equation}\label{XA}
\begin{aligned}
    X A_i(\textbf{R}) X^{\dagger}&= -A_i(\textbf{R})\\
    X E_i(\textbf{R}) X^{\dagger}&= -E_i(\textbf{R})
\end{aligned}
\end{equation} 
It is interesting to note that under the natural microscopic time-reversal symmetry of spin-$\frac{1}{2}$ denoted by $\mathcal{T}$ (see Sec.\ref{pseudo}), it follows from Eq.\eqref{spinTR} and Eqs.\eqref{ThetaA},\eqref{XA} that the gauge fields transform as:
\begin{equation}\label{XA}
\begin{aligned}
    \mathcal{T} A_i(\textbf{R}) \mathcal{T}^{-1}&= A_i(\textbf{R}) \\
    \mathcal{T} E_i(\textbf{R}) \mathcal{T}^{-1}&=-E_i(\textbf{R})
\end{aligned}
\end{equation}
and therefore, interestingly, all the magnetic fields, $B_x(\textbf{R}),B_y(\textbf{R}),B_{\text{ice}}(\textbf{R})$, are even and the electric fields are odd under this time-reversal, which is opposite to the standard situation in QED. This is a manifestation of the psedo-scalar transformation of the JW/composite-fermions under this symmetry, as discussed in Sec.\ref{pseudo} and Ref.\cite{Inti}. Similar considerations also apply to other space symmetries such as mirrors, which in order to be implemented as natural spin-$\frac{1}{2}$ symmetries need to be dressed by the hard-core boson particle-hole conjugation $X$, which would lead to transformations on gauge fields opposite to those of ordinary QED (e.g. the electric field transforming as a pseudo-vector under mirrors).

We are now in a position to write a simple bilinear Maxwell-like model Hamiltonian for the pure gauge field part invariant under all microscopic symmetries of the RK model, which we write as:

\begin{widetext}
    \begin{equation}\label{mqed}
\begin{aligned}
    H_{\rm Gauge}= \frac{\epsilon}{2}  \sum_{\textbf{R}}\sum_{i=1}^4 E_i^2(\textbf{R}) + \frac{\chi_{B}}{2}  \sum_{\textbf{R}} B_{\text{ice}}^2(\textbf{R})  +\frac{\chi_P}{2}  \sum_{\textbf{R}} (B_x^2(\textbf{R}) + B_y^2(\textbf{R}))
    \end{aligned}
\end{equation}
\end{widetext}



Here we have ignored again for simplicity the compactification of gauge fields, and $\epsilon,\chi_B,\chi_P$ are phenomenological coupling constants. The equations of motion for the Hamiltonian from Eq.\eqref{mqed} are:

     \begin{equation}\label{eqmmodqed1}
     \begin{aligned}
         \frac{d^2 B_{x}}{dt^2}(\textbf{R}) =& -2 \frac{\chi_P}{\epsilon} B_x(\textbf{R}),\\
        \frac{d^2 B_y}{dt^2}(\textbf{R}) 
        =& -2 \frac{\chi_P}{\epsilon} B_y(\textbf{R}),\\
        \frac{d^2 B_{\text{ice}}}{dt^2}(\textbf{R}) =& -\frac{2\chi_B}{\epsilon} \big[4 B_{\text{ice}}(\textbf{R})  -  \sum_{\bm{\xi} = \pm \textbf{R}_1 \atop \bm{\mu} = \pm \textbf{R}_2} B_{\text{ice}}\big(\textbf{R}+ \bm{\xi} + \bm{\mu} \big) \big].\\
     \end{aligned}
 \end{equation}

\begin{figure}
  \centering
   \includegraphics[trim={0cm 0cm 0cm 0cm}, clip, width=0.45\textwidth]{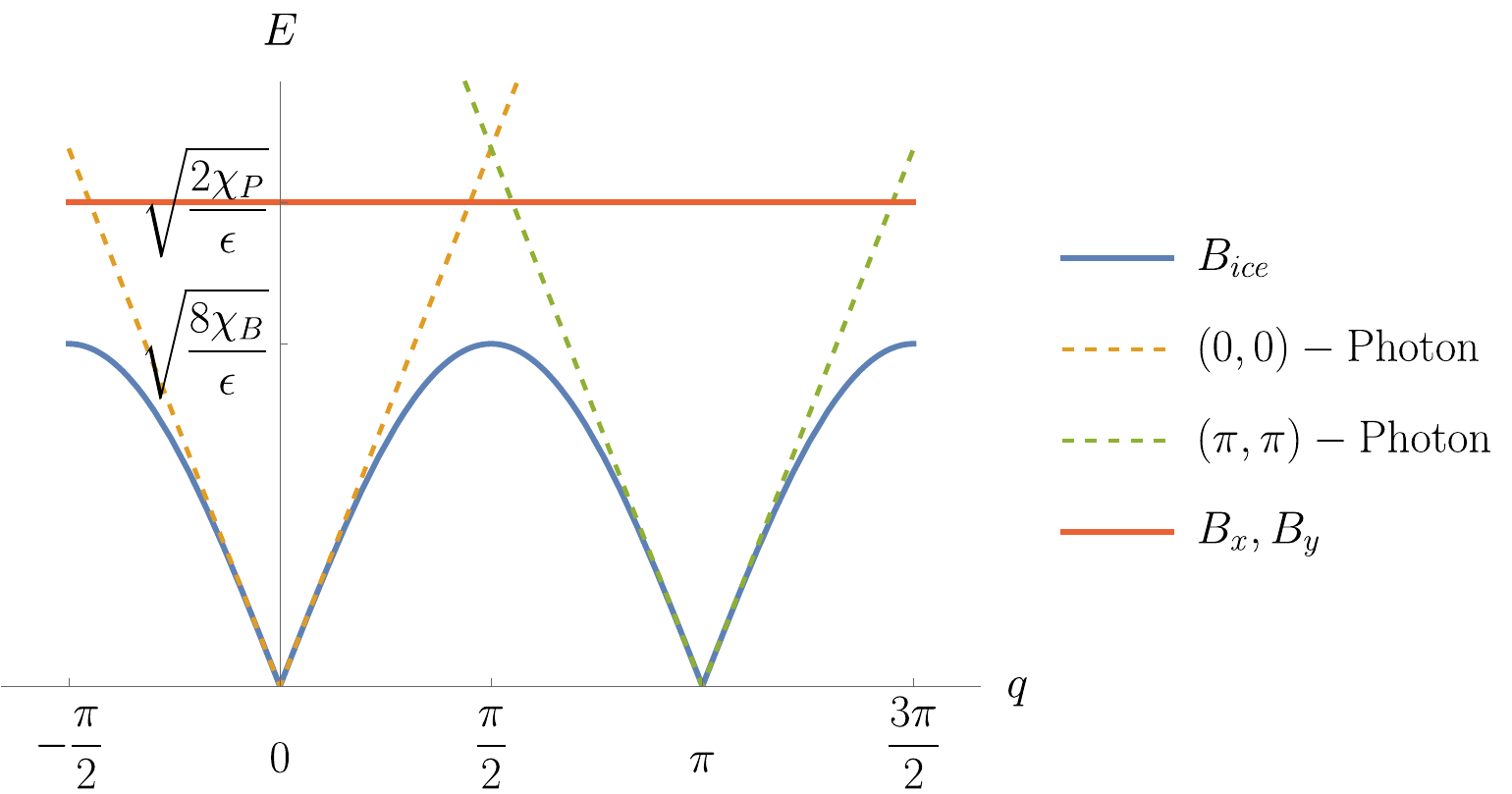}
   \caption{Dispersion relations of the gauge fields obtained from the pure Maxwell-like Hamiltonian (i.e. ignoring coupling to matter fermions), along a cut with $q_1=q_2$. There are two linearly dispersing photon-like modes near $(q_1,q_2)=(0,0)$ and $(q_1,q_2)=(\pi,\pi)$, with their dispersion shown in blue obtained from Eq.\eqref{dispmod} (the dashed lines are linearized  approximations of the photon dispersions). There are also two fully gapped gauge fluctuation modes associated with the $B_x,B_y$ fields (red lines), which are localized modes and hence have strictly flat dispersions for the ideal Maxwell-like Hamiltonian from Eq.\eqref{mqed}.}
   \label{2fot}
  \end{figure}

    Which in crystal momentum basis reduce to:
   
      \begin{equation}\label{eqmmodqed}
       \begin{aligned}
       \frac{d^2 B_x}{dt^2}(\textbf{q}) &= -\frac{2\chi_P}{\epsilon} B_x(\textbf{q}),\\
         \frac{d^2 B_y}{dt^2}(\textbf{q}) &= -\frac{2\chi_P}{\epsilon} B_y(\textbf{q}),\\
           \frac{d^2  B_{\text{ice}}}{dt^2}(\textbf{q}) &= - \omega^2(\textbf{q})B_{\text{ice}}(\textbf{q}), \\
     \end{aligned}
  \end{equation}

  where:

  \begin{equation}\label{dispmod}
      \omega^2(\textbf{q}) \doteq\frac{4\chi_B}{\epsilon} \big[2 -\cos\big( {q}_1 +{q}_2  \big) -\cos\big( {q}_1 -{q}_2 \big) \big] 
  \end{equation}

  and $q_i \doteq \textbf{q} \cdot \textbf{R}_i$ $i=1,2$. The dispersion relations show features that are crucially different with respect to the case of usual lattice QED and the emergent gauge fields discussed in Sec.\ref{3A} in the context of Abrikosov-Schwinger fermions. The $B_{x/y}$ modes display a fully gapped and dispersive-less flat band with energy $\sqrt{ 2\chi_P/\epsilon}$, as depicted in Fig.\,\ref{2fot}. While the exact flatness is a consequence of our simple model, the fact that these modes are gapped is a generic feature. Therefore the fluctuations associated with these modes are expected to be irrelevant at low energies, and they can be safely neglected from the low energy effective theory. On the other hand the $B_{\text{ice}}$ mode displays two distinct gapless photon-like \eqref{dispmod} linearly dispersing modes centered around $\textbf{q}=(0,0)$ and $\textbf{q}=(\pi,\pi)$ as depicted in Fig.\ref{2fot}.

To close this subsection, we would like to remark that the Maxwell-like Hamiltonian of Eq.\eqref{mqed} does not have the half-translational symmetry by $(\textbf{R}_1+\textbf{R}_2)/2$, that we encountered in the bare mean-field fermion Hamiltonian. This can be easily seen by noting that this symmetry would exchange the spin-ice vertices with the spin-ice plaquettes. However the $B_x(\textbf{R}) ,B_y^2(\textbf{R})$ fields are only centered around the spin-ice plaquettes, whereas the fields $B_{\text{ice}}(\textbf{R})$ are centered only around vertices, and therefore clearly the Hamiltonian of Eq.\eqref{mqed} has no symmetry relating spin-ice vertices and spin-ice plaquettes. The apparent translational symmetry in momentum space by $(\pi,\pi)$ of the pure gauge field modes that we see in Fig.\ref{2fot}, is a result of fine tunning of the model, which we have done for simplicity. For example gauge invariant terms can be easily added to the Maxwell-Hamiltonian that would delocalize the $B_x(\textbf{R}) ,B_y(\textbf{R})$ modes and make them itinerant and with dispersions that would have different energies near $(0,0)$ vs near $(\pi,\pi)$. Similarly it is possible to add gauge invariant terms to the Hamiltonian that would make the photons originating from fluctuations of $B_{\text{ice}}(\textbf{R})$ to have different speeds near $(0,0)$ vs near $(\pi,\pi)$. Therefore, the full theory of fermions coupled to gauge field fluctuations does not have the translational symmetry by  $(\textbf{R}_1+\textbf{R}_2)/2$ that we saw in the bare mean-field fermion Hamiltonian, reflecting the fact that this is not a true microscopic symmetry of the underlying RK Hamiltonian of spin-ice.

  \subsection{Gauge field and matter couplings, low energy effective field theory and dipolar nature of composite fermions}\label{25}

Let us now determine the matter coupling to the low energy gauge fields and the low-energy effective field theory. For concreteness we will focus on the case of gapless Dirac fermions obtained for the six-vertex subspace, but similar considerations would apply to the Fermi surface state of the quantum dimer model. Ignoring compactification and expanding Eq. \eqref{hami} up to linear order on the vector potentials, we obtain the following:

  \begin{widetext}
  \begin{equation}\label{hamigau}
  \begin{aligned}
      H(t, A) &= \sum_{\textbf{R}}i t^* (1 -i A_1(\textbf{R}+ \textbf{R}_2) ) f^{\dag}_{a}( \textbf{R} - \textbf{R}_1+ \textbf{R}_2)f_{b} (\textbf{R})  +i t^* (1 + i A_3(\textbf{R}))  f^{\dag}_{a} (\textbf{R})f_{b} (\textbf{R})\\
      &\,\,\,\,\,\,\,\,\,\,\,\,\,\,\,\,  + t (1 -i A_4(\textbf{R})) f^{\dag}_{a} (\textbf{R}   -  {\textbf{R}_1}) f_{b} (\textbf{R}) +  t(1 + i A_2(\textbf{R}+ \textbf{R}_2))f^{\dag}_{a} (\textbf{R} + {\textbf{R}_2}) f_{b} (\textbf{R})  + h.c + O(A^2)\\
      \end{aligned}
  \end{equation}
  \end{widetext}


  
  As discussed in Sec. \ref{C2}, the fermions have gapless Dirac nodes at the two valleys $(\pi,0)$ and $(0, \pi)$ while the gauge field has gapless photon-like modes at $(0,0)$ and $(\pi, \pi)$.  Therefore we expect that the dominant effects at low energy include:
  \begin{enumerate}
      \item Intra-valley scattering within each Dirac cone mediated by exchange of long-wavelength gauge field fluctuations with momenta near $(0,0)$.
      \item Inter-valley scattering process connecting the two Dirac cones mediated by the exchange of gauge fluctuations with momenta near $(\pi,\pi)$.
  \end{enumerate}

\begin{figure}
  \centering
  \begin{tikzpicture}[blend mode=multiply]
    \tikzset{
      clip even odd rule/.code={\pgfseteorule}, 
      invclip/.style={
          clip,insert path=
              [clip even odd rule]{
                  [reset cm](-\maxdimen,-\maxdimen)rectangle(\maxdimen,\maxdimen)
              }
      }
    } 

    \node[inner sep=0pt,minimum size=0pt,label=below:] (a) at (-1,1) {};
\node[inner sep=0pt,minimum size=0pt,label=below:] (b) at (3,3) {};
\node[inner sep=0pt,minimum size=0pt,label=above:] (c) at (7,1) {};
\node[inner sep=0pt,minimum size=0pt,label=above:] (d) at (3,-1) {};
    \draw[fill=blue!50, fill opacity=.2,draw] (-1,  1) -- (3, 3) -- (7, 1) -- (3,-1) -- cycle;

\draw [postaction={decorate},decoration={markings,
mark=between positions 0 and 1 step \UnitSegment with {
\node [inner sep=0pt,minimum size=0pt,
name=mark-1-\pgfkeysvalueof{/pgf/decoration/mark info/sequence number}]
{};}}] (a) -- (b);

\draw [postaction={decorate},decoration={markings,
mark=between positions 0 and 1 step \UnitSegment with {
\node [inner sep=0pt,minimum size=0pt,
name=mark-2-\pgfkeysvalueof{/pgf/decoration/mark info/sequence number}]
{};}}] (b) -- (c);

\draw [postaction={decorate},decoration={markings,
mark=between positions 0 and 1 step \UnitSegment with {
\node [inner sep=0pt,minimum size=0pt,
name=mark-3-\pgfkeysvalueof{/pgf/decoration/mark info/sequence number}]
{};}}] (c) -- (d);

\draw [postaction={decorate},decoration={markings,
mark=between positions 0 and 1 step \UnitSegment with {
\node [inner sep=0pt,minimum size=0pt,
name=mark-4-\pgfkeysvalueof{/pgf/decoration/mark info/sequence number}]
{};}}] (d) -- (a);

\draw (mark-1-2) -- (mark-3-8);
\draw (mark-1-3) -- (mark-3-7);
\draw (mark-1-4) -- (mark-3-6);
\draw (mark-1-5) -- (mark-3-5);
\draw (mark-1-6) -- (mark-3-4);
\draw (mark-1-7) -- (mark-3-3);
\draw (mark-1-8) -- (mark-3-2);

\draw (mark-2-2) -- (mark-4-8);
\draw (mark-2-3) -- (mark-4-7);
\draw (mark-2-4) -- (mark-4-6);
\draw (mark-2-5) -- (mark-4-5);
\draw (mark-2-6) -- (mark-4-4);
\draw (mark-2-7) -- (mark-4-3);
\draw (mark-2-8) -- (mark-4-2);
    
    \begin{scope} [xshift=0cm,yshift=0cm]
  


  \draw[thick, orange, -latex] ($(1, -1) + (30:1cm and 0.25cm)$(P) arc
  (30:340:1cm and 0.25cm);
  \draw[thick, orange, -latex] ($(5, -1) + (30:1cm and 0.25cm)$(P) arc
  (30:340:1cm and 0.25cm);
  
   \draw[->, thick, red] (1.6,3.8)  -- (4.4,3.8)   ;
   \draw[->, thick, red] (4.4,3.1) -- (1.6,3.1);
     \draw[->, thick, dashed] (1,  2) -- (5, 0);
      \draw[->, thick, dashed] (1, 0) -- (5,2);
     \node[] (x) at (1,4.5) {$(-\pi,0)$};
     \node[] (x) at (5,4.5) {$(0, \pi)$};
      \node[] (x) at (5, -0.5) {$q_1$};
      \node[] (x) at (5.35, 2.1) {$q_2$};
      \node[orange] (p) at (1,-1.5) {$(0,0)$ Photon};
       \node[orange] (p) at (5,-1.5) {$(0,0)$ Photon};

    \end{scope}
     \node[red] at (3,4) {$(\pi, \pi)$ Photon};
     \node[red] at (3,3.3) {$(\pi, \pi)$ Photon};
    
      \begin{scope}
        \begin{pgfinterruptboundingbox}
          \clip[invclip] (1,4) ellipse (.99 and .2);
          \draw[fill=red!50, fill opacity=.5] (2,4) -- (1,2) -- (0,4);
        \end{pgfinterruptboundingbox}
        \begin{pgfinterruptboundingbox}
          \clip[invclip] (1,0) ellipse (.99 and .2);
          \draw[fill=red!50, fill opacity=.5] (2,0) -- (1,2) -- (0,0);
        \end{pgfinterruptboundingbox}
      \end{scope}
      \draw[fill=red!50, fill opacity=.1,draw] (1,4) ellipse (.99 and .2);
      \begin{scope}
        \clip (0,0) rectangle ++(2,1);
        \draw[fill=red!50, fill opacity=.5,draw,dashed] (1,0) ellipse (.99 and .2);
      \end{scope}
      \begin{scope}
        \clip (0,-1) rectangle ++(2,1);
        \draw[fill=red!50, fill opacity=.5,draw] (1,0) ellipse (.99 and .2);
      \end{scope}
 
    \begin{scope}[yshift=0cm,xshift=0cm]
      \begin{scope}
        \begin{pgfinterruptboundingbox}
          \clip[invclip] (5,0) ellipse (.99 and .2);
          \draw[fill=red!50, fill opacity=.5] (4,0) -- (5,2) -- (6,0);
        \end{pgfinterruptboundingbox}
      \end{scope}
      \begin{scope}
       \begin{pgfinterruptboundingbox}
          \clip[invclip] (5,4) ellipse (.99 and .2);
          \draw[fill=red!50, fill opacity=.5] (4,4) -- (5,2) -- (6,4);
        \end{pgfinterruptboundingbox}
        \end{scope}
      \draw[fill=red!50, fill opacity=.1,draw] (5,4) ellipse (.99 and .2);
      \begin{scope}
        \clip (4,0) rectangle ++(2,1);
        \draw[fill=red!50, fill opacity=.5,draw,dashed] (5,0) ellipse (.99 and .2);
      \end{scope}
      \begin{scope}
        \clip (4,-1) rectangle ++(2,1);
        \draw[fill=red!50, fill opacity=.5,draw] (5,0) ellipse (.99 and .2);
      \end{scope}
    \end{scope}


    
  \end{tikzpicture}
  \caption{Illustration of the two type of fermion scattering processes arising from their coupling to gauge fields. The gauge modes near $q=(0,0)$ mediate ``intra-valley'' fermion scattering processes (depicted by orange circles), and the gauge modes near $q=(\pi,\pi)$ mediate ``inter-valley'' scattering processes (depicted by red straight arrows).}
   \label{graph1}
     \end{figure}
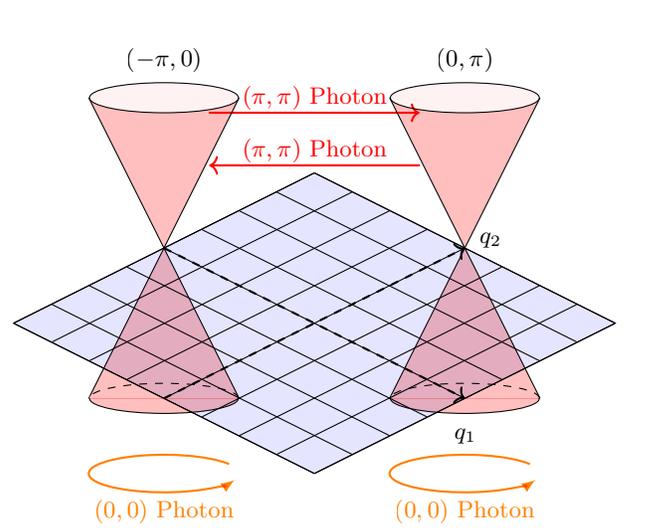

These two kind of processes are depicted  in Fig.\ref{graph1}. Therefore, in the spirit of $k \cdot p$ theory, we define the following fields by expanding the fermion and gauge fields around their different respective gapless points:
  \begin{equation*}
  \begin{aligned}
              \Psi(\textbf{q}) &\doteq \begin{pmatrix}
              f_a(\textbf{p}+ (\pi,0))\\
               f_b(\textbf{p}+ (\pi,0))\\
                f_a(\textbf{p}+ (0,\pi))\\
                 f_b(\textbf{p}+ (0,\pi))
              \end{pmatrix}
               \\
              A^{0}_j(\textbf{p})&\doteq A_j(\textbf{p})\,\,\,\,\,\,\,\,\,\,\,\,\,\,\,\,\,\,\,\,\,\,\,\,\,\,\,j \in \{1,2,3,4\} \\
              A^{\pi}_j(\textbf{p}) &\doteq A_j(\textbf{p} + (\pi,\pi))\,\,\,\,\,j \in \{1,2,3,4\} \\
  \end{aligned}
  \end{equation*}

   where $\textbf{p}$ is understood to be ``small" with respect to the size of the Brillouin zone, so that we can expand the hamiltonian \eqref{hamigau} to the first order (for the convention sublattice indices see Fig.\ref{modqed}). Details of the derivations of the small momentum expansion can be found in appendix \ref{appd}, we will here summarize the final results next. 
   
  

\subsubsection{$\textbf{p}= (0,0)$ scattering terms}\label{ss25}
  The hamiltonian density describing processes of the first type for the state with $\Theta_x$ extended PSG ($t \in \mathbb{R}$) is:

\begin{equation*}
               H = v  \Psi^\dagger({\bf x}) \left[(p_x-A_x^0({\bf x}))\tau^x+(p_y-A_y^0({\bf x}))\tau^y\rho^z \right]\Psi({\bf x}), 
\end{equation*}

and for the state with $\Theta_y$ extended PSG ($t \in i \mathbb{R}$) is:

\begin{equation*}
               H = v  \Psi^\dagger({\bf x}) \left[(p_x-A_x^0({\bf x}))\tau^y+(p_y-A_y^0({\bf x}))\tau^x\rho^z \right]\Psi({\bf x}),
\end{equation*}


where the convention of momenta is the same as in Eq.\eqref{linhamire}, and $\tau^i$, $\rho^i$ denote Pauli matrices in $\{a,b\}$ sub-lattice and on $\{(\pi,0), (0, \pi)\}$ valley spaces respectively, and we have defined continuum vector potential fields as follows: 
 \begin{equation}
              \begin{aligned}
      A^{0}_x({\bf x}) &\doteq \frac{A^{0}_{1}({\bf x}) + A^{0}_{3}({\bf x})}{\sqrt{2} |{\bf R}_1|},\\
       A^{0}_y({\bf x}) &\doteq \frac{A^{0}_{2}({\bf x}) + A^{0}_{4}({\bf x})}{\sqrt{2} |{\bf R}_1|}.
      \end{aligned}
  \end{equation}

Therefore, we see that the long-wavelength fluctuations gauge fluctuations that are gapless near $(0,0)$, simply behave as the standard minimal coupling of a photon-like mode to the matter fields (compare with mean field Hamiltonian from Eq.\eqref{linhamire}).

  \subsubsection{$\textbf{p}=(\pi,\pi)$ scattering terms}\label{ss26}
  
  The contribution to the hamiltonian density accounting processes of the second type, for the state with $\Theta_x$ extended PSG ($t \in \mathbb{R}$) is:

 \begin{equation}
               \delta H = - v \Psi^\dagger({\bf x}) \left[B^{\pi}_x({\bf x}) {\tau^x} \rho^1 + B^{\pi}_y({\bf x}) \tau^x \rho^2 \right]\Psi({\bf x}), 
\end{equation}

and for the state with $\Theta_y$ extended PSG ($t \in i \mathbb{R}$) is:

\begin{equation}
               \delta H = - v \Psi^\dagger({\bf x}) \left[B^{\pi}_x({\bf x}) {\tau^y} \rho^1 - B^{\pi}_y({\bf x}) \tau^y \rho^2 \right]\Psi({\bf x}), 
\end{equation}

where the continuum vector potential fields are defined as follows:
\begin{equation}
    \begin{aligned}
          B^{\pi}_x(\textbf{x}) &= \frac{A^{\pi}_{3}(\textbf{x}) - A^{\pi}_{1}(\textbf{x})}{\sqrt{2}|\textbf{R}_1|}, \\  
          B^{\pi}_y(\textbf{x}) &= \frac{A^{\pi}_{4}(\textbf{x}) - A^{\pi}_{2}(\textbf{x})}{\sqrt{2}|\textbf{R}_1|}. \\
    \end{aligned}
\end{equation}

Notice that the fields $B^{\pi}_x, B^{\pi}_y$ are the continuum limits of the fields defined in Eq. \eqref{gaugefildi} expanded around momentum $(\pi, \pi)$. Therefore, remarkably, what we are finding here is that to linear order in vector potentials, there is no coupling to the linearly dispersing gapless photon modes near $(\pi,\pi)$, but instead the inter-valley scattering process are only mediated by gauge fields associated with the $B_x$ and $B_y$ modes, which are fully gapped throughout the entire Brillouin. Therefore, at low energies compared to the gap of the $B_x, B_y$ modes and the band-width of the photon modes, we have two emergent massless photon modes, and two massless Dirac fermions. But the fermions are only carry gauge charge under the $(0,0)$ photon but appear as gauge neutral under the $(\pi,\pi)$ photon.

\begin{figure}
  \centering
 \includegraphics[trim={3cm 0cm 0cm 0cm}, clip, width=0.5\textwidth]{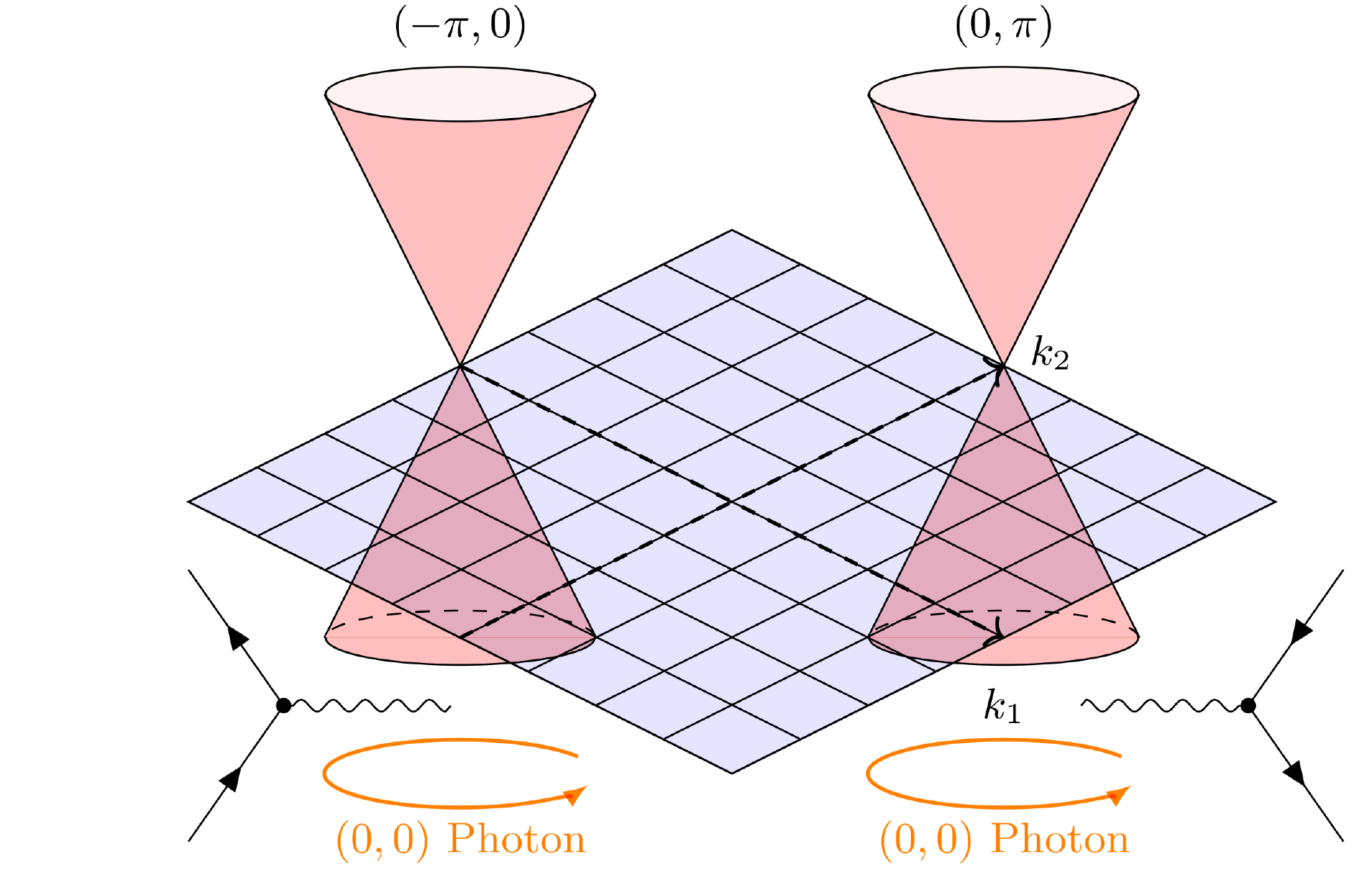}
   \caption{Depiction of the expected infrared effective low energy theory, which is a U(1) compact QED in 2+1 dimensions minimally coupled to two massless Dirac fermions. The Dirac fermions are centered at $(0,\pi)$ and $(\pi,0)$, and they are minimally coupled to the single U(1) photon gapless at $(0,0)$. The photon at $(\pi,\pi)$ likely undergoes Polyakov style confinement, hence disappearing at low energies.}
   \label{lowenergy}
  \end{figure}

\subsection{$U(1)\times U(1)$ Gauge structure}\label{esa}

  In this section we will explain why the occurrence of two gapless photon modes and the gauge coupling of the Dirac composite fermions to only one of them, that we encountered in sections \ref{3B} and \ref{3B}, is not accidental. We will show there are two emergent $U(1)$ gauge structures with independent local Gauss laws and two global flux conservation, as if we had two copies of ordinary lattice QED. 
  
\begin{figure*}
  \centering
   \includegraphics[trim={0cm 1cm 0cm 0cm}, clip, width=0.6\textwidth]{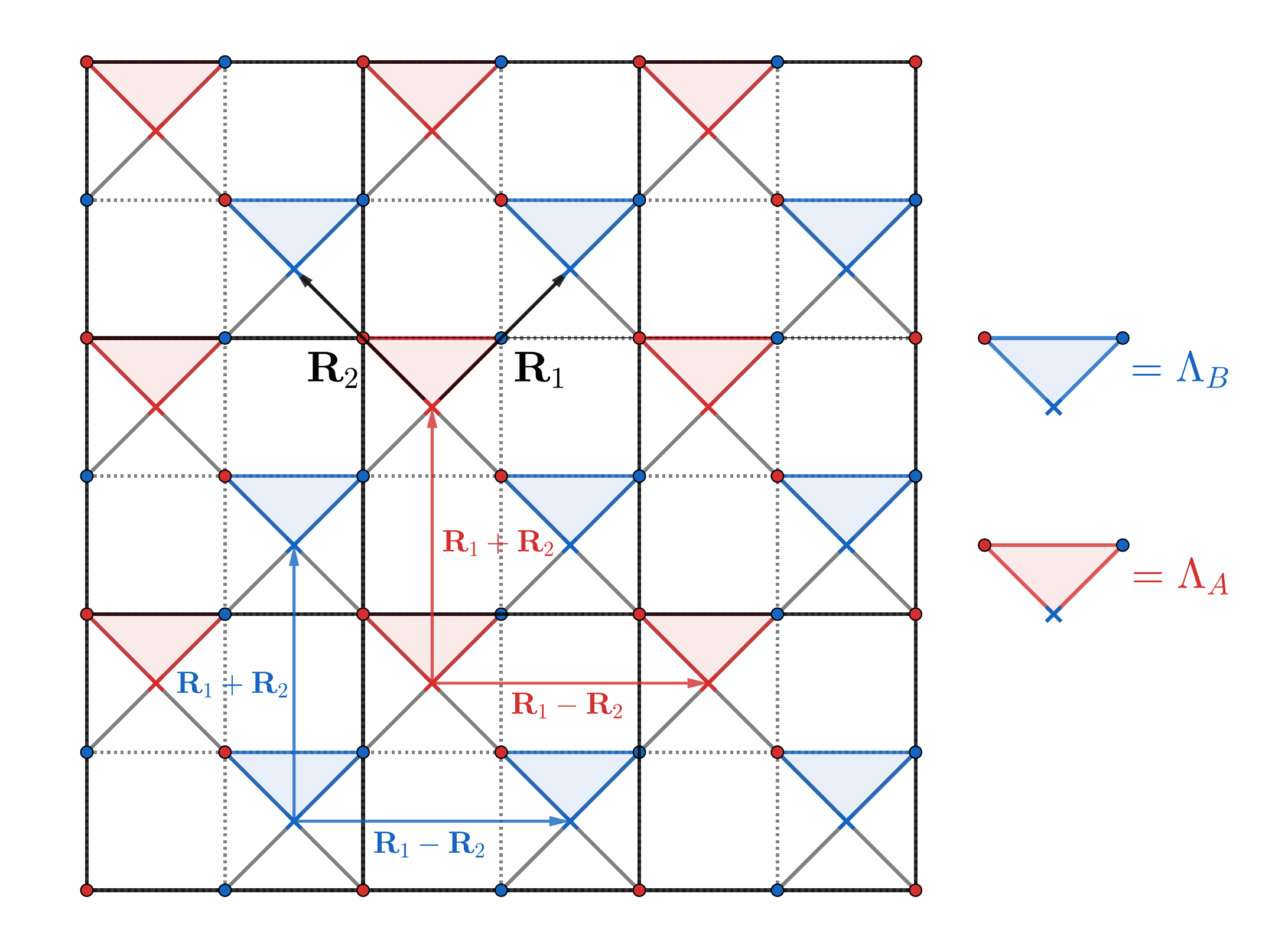}
   \caption{Separation of the lattice of vertices into $\Lambda_A$ and $\Lambda_B$ sub-lattices, which allows to understand the $U(1)_A \times U(1)_B$ gauge structure. The JW/Composite-fermions are located at the dots and carry equal charge $(q_A,q_B)=(1,1)$ under these $U(1)_A \times U(1)_B$ gauge transformations. The photon that is gapless near $(0,0)$ (see Fig.\ref{2fot}) corresponds to the sublattice symmetric gauge transformation, for which the fermions are charged. This is why the JW/composite-fermions are minimimally coupled to this photon at low energies (see Fig.\ref{lowenergy}). The photon that is gapless near $(\pi,\pi)$ (see Fig.\ref{2fot}) corresponds to the sublattice asymmetric gauge transformations (i.e. staggered). For which these asymmetric gauge transformations the JW/composite-fermions is not charged (behaves instead as a gauge dipole) and this is why it is not minimally coupled to the $(\pi,\pi)$ photon. }
   \label{sublatqed}
  \end{figure*}

  To see this it is convenient to split the Bravais lattice of vertices of the spin-ice model, which are located at vectors ${\bf R}$, into two sublattices denoted by $\Lambda_A$ and $\Lambda_B$, as depicted in Fig.\,\ref{sublatqed}. The sublattice $\Lambda_B$ can be obtained by displacing the $\Lambda_A$ by either the Bravais vector ${\bf R}_1$ or ${\bf R}_2$, and vice-versa. Therefore, the Bravais unit vectors of the lattice $\Lambda_A$ can be taken to be $\{{\bf R}_1-{\bf R}_2,{\bf R}_1+{\bf R}_2\}$, and similarly for $\Lambda_B$ (see Fig.\,\ref{sublatqed}). Notice that the operators that measure the divergence of the dynamical emergent fields, $(\nabla \cdot E)_{\text{ice}} (\textbf{R})$ defined in Eq. \eqref{divmodqed} and illustrated in Fig.\ref{divmod}, behave as two independent divergences obeying separate Gauss' laws. Namely when we sum $(\nabla \cdot E)_{\text{ice}} (\textbf{R})$ over $\textbf{R}$ restricted to region of points residing only on sublattice $\Lambda_A$, we will get a sum of electric fields residing only at the boundary of such region and normal to the boundary, as expected for a lattice divergence, and similarly for regions of points contained only in sublattice $\Lambda_B$. Moreover, we can also restrict the operators $G_{\text{ice}}(\textbf{R})$ to reside over either of the sublattices and in this way we can view the $U(1)$ gauge group as a product $U(1)_A \times U(1)_B$. We can assign a pair charges $(q_A,q_B)$ with $q_{A,B} \in \mathbb{Z}$ to matter operators (namely those constructed as products of fermion creation/destruction operators) under this $U(1)_A \times U(1)_B$ gauge group. In particular the JW/composite-fermion creation operator, would transforms as a charge $(q_A,q_B)=(1,1)$ under such sublattice gauge groups.
  
 There are also two independent global flux conservation symmetries (when ignoring gauge field compactification), one associated with sublattice $\Lambda_A$ and the other with sublattice $\Lambda_B$, which are responsible for the gaplessness of the two photons. This can be seen by adding the operators $d B_{\text{ice}}(\textbf{R})/dt$, defined in Eq.\eqref{defBice} and illustrated in Fig.\ref{divmod}, over some region of $\textbf{R}$ that only contains points in the sublattice $\Lambda_A$, resulting into a boundary operator, that can be viewed as a line integral of the operator $(\nabla \times E)_{\text{ice}} (\textbf{R})$ from Eq.\eqref{defBice}. This can be interpreted as a conservation law analogous to the lattice Faraday law of QED from Eq.\eqref{Faraday}, except that now there are two such conservation laws, one for the $\Lambda_A$ and another one for the $\Lambda_B$ sublattice.

   Moreover, our choice of Maxwell Hamiltonian in Eq.\eqref{mqed} has actually been made so that the two photons of the $U(1)_A \times U(1)_B$ gauge structure, are also dynamically decoupled. This can be seen by noticing that the commutator of $B_{\text{ice}}(\textbf{R})$ and $(\nabla \times E)_{\text{ice}}(\textbf{R}')$ vanishes whenever $\textbf{R}$ and $\textbf{R}'$ belong to different $\Lambda_A$,$\Lambda_B$ sublattices. The set of operators $B_{\text{ice}}(\textbf{R})$ and their canonical partner $(\nabla \times E)_{\text{ice}}(\textbf{R})$ with $\textbf{R}$ restricted to a given sublattice $\Lambda_A$,$\Lambda_B$ have indeed exactly the same equations of motion of ordinary QED with a single photon that we reviewed in Sec.\ref{3A}. If we expand these operators in the crystal momentum basis of each $\Lambda_A$, $\Lambda_B$ sublattice, associated with Bravais vectors $\{{\bf R}_1-{\bf R}_2,{\bf R}_1+{\bf R}_2\}$, then we would obtain the following decoupled equations of motion:

   \begin{equation*}
      \begin{aligned}
            \frac{d^2 B_{\text{ice}}^A (\textbf{k})}{dt^2} = -\omega^2(\textbf{k}) B_{\text{ice}}^A (\textbf{k})  \\
            \frac{d^2 B_{\text{ice}}^B (\textbf{k})}{dt^2 } = -\omega^2(\textbf{k})  B_{\text{ice}}^B (\textbf{k})  \\
      \end{aligned}
  \end{equation*}

  where each of the above equations is now identical to the ordinary Maxwell theory in the square lattice from Eq.\eqref{LatticeMaxwell}, with the dispersion $\omega^2(\textbf{k})$ given by the same expression as in Eq.\eqref{dis}. Here the wavevector $k$ is defined for the Bravais vectors spanned by vectors $\{{\bf R}_1-{\bf R}_2,{\bf R}_1+{\bf R}_2\}$, and therefore its Brillouin zone is half of the size of the Brillouin zone associated with the full translational symmetry of the lattice.
  

  We therefore see that we have two decoupled copies of standard lattice QED, featuring linearly dispersing photon modes at $\textbf{k}=(0,0)$ for each of the sublattices $\Lambda_A$ and $\Lambda_B$. The underlying model has a translational symmetry that exchanges these two sublattices. Therefore in the lattice momentum convention that exploits the full lattice translational symmetry that was employed in deriving the dispersions from Eq.\eqref{mqed}, those two modes combines into a symmetric one and an  antisymmetric one \footnote{Namely with staggered alternating signs $+1,-1$ the $\Lambda_A$ and $\Lambda_B$ sublattices.}, to give rise respectively to the $\textbf{q}=(0,0)$ mode and $\textbf{q}=(\pi,\pi)$ mode in Fig.\ref{2fot}. Now since the femrion carries charge $(q_A,q_B)=(1,1)$ for the gauge fields associated with the two sublattices, it will therefore carry total gauge charge under the sublattice symmetric combination of those fields, associated with the photon $\textbf{q}=(0,0)$, and carry zero charge under their sublattice antisymmetric (staggered) combination, associated with the photon $\textbf{q}=(\pi,\pi)$, explaining the result we encountered in the previous section by direct calculation.

While the above structure is certainly remarkable, its appearance can be intuitively understood by simply appealing to the interplay of the local conservation laws of the spin-ice models and the nature of the Jordan-Wigner composite fermion. Notice that the creation of a Jordan-Wigner composite fermion, which involves the reversal of the z-direction of a single spin, violates necessarily two ice rules associated to two vertices that are connected to by the link in which such spin resides. One of these vertices is located in the $\Lambda_A$ sublatice and the other in the $\Lambda_B$ sublatice. Thus it is natural to see the Jordan-Wigner composite fermion as an extended dipole-like object which has two charges located at the end of the link that connects the two vertices (see Fig.\eqref{modqed}), which will be charged under the sublattice symmetry local gauge transformations, but will be a neutral dipole under the staggered antisymmetric gauge transformation\footnote{Notice that interestingly the global operator associated with the staggered sum of the spin-ice charges over all the lattice in a periodic torus is identically zero. This global subgroup of the staggered gauge group acts therefore trivially within the physical Hilbert space.}. This is why we have called it an ``extended parton", to emphasize the distinction with a ``point-like parton", such as the Abrikosov-Schwinger fermion.

\section{Summary and discussion}
  We have build upon the idea that the standard Jordan-Wigner transmutation that maps spin-$\frac{1}{2}$ degrees of freedom onto spinless fermions in a 2D lattice is exactly equivalent to another celebrated statistical transmutation of attaching a $2\pi$ flux to a spinless hardcore boson that maps these onto spinless composite fermions. In one dimensional chains, such Jordan-Wigner transformation has the property that it maps local Hamiltonians of spins that are symmetric under a global parity onto local Hamiltonian of fermions. However, in 2D models simply imposing a global symmetry is not enough to preserve locality on both the {\it physical side} (the spin representation) and the {\it dual side} (the fermion representation). Nevertheless, this should not be viewed as a {\it bug} but rather as a {\it feature} of the mapping: the non-locality is expressing the fact that the fermion is not the underlying microscopic local particle of the Hilbert space of interest, but instead it is a non-local composite fermion object obtained from attaching a $2\pi$ flux to the underlying microscopic particles.

One ad-hoc approach to handle the above inherent non-locality of Jordan-Wigner/Composite-Fermions in 2D, that is often used in mean-field treatments, is to simply ignore the detailed structure of non-locality by replacing the gauge fields associated with the flux attachment by averaged ``smeared" values that can be chosen to match the net background magnetic field which is given by the composite fermion density. However, in this work we have advanced a completely different route to capture this non-locality of the Jordan-Wigner/Composite-fermions. Namely, we have exploited the fact that Hamiltonians of spin-$\frac{1}{2}$ degrees of freedom that respect certain local symmetries do remain local in their dual Jordan-Wigner/Composite-Fermions representation. The local symmetries that we have focused on are the $U(1)$ symmetries associated with ice rules in 2D quantum spin-ice models which allow to map Rokshar-Kivelson-like models of spins onto local models of Jordan-Wigner/Composite-fermions. The local gauge symmetry structure in these 2D models therefore plays an analogue role to the global symmetries in 1D that allows to keep the models local in the {\it physical} (spin) and {\it dual} (fermion) representations. 

The main difficulty for constructing interesting quantum disordered 2D states within our approach, is that quantum spin-ice models with RK-like Hamiltonians would necessarily map onto interacting Hamiltonians of fermions (e.g. the plaquette resonance term maps onto a quartic fermion interaction). Therefore, we don't have the luxury of 1D where non-trivial spin models can be exactly mapped onto purely free fermion models. More fundamentally, we have seen that even though Slater determinants of fermions can be viewed as a zeroth order mean field approximations to the ground states of quantum spin-ice Hamiltonians (which only satisfy the ice rules in a global averaged sense), such Slater determinants necessarily violate the exact local ice rules, and therefore are not satisfactory approximations to their true ground states satisfying the local ice rules. This obstacle can, however, be naturally overcome by acting on these Slater determinants with a Gutzwiller projector that enforces the local ice rules, making such projected states satisfactory trial ground states of 2D quantum spin-ice Hamiltonians. Computing local spin operators exactly, such as those that enter the RK Hamiltonian, is however a hard analytic task, but it should be possible to efficiently implement these constraints numerically, as it has been done succesfully in previous studies of the more common Gutzwiller projected states of Abrikosov-Schwinger fermions (see e.g.  \cite{PhysRevLett.98.117205, PhysRevLett.101.027204, PhysRevB.87.060405, PhysRevB.104.144406}). This is an interesting direction that we hope future studies will further explore.

However, while explicit analytic calculations of ground state energies for these states is challenging, it is possible to develop a precise understanding of the implementation of the global physical symmetries of the spin model in their dual Jordan-Wigner/Composite-Fermion representation, which is one of the central themes in this study. For the RK-like models such global symmetries include lattice space symmetries, time reversal and onsite spin symmetries (e.g. unitary particle-hole conjugation of hard-core bosons). While the implementation of these symmetries is simple and standard in the physical spin-$\frac{1}{2}$ degrees of freedom, their implementations in the dual Jordan-Wigner/composite-Fermion degrees of freedom can look fairly unusual, which is not surprising because of the non-local nature of the operators creating the composite-Fermion particles. However, because the Jordan-Wigner transformation is an explicit operator map, it is straightforward to determine the exact symmetry action on the Jordan-Wigner/composite-Fermions.

Nevertheless, as a result of the additional local symmetry structure that we have imposed on the Jordan-Wigner/composite-fermions in spin-ice models, a kind of freedom appears on how the symmetry is implemented that bears a resemblance to the problem of implementing physical symmetries on the standard parton constructions of  Abrikosov-Schwinger fermions. Such implementation of symmetries on Gutzwiller projected states of Abrikosov-Schwinger fermions leads naturally to the notions the projective-symmetry groups \cite{WenA02, WenB02}. A remarkable fact about such projective symmetry group implementations is that a given specific microscpic symmetry acting on the physical spins can be implemented in many inequivalent ways on the parton fermions, but these distinct implementations can lead to sharply physically distinct quantum disordered spin liquids of the underlying physical spins (all still obeying the same microscopic symmetries) \cite{WenA02, WenB02}. We have seen that an analogous situation arises in our construction of Jordan-Wigner/composite-fermions states that are Gutzwiller projected to satisfy the ice rules of 2D RK-like models. Namely, gauge inequivalent symmetry implementations on the Jordan-Wigner/composite-fermions can act identically on all the gauge invariant operators within a given subspace with definite values of the ice rules, but which will lead to sharply physically distinct quantum disordered states of the Jordan-Wigner/composite-fermions. This freedom of symmetry implementations turns also to be a very valuable resource.  For example, it is very difficult to enforce the $\pi/2$ rotational symmetry on the mean field states, by using the fully microscopically explicit ``bare" action of this symmetry on the Jordan-Wigner creation operators that include the  full string ordering of the 2D lattice. However we have seen that there are alternative projective symmetry implementations of the $\pi/2$ rotation symmetry that act as effectively local operations on the Jordan-Wigner/composite-fermions, and which have exactly the same action on all the spin-ice gauge invariant operators, which lead therefore to satisfactory and much simpler implementations of this microscopic symmetry on the Gutzwiller projected states.

We have not attempted to classify all the possible spin liquid states that can result from this {\it extended parton construction} of Jordan-Wigner/composite-fermions. From the precedents in Abrikosov-Schwinger fermions \cite{WenA02, WenB02} it is only natural to expect that it will also have a diverse and colorful variety of possibilities, which we hope future studies can investigate. We have instead focused on constructing interesting concrete examples that satisfy the following criteria: (1) a projective symmetry implementation  of all the physical global symmetries of the classic RK model for 2D quantum spin-ice applicable to the six-vertex and quantum dimer subspaces, (2) that the implementation allows for non-zero value of the nearest neighbor hopping of fermions. The first demand guarantees that the composite fermion liquid is a fully symmetric spin liquid that does not break any of the symmetries of the model. Notice in particular that we have enforced time reversal for both six-vertex and quantum dimers and also the particle-hole symmetry of the six-vertex model, which often are neglected in ad-hoc mean-field constructions of composite fermions based on flux smearing at fractional filling of the lattice. The second requirement is a desirable requirement to make the states potentially energetically competitive trial ground states of microscopic RK-like Hamiltonians with short-range couplings, since in these models a big portion of the energy density is determined typically by optimal short distance correlations.

We have successfully constructed two explicit examples of projective symmetry implementations of Jordan-Wigner/composite-fermions based on this {\it extended parton construction}. For the quantum six-vertex model (realized when the Jordan-Wigner/composite-fermions are at half-filling of the lattice) these states feature two massless Dirac cones centered at $(\pi,0)$ and $(0,\pi)$, and thus the state is a putative composite fermion Dirac spin liquid. For the quantum dimer model (realized when the Jordan-Wigner/composite-fermions are at quarter-filling of the lattice) these states display a Fermi-surface of the size of half of the Brillouin zone, and thus the state is a putative composite fermi liquid state. This Fermi surface is perfectly nested when the mean field state only includes nearest neighbor composite fermion hopping, but further neighbor hoppings remove the perfect nesting and could stabilize this state. Because of this strong tendency to being unstable from nesting, this composite Fermi liquid could be a useful parent state to understand the descending ordered states and their competitions in the RK model, which is another interesting direction for future studies.

We have also developed a simplified description of the gauge field fluctuations around these mean field states, aimed at qualitatively capturing the nature of the low energy field theories emerging in the infrared limit (i.e. low energy and long wave-lengths compared to lattice scales) and particularly the nature of the potentially deconfined low energy gauge structure. As it is well known from Abrikosov Schwinger fermions, the low energy gauge structure can be different from the UV parton gauge. We have seen that the low energy gauge structure differs from the UV spin-ice gauge structure, although they have some precise relations. To determine this structure, we have performed an analysis in two stages. 

In the first stage, for a given bare mean field hamiltonian (not Gutzwiller projected) which has some specific non-zero hopping elements Jordan-Wigner/composite-fermions in the lattice, we consider trial Hamiltonians in which only the phases of these non-zero hopping elements are allowed to fluctuate. The spirit behind this is that the amplitude fluctuations should be generically gapped but the fluctuations of the phases could possibly be soft. We then promote these fluctuating phases to become local quantum bosonic degrees of freedom residing on the links connecting the spin sites (or equivalently the links associated with Jordan-Wigner/composite-fermion hopping). Such fluctuating phases of hopping can be viewed as an emergent vector potentials and their canonically conjugate momentum as emergent electric fields. We then generalize the action of UV gauge symmetry group, which is generated by the local operators determining the spin-ice rule constraints, to act not only on the Jordan-Wigner/composite-fermions but also on these bosonic phases/vector potential degrees of freedom, by demanding that the combinantion of the fermion bilinear operators and exponential of the vector potential degrees of freedom associated with hoppings are invariant under the UV spin-ice gauge group, and therefore, in a sense, the phase fluctuations dress the mean field state so as to become locally gauge invariant and thus become a more satisfactory approximation to the fully Gutzwiller projected trial state. With these gauge transformation rules, we then write a simple lattice model for the leading bilinear order Hamiltonian in powers of gauge fields, namely the analogue of the usual Maxwell action, that is consistent with the microscopic global symmetries of the model, while neglecting their compactification (whose potential impact is to be re-considered at the end of the analysis, see below.). For the RK-like models of 2D spin-ice this pure gauge field Hamiltonian features four vector potentials per Bravais unit cell, but there is a local constraint analogous to the zero divergence of electric field, leading to three truly dynamical gauge fields with associated energy bands. Out of these three, two are fully gapped over the entire Brillouin zone (and thus unimportant at low energies) while one band features two linearly dispersing $U(1)$ photon-like modes that are gappless at $(0,0)$ and $(\pi,\pi)$ momentum in the Brillouin zone. This suggests a $U(1) \times U(1)$ low energy gauge structure when compactification is neglected (but see below for discussion of its important impacts).

In the second stage we determine the minimal coupling of the Jordan-Wigner/composite-fermion to this $U(1) \times U(1)$ low energy gauge structure. To do so, we Taylor expand the exponential coupling of the Jordan-Wigner/composite-fermions to the gauge fluctuation fields to linear order in vector potentials, giving us the lattice analogue of the minimal $j\cdot A$ coupling. We have found that the fermions couple minimally to only one of the $U(1)$ gauge fields associated with the massless photon at $(0,0)$ momentum, while they have zero gauge charge under the $(\pi,\pi)$ photon. There is a simple intuitive picture, closely related to the UV gauge structure of the spin-ice, that sheds light on this seemingly peculiar low energy structure. The creation of a composite fermion, which locally reverses one spin along z, violates the two ice rules associated to the vertices connected to such reversed spin. In the lattice gauge theory convention, the lattice of spin-ice vertices is separated into two sublattices and the Gauss law is conventionally taken as a staggered ice rule with alternating signs in the sublattices. In this convention, the spin reversal is viewed as creating a dipole pair of gauge charges but which is in total gauge neutral. Similarly, in our construction we could separate the spin-ice rules into two sublattices, and we could say that the Jordan-Wigner/composite-fermion is charged under a symmetric combination of the ice rules of the two sublattices and neutral dipole-pair object under a staggered antisymmetric combination. These two sublattices are related by a lattice translational symmetry, and since we enforce such symmetry, we encounter that the photon at $(0,0)$ is associated with the sublattice symmetric combination of these ice rules while the sublattice staggered antisymmetric combination is associated with the photon at $(\pi,\pi)$. Because the creation of the Jordan-Wigner/composite-fermion violates the ice rules symmetrically, the particle only carries gauge charge under the symmetric photon at $(0,0)$, but behaves as a neutral gauge dipolar object with respect to the antisymmetric photon $(\pi,\pi)$.

We have therefore obtained a low energy gauge structure of two massless Dirac fermions minimally coupled to a $U(1)$ massless photon, and neutral under another $U(1)$ massless photon. We would like now re-consider qualititatively the impact of gauge field compactification on this low energy structure. The photon that ``sees" the fermions as neutral dipoles, propagates in this medium as if it was an insulating dielectric liquid. This is not very different from a how a photon would ``see" a gapped insulating charged fermionic matter. Monopole fluctuations are therefore expected to be relevant and lead to Polyakov-style confiment for this $U(1)$ sublattice antisymmetric photon at $(\pi,\pi)$, as it is generically expected in 2+1D for a compact $U(1)$ gauge field when the matter that carries its gauge charge is gapped. We thus expect that compactification gaps out this $(\pi,\pi)$ photon and that it is not relevant at low energies. However, since the fermion is already a short distance neutral dipole object with respect to this gauge field, such gauge confinement is not expected to confine the fermions themselves. We are thus left with a low energy structure in which we have two gapless Dirac nodes minimally coupled to a single $U(1)$ photon, the one gapless at momentum $(0,0)$. The  ultimate details of the infrared behavior and strict stability of such $N=2$ QED in 2+1 dimensions, is still an open problem \cite{PhysRevD.105.085008,  PhysRevD.100.054514, Chester_2016}, but provided such theory flows to a stable fixed point, our state would be therefore an analogue of a Dirac spin liquid of composite fermions (but with a pseudoscalar symmetry implementation \cite{Inti}, see below). We hope that future studies can investigate in more detail this qualitative arguments on the nature of the low energy effective field theory.

Finally, we would like comment on the connections between our construction and the pseudo-scalar spin liquids introduced in Ref. \cite{Inti}. The Jordan-Wigner/composite-fermion can be naturally understood to behave as a pseudo-scalar spinon under symmetries that reverse the direction of the $z$-spin, because this is equivalent to the Jordan-Wigner/composite-fermion occupation on a site. Therefore the ordinary spin time reversal symmetry that squares to $-1$ and space operations such as mirrors that reverse the $z$-spin, would act as particle-hole conjugations on the Jordan-Wigner/composite-fermion. The emergent magnetic and electric fields would also have opposite transformation laws to those of ordinary QED, for example being even and odd respectively under such spin time-reversal operation. Therefore we see that the $U(1)$ gauge structure is pseudo-scalar in the sense described in Ref.\cite{Inti}. This can also be understood in very direct microscopic terms. For example, magnetic field operator associated with the simplest gauge invariant loop of quantum spin-ice, is the four spin operator composed of alternating spin raising and lowering operators in a plaquette (properly symmetrized so as to make the analogue of $\sin(B)$ combination of lattice gauge theory). It is easy to see that this operator is even under the previously mentioned mirrors and time reversal operations. This magnetic field operator can be viewed as a measure of a local correlation for the XY projection of the spins to spiral around a small closed loop, and thus is physically very different from those of more traditional U(1) spin liquids, such as the triple product correlator associated spin chirality around triangles \cite{PhysRevB.51.1922, PhysRevB.73.155115, PhysRevLett.104.066403} which transforms in the similar way as the usual magnetic field experienced by electrons under point group and time-reversal. Therefore, our current construction of Jordan-Wigner/composite-fermion provides a different and perhaps more intuitive way to describe certain pseudoscalar spin liquids which illuminates on the kind of correlations associated the appearance of their magnetic fields , namely a kind of tendency towards short-distance spin-spiraling on loops. Understanding more precisely these connections is another interesting avenue of future research, which could help understand how to realize such spin liquids in real materials, such as $\alpha$-RuCl$_3$, where the oscillations of thermal conductivity seen in experiments \cite{czajka2021oscillations,bruin2022origin,suetsugu2022evidence,lefranccois2023oscillations} are consistent with the expected quantum oscillations of pseudo-scalar U(1) spin-liquid \cite{Inti}.

\vspace{3mm}
\begin{acknowledgments}
 We are thankful to Hong-Hao Tu, Debasish Banerjee, Karlo Penc, Nic Shannon, Xue-Feng Zhang and Roderich Moessner for stimulating discussions. I.S. is specially thankful to Zhenjiu Wang for discussions and performing unpublished analysis in a preeliminary stage of the project. L.G. would like to thank \href{https://davidemorgante.github.io/}{Davide Morgante} for his help with Ti\textit{k}Z pictures. We acknowledge
support by the Deutsche Forschungsgemeinschaft (DFG) through research grant project number 518372354. 
\end{acknowledgments}

\bibliography{BCD-ingap-rectification}

\clearpage

\renewcommand{\theequation}{S-\arabic{equation}}
\renewcommand{\thefigure}{S-\arabic{figure}}
\renewcommand{\thetable}{S-\Roman{table}}
\makeatletter
\renewcommand\@biblabel[1]{S#1.}
\setcounter{equation}{0}
\setcounter{figure}{0}

\onecolumngrid

\appendix

\section{$2 \pi$ Flux attachment equivalence}
\label{2DfluxJWequiv}
We adopt the same labeling convention defined in the main text in Sec. \ref{equivJWCF}, namely sites are ordered according to western typing convention. We take the lattice sites of the original square lattice to be labeled by $\textbf{r}=(x,y)$, where $x,y$ are integers running from $1,...,N$. We would like to show here for any site $\textbf{r}$ in the lattice, the following equations hold:
\begin{enumerate}
    \item \begin{equation*} 
   {f}^{\dag}(\textbf{r} + \textbf{e}_x + \textbf{e}_y) \, e^{i \pi \sum_{ \textbf{r} \preceq \textbf{r''} \prec \textbf{r} + \textbf{e}_x + \textbf{e}_y} {n}(\textbf{r''})} \, {f}(\textbf{r}) =     {f}^{\dag}(\textbf{r} + \textbf{e}_x + \textbf{e}_y) \exp\bigg(i \int_{\textbf{r}}^{\textbf{r} + \textbf{e}_x + \textbf{e}_y} \textbf{A}(\textbf{x}) \cdot d \textbf{x}\bigg)  {f}(\textbf{r})
\end{equation*}
\item \begin{equation*} 
  {f}^{\dag}(\textbf{r} - \textbf{e}_x + \textbf{e}_y) \, e^{i \pi \sum_{ \textbf{r} \preceq \textbf{r''} \prec \textbf{r} - \textbf{e}_x + \textbf{e}_y} {n}(\textbf{r''})} \, {f}(\textbf{r}) =     {f}^{\dag}(\textbf{r} - \textbf{e}_x + \textbf{e}_y) \exp\bigg(i \int_{\textbf{r}}^{\textbf{r} - \textbf{e}_x + \textbf{e}_y} \textbf{A}(\textbf{x}) \cdot d \textbf{x}\bigg)  {f}(\textbf{r})
\end{equation*}
\item\begin{equation*} 
{f}^{\dag}(\textbf{r}) \, e^{i \pi \sum_{ \textbf{r}  + \textbf{e}_x - \textbf{e}_y \preceq \textbf{r''} \prec \textbf{r}} {n}(\textbf{r''})} \, {f}(\textbf{r}  + \textbf{e}_x - \textbf{e}_y) =     {f}^{\dag}(\textbf{r}) \exp\bigg(i \int_{\textbf{r}  + \textbf{e}_x - \textbf{e}_y}^{\textbf{r}} \textbf{A}(\textbf{x}) \cdot d \textbf{x}\bigg)  {f}(\textbf{r}  + \textbf{e}_x - \textbf{e}_y)
\end{equation*}
\item \begin{equation*} 
  {f}^{\dag}(\textbf{r}) \, e^{i \pi \sum_{ \textbf{r}  - \textbf{e}_x - \textbf{e}_y \preceq \textbf{r''} \prec \textbf{r}} {n}(\textbf{r''})} \, {f}(\textbf{r}  - \textbf{e}_x - \textbf{e}_y) =     {f}^{\dag}(\textbf{r}) \exp\bigg(i \int_{\textbf{r}  - \textbf{e}_x - \textbf{e}_y}^{\textbf{r}} \textbf{A}(\textbf{x}) \cdot d \textbf{x}\bigg)  {f}(\textbf{r} - \textbf{e}_x - \textbf{e}_y)
\end{equation*}
\end{enumerate}

To demonstrate the above, it is sufficient to demonstrate only $1.$ and $2.$, because the other two cases follows from the previous ones by globally translating the initial and final sites of the hopping. Below we show the arguments then for $1.$ and $2.$:

\begin{enumerate}
    \item The sites involved in the 2D Jordan-Wigner transformation for the hopping term ${b}^{\dag}(\textbf{r}+\textbf{e}_x+\textbf{e}_y) {b}_{\textbf{r}}$ are, by \eqref{bdagb}:

\begin{equation*}
    (r_x,r_y) \to (r_x+1, r_y) \to \cdots \to  (N,r_y) \to (1,r_y+1) \to \cdots \to (r_x,r_y+1)
\end{equation*}
 
 Thus the phase gained is:

\begin{equation*}
    \exp\bigg[i \pi \big(n(r_x,r_y) + n(r_x+1, r_y)+ \cdots + n(N,r_y) + n(1,r_y+1) + \cdots + n(r_x-1,r_y+1) + n(r_x,r_y+1) \big) \bigg]
\end{equation*}
Let's focus now on the gauge potential. We are free to choose the  path $\textbf{r} \to \textbf{r}+ \textbf{e}_y \to \textbf{r}+ \textbf{e}_y + \textbf{e}_x$. Since:

\begin{equation*}
    \exp\bigg(i \int_{\textbf{r}}^{\textbf{r}+\textbf{e}_x+\textbf{e}_y} \textbf{A}(\textbf{x}) \cdot d \textbf{x} \bigg) = \exp \bigg(i \int_{\textbf{r}}^{\textbf{r}+\textbf{e}_y} \textbf{A}(\textbf{x}) \cdot d \textbf{x} \bigg) \exp \bigg(i \int_{\textbf{r}+\textbf{e}_y}^{\textbf{r}+\textbf{e}_y + \textbf{e}_x} \textbf{A}(\textbf{x}) \cdot d \textbf{x} \bigg)
\end{equation*}

we have, using the previously established equations, that:

\begin{equation*}
\begin{aligned}
   \exp\bigg(i \int_{\textbf{r}}^{\textbf{r}+\textbf{e}_x+\textbf{e}_y} \textbf{A}(\textbf{x}) \cdot d \textbf{x} \bigg) &=  \bigg( e^{i \pi \sum_{ \textbf{r} \preceq \textbf{r''} \prec \textbf{r}+\textbf{e}_y - \textbf{e}_x} {n}(\textbf{r''})} \bigg) \bigg( e^{i \pi n(\textbf{r}+\textbf{e}_y)} \bigg)\\
    &=    \exp \bigg[i \pi \sum_{ \textbf{r} \preceq \textbf{r''} \prec \textbf{r}+\textbf{e}_y } {n}(\textbf{r''}) \bigg]
    \end{aligned}
\end{equation*}

    \item The sites involved in the 2D Jordan-Wigner transformation for the hopping term ${b}^{\dag}(\textbf{r}-\textbf{e}_x + \textbf{e}_y) {b}(\textbf{r})$ are, by \eqref{bdagb}:

\begin{equation*}
    (r_x,r_y) \to (r_x+1, r_y) \to \cdots \to  (N,r_y) \to (1,r_y+1) \to (r_x-2,r_y+1)
\end{equation*}
 
 Thus the phase gained is:

\begin{equation*}
    \exp\bigg[i \pi \big(n(r_x,r_y) + n(r_x+1, r_y)+ \cdots + n(N,r_y) + n(1,r_y+1) + \cdots + n(r_x-2,r_y+1) \big) \bigg]
\end{equation*}
Let's focus now on the gauge potential. We are free to choose the  path $\textbf{r} \to \textbf{r}+\textbf{e}_y \to  \textbf{r}+\textbf{e}_y -\textbf{e}_x$. The result is not affected by the orientation in which the integral is taken, as a simple change in orientation of the contour would just give the complex conjugate of the phase gained on any link and each of them can only be $\pm 1$. Thus:

\begin{equation*}
    \exp\bigg(i \int_{\textbf{r}}^{\textbf{r}-\textbf{e}_x+\textbf{e}_y} \textbf{A}(\textbf{x}) \cdot d \textbf{x} \bigg) = \exp \bigg(i \int_{\textbf{r}}^{\textbf{r}+\textbf{e}_y} \textbf{A}(\textbf{x}) \cdot d \textbf{x} \bigg) \exp \bigg(i \int_{\textbf{r}+\textbf{e}_y}^{\textbf{r}+\textbf{e}_y -\textbf{e}_x} \textbf{A}(\textbf{x}) \cdot d \textbf{x} \bigg)
\end{equation*}

Using the previously established equations, we get:

\begin{equation*}
\begin{aligned}
     \exp\bigg(i \int_{\textbf{r}}^{\textbf{r}-\textbf{e}_x+\textbf{e}_y} \textbf{A}(\textbf{x}) \cdot d \textbf{x} \bigg)  &=  \bigg( e^{i \pi \sum_{ \textbf{r} \preceq \textbf{r''} \prec \textbf{r}+\textbf{e}_y} {n}(\textbf{r''})} \bigg) \bigg( e^{i \pi n(\textbf{r} -\textbf{e}_x +\textbf{e}_y)} \bigg)\\
    &=  \exp\bigg[i \pi (n(r_x,r_y) + \cdots + n(r_x-2,r_y+1) + n(r_x-1,r_y+1) + n(r_x-1,r_y+1) ) \bigg]\\
    &= \exp\bigg[i \pi \big(n(r_x,r_y)  + \cdots + n(r_x-2,r_y+1) + \cancel{2n(r_x-1,r_y+1)} \big) \bigg]\\
    &= \exp\bigg[i \pi  \sum_{\textbf{r} \preceq \textbf{r}'' \prec \textbf{r} + \textbf{e}_y-\textbf{e}_x} n(\textbf{r}'') \bigg]\\
    \end{aligned}
\end{equation*}
\end{enumerate}

We are now ready to prove Eq.\eqref{clasgau} of the main text, which states that for any sites $\textbf{r}, \textbf{r}'$ in the lattice, it holds that:

\begin{equation*}
     {f}^{\dag}(\textbf{r}')  e^{i \pi \sum_{ {\textbf{r}} \preceq \textbf{r''} \prec {\textbf{r}}'} {n}(\textbf{r''})}  {f}(\textbf{r}) =  {f}^{\dag}(\textbf{r}')  \exp\bigg[i \int_{\textbf{r}}^{\textbf{r}'} \textbf{A}(\textbf{x}) \cdot d \textbf{x}\bigg]  {f}(\textbf{r})
 \end{equation*}

\begin{proof}

The sites involved in the 2D Jordan-Wigner transformation for the hopping term ${b}^{\dag}(\textbf{r}') {b}(\textbf{r})$ are (without loss of generality we will assume $r_y'>r_y+1$), by \eqref{bdagb}:

\begin{equation*}
    (r_x,r_y) \to (r_x+1, r_y) \to \cdots \to  (N,r_y) \to (1,r_y+1) \to \cdots \to (N,r_y+1) \to \cdots \to (N,r'_y-1) \to (1,r'_y) \to \cdots \to (r'_x-1,r'_y)
\end{equation*}
 
 Thus the phase gained is:
 
\begin{equation*}
    \exp\bigg[i \pi \sum_{\textbf{r} \preceq \textbf{r}'' \prec \textbf{r}'} n(\textbf{r}'') \bigg]
\end{equation*}

    Let's focus now on the gauge potential. We are free to choose the $L$-shaped path that connects $\textbf{r}$ to $\textbf{r}'$ for $r'_x \geq r_x$ (figure \ref{lpTH}):

\begin{figure}[H]
  \centering
   \includegraphics[trim={0cm 8cm 0cm 6cm}, clip, width=\textwidth]{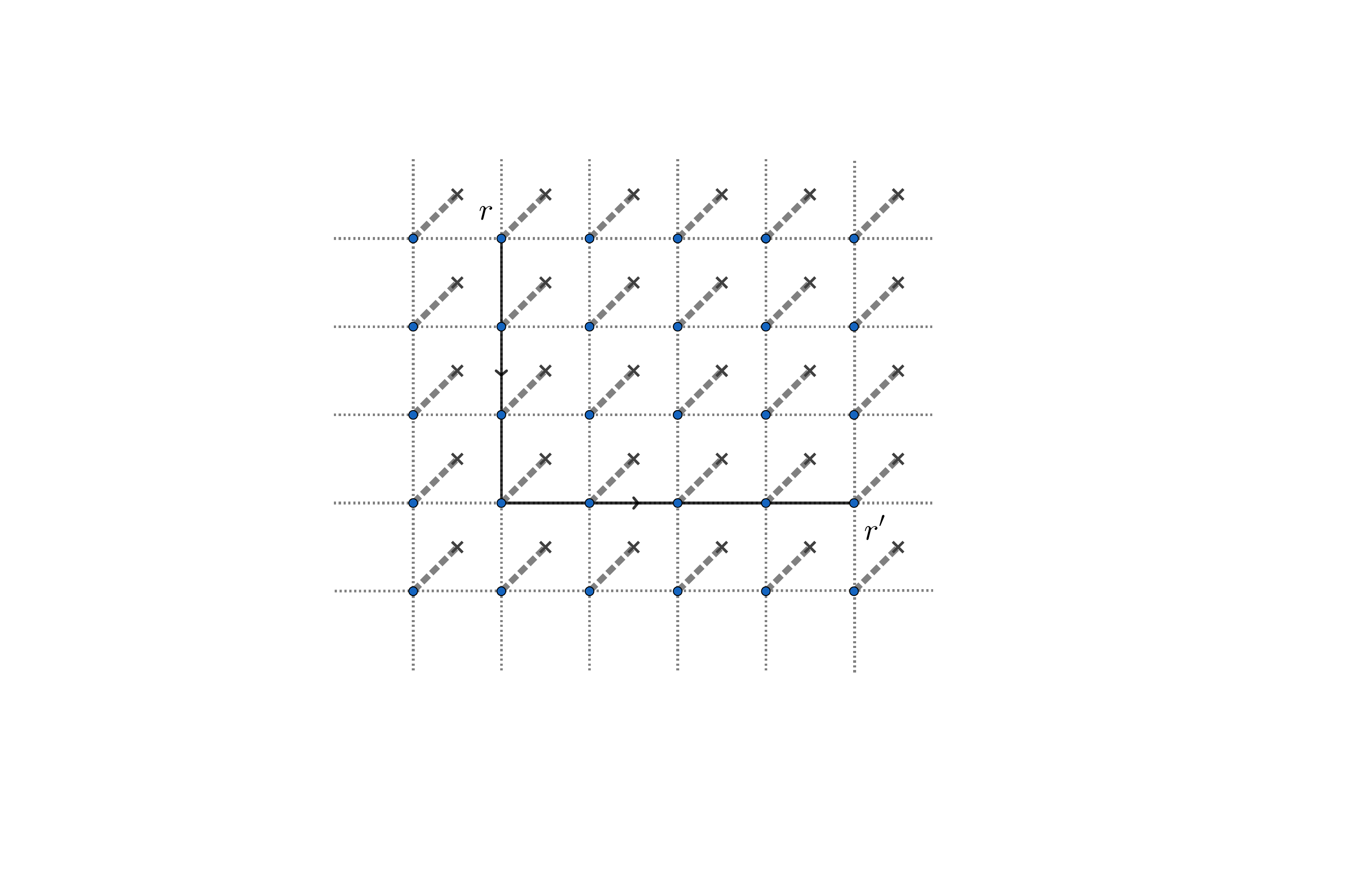}
   \caption{$L$-shaped path that connects $\textbf{r}$ to $\textbf{r}'$ for $r_x \leq r'_x$.}
   \label{lpTH}
  \end{figure}

Since the phases factorizes, we have:

\begin{equation*}
\begin{aligned}
    \exp\bigg(i \int_{(r_x,r_y)}^{(r'_x,r'_y)} \textbf{A}(\textbf{x}) \cdot d \textbf{x} \bigg) &= \exp \bigg(i \int_{(r_x,r_y)}^{(r_x,r'_y)} \textbf{A}(\textbf{x}) \cdot d \textbf{x} \bigg) \exp \bigg(i \int_{(r_x,r'_y)}^{(r'_x,r'_y)} \textbf{A}(\textbf{x}) \cdot d \textbf{x} \bigg)\\
    &=\bigg( e^{i \pi \sum_{ \textbf{r} \preceq \textbf{r''} \prec (x,y')} {n}(\textbf{r''})} \bigg) \bigg( e^{i \pi \sum_{x \leq x'' < x'} n(r''_x,r'_y )} \bigg)\\
    &=  \exp\bigg[i \pi \sum_{\textbf{r} \preceq \textbf{r}'' \prec \textbf{r}'} n(\textbf{r}'')\bigg]\\
    \end{aligned}
\end{equation*}

For $r'_x<r_x$, the phase gained is:

\begin{equation*}
    \exp\bigg[i \pi \sum_{\textbf{r} \preceq \textbf{r}'' \prec \textbf{r}'} n(\textbf{r}'') \bigg]
\end{equation*}

We can argue similarly, choosing the path \ref{L2}:

\begin{figure}[H]
  \centering
   \includegraphics[trim={0cm 8cm 0cm 6cm}, clip, width=\textwidth]{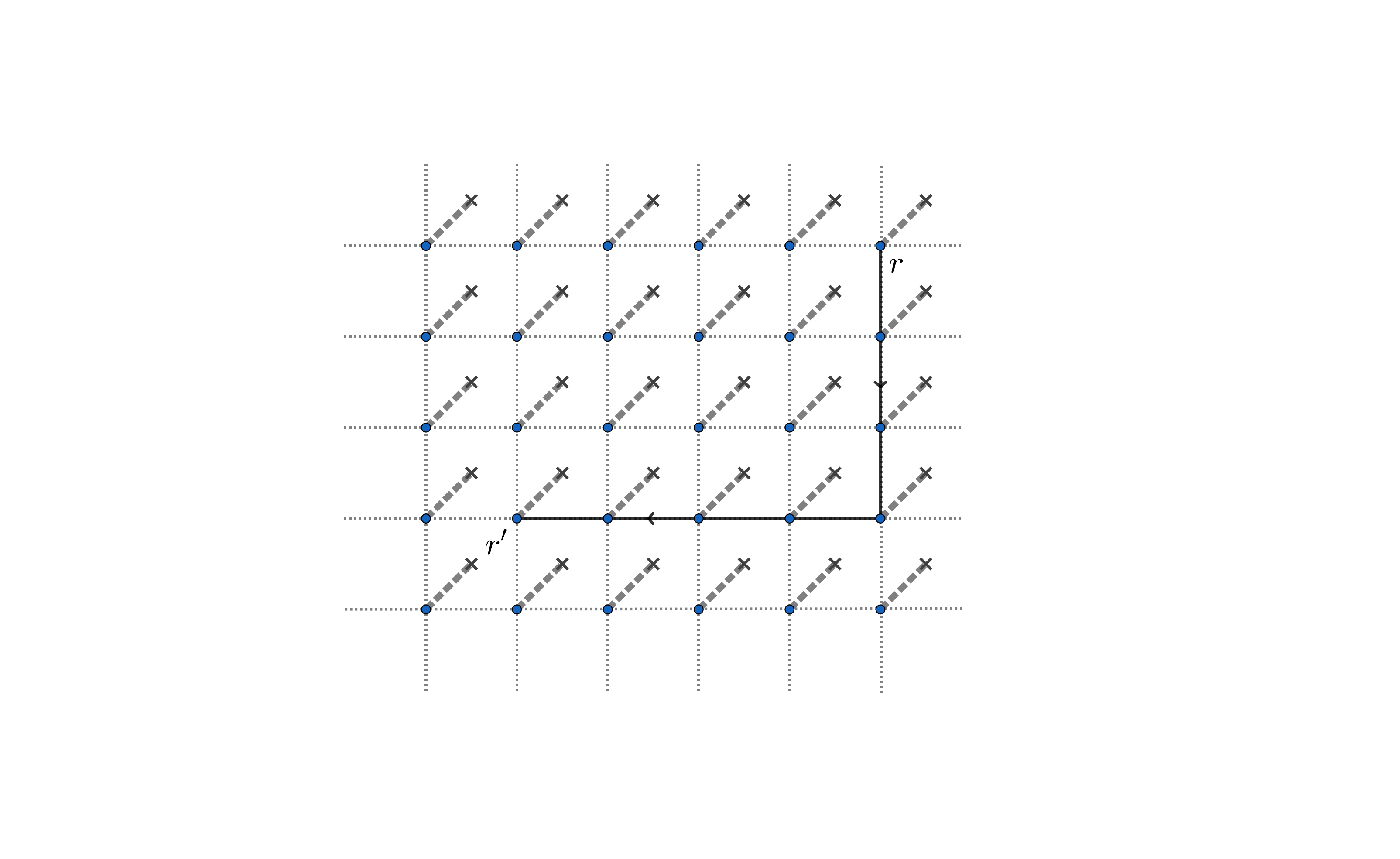}
   \caption{$L$-shaped path that connects $\textbf{r}$ to $\textbf{r}'$ for $r'_x<r_x$.}
   \label{L2}
  \end{figure}
  
 and we find:

\begin{equation*}
\begin{aligned}
    \exp\bigg(i \int_{(r_x,r_y)}^{(r'_x,r'_y)} \textbf{A}(\textbf{x}) \cdot d \textbf{x} \bigg) &= \exp \bigg(i \int_{(r_x,r_y)}^{(r_x,r'_y)} \textbf{A}(\textbf{x}) \cdot d \textbf{x} \bigg) \exp \bigg(i \int_{(r_x,r'_y)}^{(r'_x,r'_y)} \textbf{A}(\textbf{x}) \cdot d \textbf{x} \bigg)\\
    &=\bigg( e^{i \pi \sum_{ \textbf{r} \preceq \textbf{r''} \prec (x,y')} {n}(\textbf{r''})} \bigg) \bigg( e^{i \pi \sum_{x' \leq x'' < x} n(r''_x,r'_y )} \bigg)\\
    &=   \exp\bigg[i \pi \sum_{\textbf{r} \preceq \textbf{r}'' \prec \textbf{r}'} n(\textbf{r}'') \bigg]
    \end{aligned}
\end{equation*}

\end{proof}

   \section{Projective symmetry implementations}\label{PSI}

   \subsection{$\frac{\pi}{2}$ rotation enforcement}
    In section \ref{C1}, we worked out the form of fermion transformation under $U_{\frac{1}{4}} R_{\frac{\pi}{2}}$ (), then for the particular case of the mean field nearest neighbours hoppings:

   
      \begin{equation*}
    t_{31} f_3^{\dag} f_1 + t_{12} f_1^{\dag} f_2 + t_{24} f_2^{\dag} f_4 +t_{43} f_4^{\dag} f_3 \xrightarrow[]{\text{adj}_{U_{\frac{1}{4}}R_{\frac{\pi}{2}}}(\cdot)}   -it_{31} f_4^{\dag} f_3 +i t_{12} f_3^{\dag} f_1 -i t_{24} f_1^{\dag} f_2 + it_{43} f_2^{\dag} f_4 
   \end{equation*}
   
   which gives the following system of constraints:
   
   \begin{equation*}
       \begin{cases}
          it_{31} = -t_{43}\\
          i t_{12} = t_{31}\\
          it_{24} = -t_{12}\\
          it_{43} = t_{24}
       \end{cases}
   \end{equation*}
   
   If we call $t_{31} \doteq t$, the solution of the system is given by the hoppings depicted in Fig.\ref{fig121} below. Therefore, we see that the $\pi/2$ rotation is highly constraining as it combination with translations implemented in a non-projective way, would fix all the hoppings up to a global complex phase.

    \begin{figure}[H]
  \centering
   \includegraphics[trim={0cm 3cm 0cm 3cm}, clip, width=0.75\textwidth]{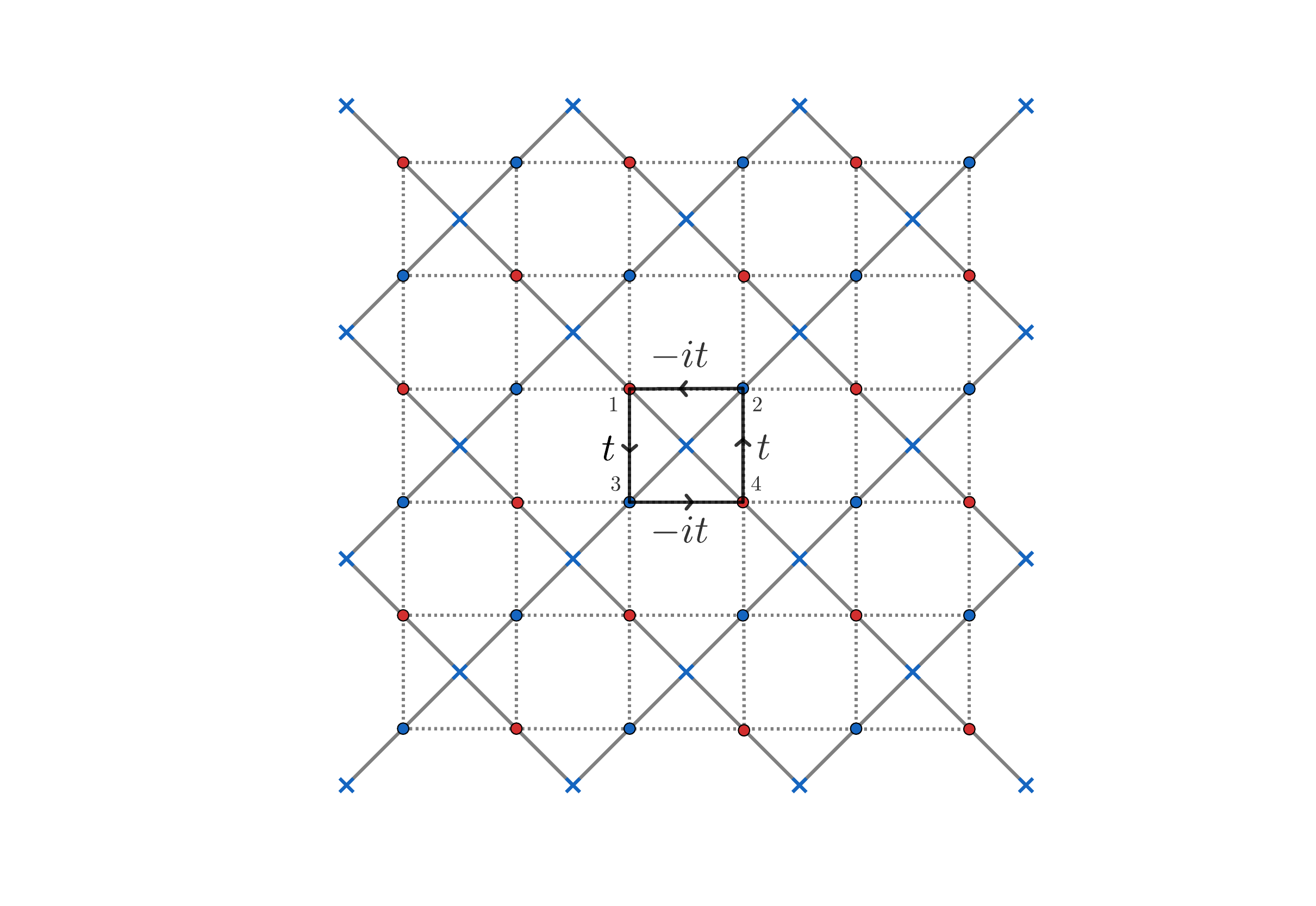}
   \caption{The arrow indicates in which direction the hopping happens, namely in the site where the arrow points a fermion just created.}
   \label{fig121}
  \end{figure}
  
   
   
   \subsection{Time-reversal enforcement}
   The bare time reversal operation, denoted by $\Theta$, is defined in Table \ref{symtablejw} and Eq.\eqref{parton}. If we then try to enforce invariance under such bare time reversal operation on the hopping terms, we get the following set of constraints:
   
   \begin{equation*}
       \begin{cases}
           t=t^*\\
         -it = i t^*
       \end{cases}
   \end{equation*}
   
   which can be solved only by $t=0$. We are then led to conclude that the bare representation of time reversal symmetry makes all the nearest neighbour hopping terms vanish. \\

   Inspired by projective symmetries group construction, we can try to represent differently the time reversal symmetry, dressing it with an unitary map in the gauge Group. Let's consider for example the following gauge transformations:
   
   \begin{equation*}
 \begin{aligned}
              G_x b^{\dag}(\textbf{r}) G^{\dag}_x &=  (-1)^{r_x} b^{\dag}(\textbf{r})\\
              G_y b^{\dag}(\textbf{r}) G^{\dag}_y &=  (-1)^{r_y} b^{\dag}(\textbf{r})\\
 \end{aligned}
   \end{equation*}
   
   It's clear that on boson plaquette operators, $\text{adj}_{G_i}(L_{\textbf{R}})= L_{\textbf{R}}$ ($i=x,y$) because it is built by two terms with even $i$ coordinate and two terms with odd $i$ coordinate.  Moreover since:
   
   \begin{equation*}
       G_i \, \sigma^z(\textbf{r}) \, G_i^{\dag} =   G_i [b(\textbf{r}), b^{\dag}(\textbf{r})] G_i^{\dag} = [G_i b(\textbf{r}) G_i^{\dag}, G_i b^{\dag}(\textbf{r})G_i^{\dag}] = [b(\textbf{r}), b^{\dag}(\textbf{r})] = \sigma^z(\textbf{r})
   \end{equation*}
   
   the invariance under the action of $G_i$ applies to any gauge invariant operator, thus it is a UV gauge transformation. The two gauge transformations have the following action on fermion operators:

   \begin{equation*}
 \begin{aligned}
              G_x f^{\dag}(\textbf{r}) G^{\dag}_x &=  (-1)^{r_x} f^{\dag}(\textbf{r})\\
              G_y f^{\dag}(\textbf{r}) G^{\dag}_y &=  (-1)^{r_y} f^{\dag}(\textbf{r})\\
 \end{aligned}
   \end{equation*}
   
   Let's see now the dressed time reversal transformations $\Theta_i \doteq G_i \,\Theta$ constraint the hoppings. For the horizontal hoppings, the additional actions of $G_x$ and $G_y$ respectively does not change the value (as both sites have the $x$ coordinate) and make them acquire a minus sign (as the sites have a different parity $y$ coordinate). For the vertical hoppings, conversely, the additional actions of $G_x$ and $G_y$ respectively make them acquire a minus sign (as the sites have a different parity $x$ coordinate) and does not change the value (as both sites have the $y$ coordinate). In the end, the constraints imposed by $\Theta_x$ and $\Theta_y$ are respectively:
   
  \begin{equation*}
       \begin{cases}
           t=t^*\\
         -it = -i t^*
       \end{cases}
   \end{equation*}
   
   i.e. $t \in \mathbb{R}$, and
   
   \begin{equation*}
       \begin{cases}
           t=-t^*\\
         -it = i t^*
       \end{cases}
   \end{equation*}
   
   i.e. $t \in i\, \mathbb{R}$. And therefore we conclude that time-reversal symmetry $\Theta$ can be implemented on the JW/composite-fermions by two different extended projective symmetry groups $\Theta_x$ and $\Theta_y$.

  \subsection{Reflection symmetries enforcement}
    
   We enforce now reflection symmetries on $\Theta_x$ and $\Theta_y$ extended projective symmetry groups, with the condition that all nearest neighbour hopping (already fixed by in the previous two subsections) are non vanishing.

For the $\Theta_x$ extended projective symmetry group ($t \in \mathbb{R}$), reflection symmetries are realized in the following way on the JW/composite fermions (see Tables \ref{symtable},\ref{symtablejw},\ref{implsym} and Fig.\ref{D8} for conventions):

   \begin{equation}
   \begin{aligned}
     S_x &\leftrightarrow G_x \Sigma_x\\
     S_y &\leftrightarrow G_x \Sigma_y\\
     S_{1} &\leftrightarrow G_yU_{\frac{1}{4}} \Sigma_1\\
     S_{2} &\leftrightarrow G_y  U_{\frac{1}{4}} \Sigma_2.
     \end{aligned}
  \end{equation}

\begin{figure}[H]
  \centering
   \includegraphics[trim={0cm 0cm 0cm 0cm}, clip, width=\textwidth]{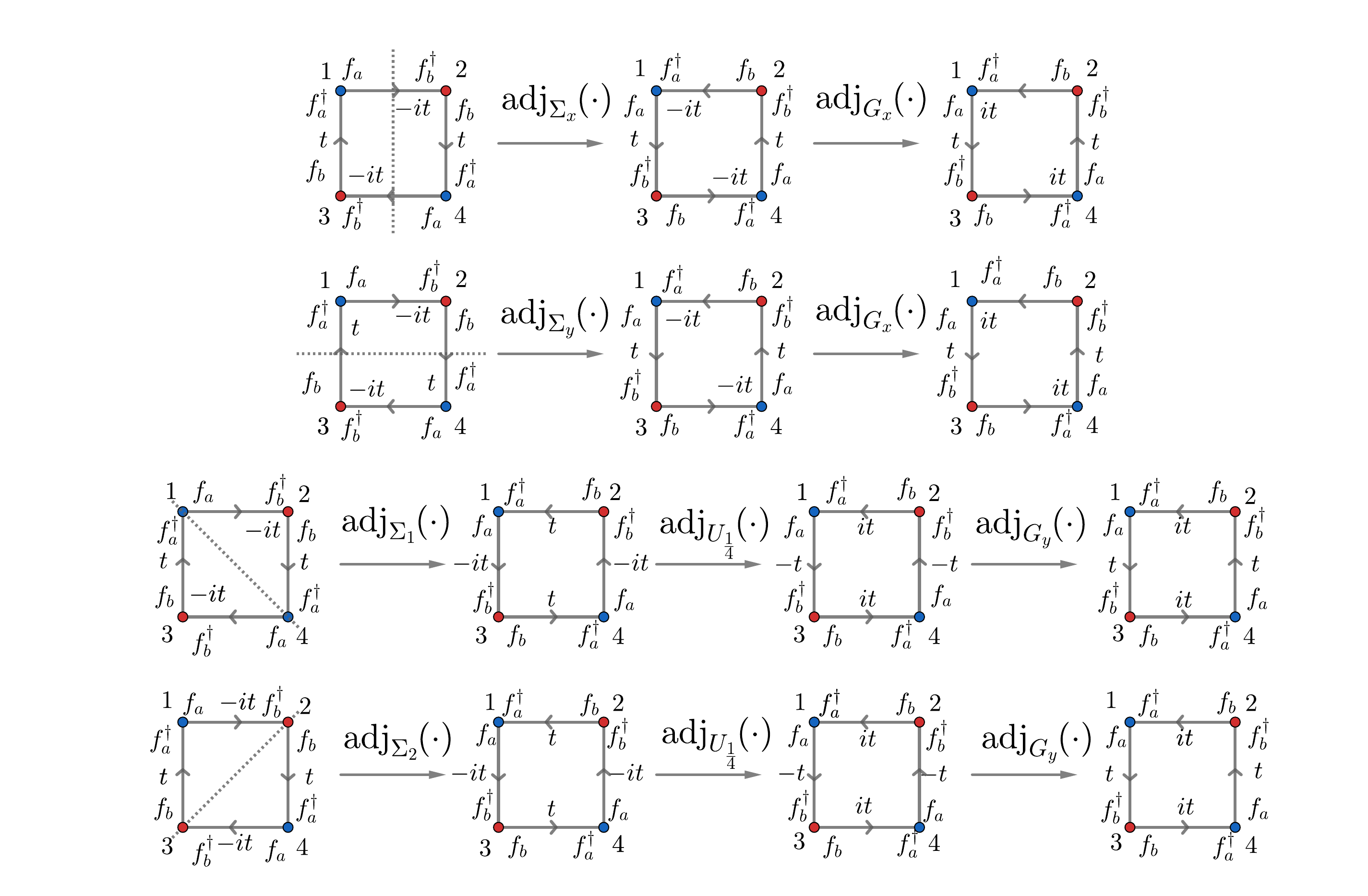}
   \caption{Reflection action on nearest neighbour hoppings for $\Theta_x$ extended projective symmetry group.}
   \label{mirror real}
  \end{figure}

Figure \ref{mirror real} shows why the above action allows for the nearest neighbor hopping  of JW/composite-fermion for the $\Theta_x$ extended projective symmetry group.
  

 On the other hand, for the $\Theta_y$ extended projective symmetry group ($t \in i \mathbb{R}$), reflection symmetries are realized in the following way:
    
    \begin{equation}
    \begin{aligned}
     S_x &\leftrightarrow G_y \Sigma_x\\
   S_y &\leftrightarrow G_y \Sigma_y \\
   S_{1} &\leftrightarrow G_x U_{\frac{1}{4}} \Sigma_1\\
  S_{2} &\leftrightarrow G_x U_{\frac{1}{4}} \Sigma_2.
   \end{aligned}
  \end{equation}

\begin{figure}[H]
  \centering
   \includegraphics[trim={0cm 0cm 0cm 0cm}, clip, width=\textwidth]{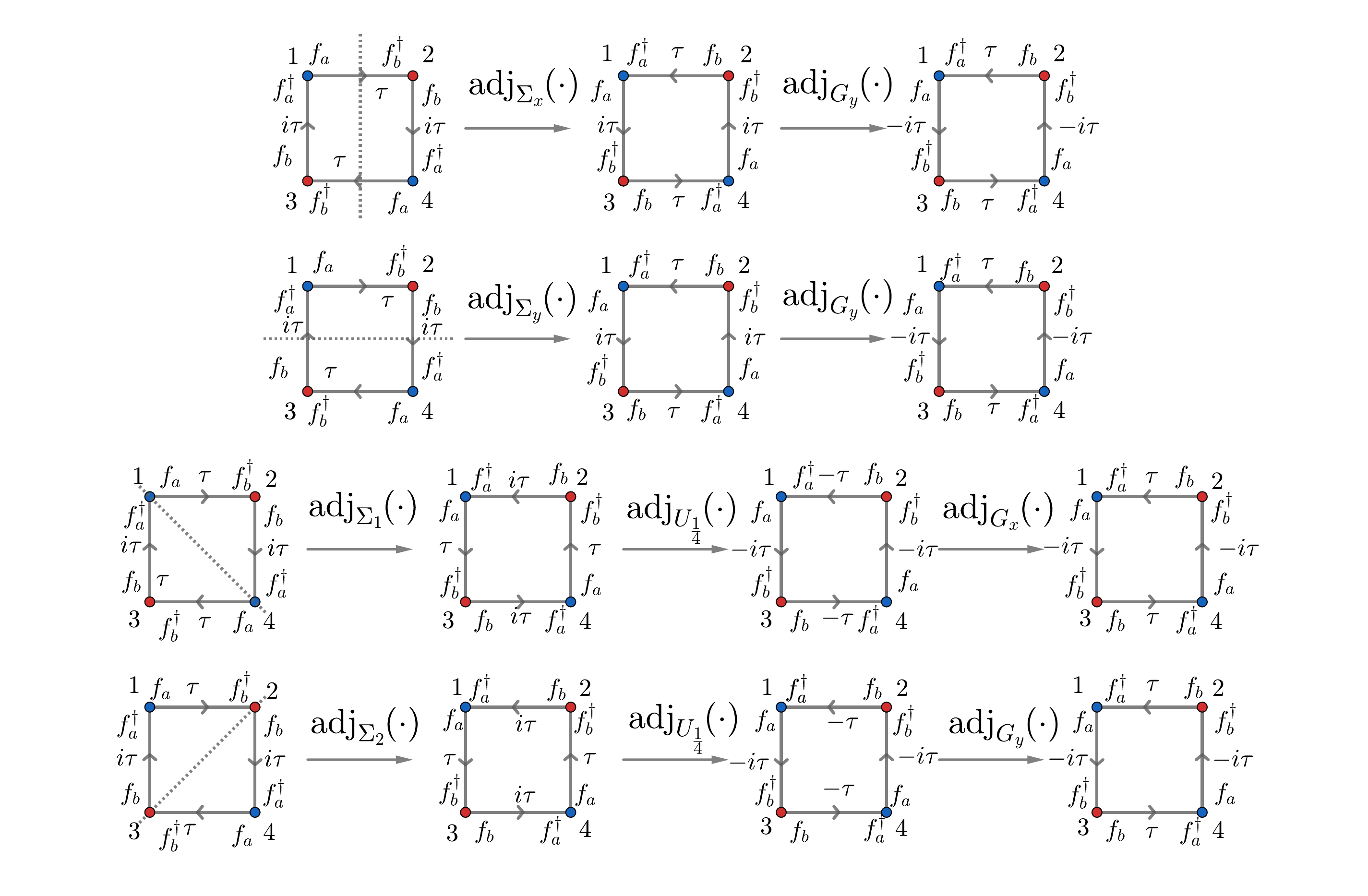}
   \caption{Reflection action on nearest neighbour hoppings for $\Theta_y$ extended projective symmetry group ($t= i \tau$ with $\tau \in \mathbb{R}$).}
   \label{mirror imaginary}
  \end{figure}

  Figure \ref{mirror imaginary} shows why the above action allows for the nearest neighbor hopping  of JW/composite-fermion for the $\Theta_y$ extended projective symmetry group.

  \subsection{Particle-Hole enforcement}
  
  Similarly to Reflection symmetries, particle-hole symmetry is enforced on $\Theta_x$ and $\Theta_y$ extended projective symmetry groups, with the condition that all nearest neighbour hopping are non vanishing.\\

  For the $\Theta_x$ and $\Theta_y$ extended projective symmetry group ($t \in \mathbb{R}$), particle-hole symmetry is respectively realized in the following way on the JW/composite fermions (see Tables \ref{symtable},\ref{symtablejw},\ref{implsym} and Fig.\ref{D8} for conventions):

  \begin{equation*}
  \begin{aligned}
      X &\leftrightarrow G_x 
      \Xi\\
      X &l\leftrightarrow G_y \Xi
      \end{aligned}
  \end{equation*}

\begin{figure}
  \centering
   \includegraphics[trim={0cm 6cm 0cm 0cm}, clip, width=\textwidth]{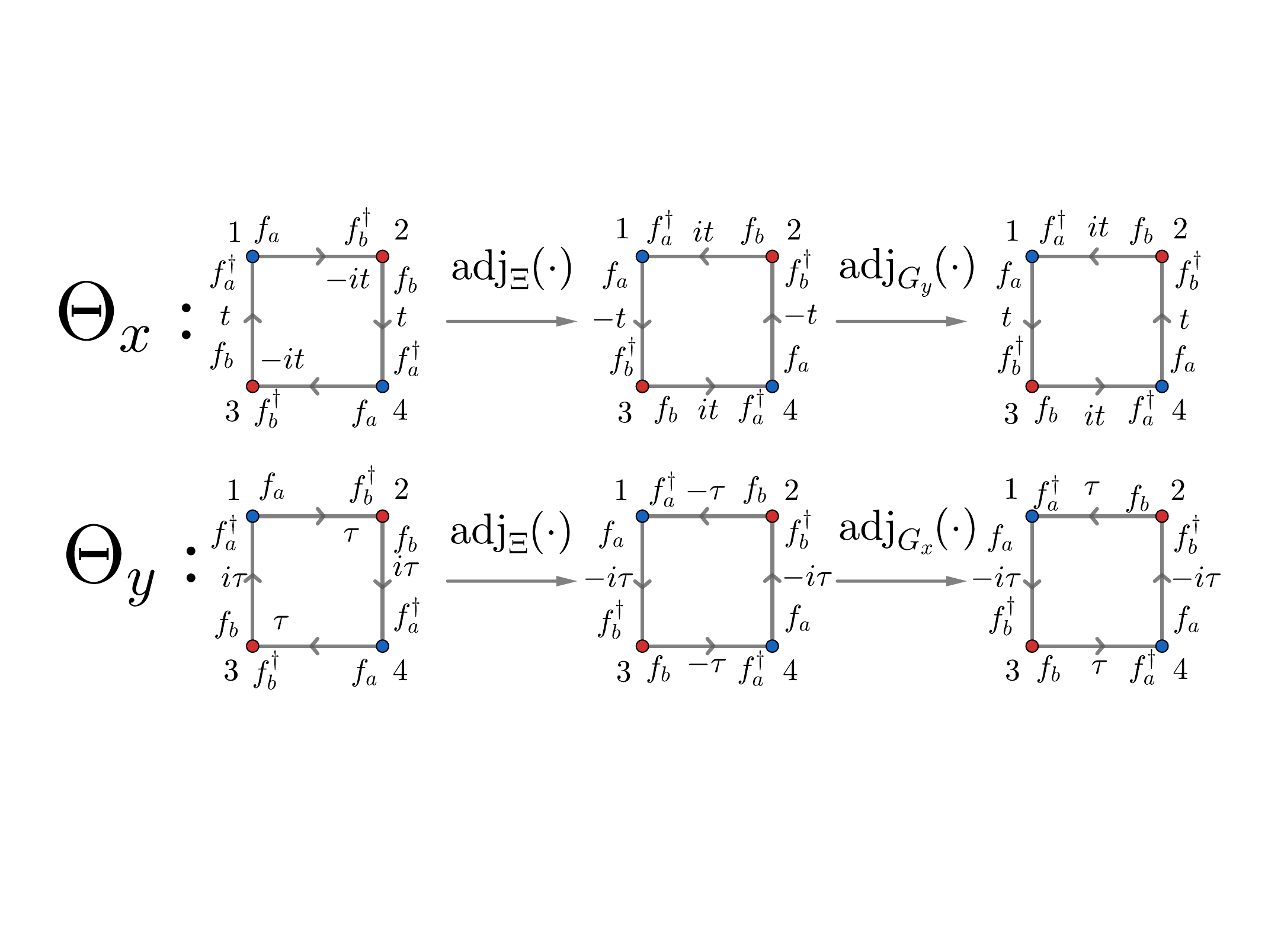}
   \caption{Particle-Hole action on nearest neighbour hoppings for $\Theta_x$ ($t \in \mathbb{R}$) and $\Theta_y$ extended projective symmetry group ($\tau \in \mathbb{R}$}
   \label{ph}
  \end{figure}

  Figure \ref{ph} shows why the above action allows for the nearest neighbor hopping  of JW/composite-fermion for the $\Theta_x$ and $\Theta_y$ extended projective symmetry group.

  \section{Low energy effective theory derivation}\label{appd}

   After linearizing in vector potentials, the Fourier transform of Eq.\eqref{hami} is given by:
 \begin{equation}\label{hamigau1}
      \begin{aligned}
            H(t,A) = &  \sum_{\textbf{q}, \textbf{p} \in \text{BZ}} i t^* \bigg( \big(\delta_{\textbf{p}\textbf0} + i A_3(\textbf{p}) \big) + \big(\delta_{\textbf{p}\textbf0}  - i A_1(\textbf{p}) e^{-i \textbf{p} \cdot \textbf{R}_2} \big) e^{-i \textbf{q} \cdot (\textbf{R}_2 - \textbf{R}_1)} \bigg) f^{\dag}_a(\textbf{q}) f_b(\textbf{q}+ \textbf{p})  \\
            &+ t \bigg( \big(\delta_{\textbf{p}\textbf0} - i A_4(\textbf{p}) \big)e^{i \textbf{q} \cdot \textbf{R}_1} + \big(\delta_{\textbf{p}\textbf0} + i A_2(\textbf{p}) e^{-i \textbf{p} \cdot \textbf{R}_2} \big) e^{-i \textbf{q}_2}\bigg) f^{\dag}_a(\textbf{q}) f_b(\textbf{q}+ \textbf{p}) 
      \end{aligned}
  \end{equation}
  
  Following discussion in Sec. \ref{25}, we define the following set of operators:

    \begin{equation*}
  \begin{aligned}
              \Psi(\textbf{q}) &\doteq \begin{pmatrix}
              f_a(\textbf{p}+ (\pi,0))\\
               f_b(\textbf{p}+ (\pi,0))\\
                f_a(\textbf{p}+ (0,\pi))\\
                 f_b(\textbf{p}+ (0,\pi))
              \end{pmatrix}
               \\
              A^{0}_j(\textbf{p})&\doteq A_j(\textbf{p})\,\,\,\,\,\,\,\,\,\,\,\,\,\,\,\,\,\,\,\,\,\,\,\,\,\,\,j \in \{1,2,3,4\} \\
              A^{\pi}_j(\textbf{p}) &\doteq A_j(\textbf{p} + (\pi,\pi))\,\,\,\,\,j \in \{1,2,3,4\} \\
  \end{aligned}
  \end{equation*}

  where $\textbf{p}$ has to be understood as ``small" momentum with respect to Brillouin zone size, so that we can expand the hamiltonian \eqref{hamigau} to the first order in $\bm{R}_i \cdot \bm{p}$. 
  
  \subsubsection{$\textbf{q}= (0,0)$ scattering processes analysis}
 In the limit of $\textbf{q} \cdot \textbf{R}_i \ll 1$ and to first order in crystal momenta, The hamiltonian describing $\textbf{q}= (0,0)$ scattering processes is:

\begin{equation*}
      \begin{aligned}
            H(t,A^0)|_{(0,0)} 
            & \simeq 
              \sqrt{2 }  |\textbf{R}_1|  \sum_{\textbf{q}, \textbf{p} \in \text{BZ}} 
             t^* \big[   \textbf{q}_x \delta_{\textbf{p}\textbf0}  - A^0_{x}(\textbf{p} )    \big] \bigg( f^{\dag}_a ((\pi,0) + \textbf{q}) f_b((\pi,0)+\textbf{q}+ \textbf{p}) + f^{\dag}_a((0,\pi)+\textbf{q}) f_{b}((0,\pi)+ \textbf{q}+ \textbf{p}) \bigg) \\
            &+ i t \big[  \textbf{q}_y \delta_{\textbf{p}\textbf0}  - A^0_{y}(\textbf{p})   \big] \bigg( f^{\dag}_a((0,\pi) + \textbf{q}) f_{b}((0,\pi)+\textbf{q}+ \textbf{p}) - f^{\dag}_{a}((\pi,0)+\textbf{q}) f_b((\pi,0)+\textbf{q}+ \textbf{p})  \bigg) +h.c.
            \end{aligned}
            \end{equation*}

           where:

  \begin{equation*}
  \begin{aligned}
      A^{0}_x(\textbf{p}) &\doteq \frac{A^{0}_{1}(\textbf{p}) + A^{0}_{3}(\textbf{p})}{\sqrt{2} |\textbf{R}_1| }\\
       A^{0}_y(\textbf{p}) &\doteq \frac{A^{0}_{2}(\textbf{p}) + A^{0}_{4}(\textbf{p})}{\sqrt{2} |\textbf{R}_1|}
      \end{aligned}
  \end{equation*}

  
  
  
  
  Thus:
  
            \begin{equation*}
               \begin{aligned}
          H(t,A^0)|_{(0,0)}   &=  \sum_{\textbf{q}, \textbf{p}   \in \text{BZ}}  v \begin{cases}
               
               \Psi^{\dag}(\bm{q})\bigg[ \big(  \textbf{q}_x \delta_{\textbf{p}\textbf0}  - A^0_{x}(\textbf{p})   \big) \mathbb{1}  \otimes  \tau^x +      \big(  \textbf{q}_y \delta_{\textbf{p}\textbf0} -  A^0_{y}(\textbf{p})     \big)  \rho^z \otimes  \tau^y\bigg]   \Psi(\bm{q}+\bm{p}) \,\,\,\,\,    t \in \mathbb{R}\\
                       \Psi^{\dag}(\bm{q})\bigg[ \big(  \textbf{q}_x \delta_{\textbf{p}\textbf0}  - A^0_{x}(\textbf{p})   \big)  \mathbb{1}  \otimes  \tau^y +      \big(  \textbf{q}_y \delta_{\textbf{p}\textbf0}  -  A^0_{y}(\textbf{p})     \big) \rho^z \otimes  \tau^x \bigg]   \Psi(\bm{q}+\bm{p}) \,\,\,\,\,      t \in i \mathbb{R}\\
            \end{cases}
      \end{aligned}
  \end{equation*}

  where $v \doteq \sqrt{2}|t||{\bf R}_1|$ and the sets of $\tau^{i}$ and $\rho^{j}$ Pauli matrices are understood to act respectively on $\{a,b\}$\, sub-lattice spaces and on $\{(\pi,0), (0, \pi)\}$ valley space. \\

  \subsubsection{$\textbf{q}=(\pi,\pi)$ momentum scattering processes}
  
  To first order in crystal momenta, The hamiltonian describing $\textbf{q}= (\pi,\pi)$ scattering processes is:

\begin{equation*}
      \begin{aligned}
            H(t,A^{\pi})|_{(\pi,\pi)} 
            &\simeq \sqrt{2}|\textbf{R}_1| \sum_{\textbf{q}, \textbf{p} \in \text{BZ}} \big(-t^* B^{\pi}_{x}(\textbf{p})  +it  B^{\pi}_y(\textbf{p} ) \big) f^{\dag}_a((\pi,0)+\textbf{q}) f_{b}((0,\pi)+\textbf{q}+ \textbf{p})\\
            &- \big(t^* B^{\pi}_x(\textbf{p}) 
            +it B^{\pi}_y(\textbf{p}) \big) f^{\dag}_a((0,\pi)+\textbf{q}) f_{b}((\pi,0)+\textbf{q}+ \textbf{p}) + h.c.
      \end{aligned}
  \end{equation*}

  where:

  \begin{equation}
    \begin{aligned}
          B^{\pi}_x(\textbf{x}) &= \frac{A^{\pi}_{3}(\textbf{x}) - A^{\pi}_{1}(\textbf{x})}{\sqrt{2}|\textbf{R}_1| }, \\  
          B^{\pi}_y(\textbf{x}) &= \frac{A^{\pi}_{4}(\textbf{x}) - A^{\pi}_{2}(\textbf{x})}{\sqrt{2}|\textbf{R}_1| }. \\
    \end{aligned}
\end{equation}

  
  In the limit of $\textbf{q} \cdot \textbf{R}_i \ll 1$, and following a similar procedure as before we can write the hamiltonian above in a matrix form:
  
  \begin{equation*}
             H(t,A^{\pi})|_{(\pi,\pi)} =-v \sum_{\textbf{q},\textbf{p} \in \text{BZ}}  \begin{cases} 
             \Psi^{\dag}(\bm{q}) \rho^x \otimes \bigg[  B^{\pi}_x(\textbf{p})   \tau^x + B^{\pi}_y(\textbf{p})   \tau^y \bigg] \Psi(\bm{q}+\bm{p}) \,\,\,\,t \in \mathbb{R} \\
                 \Psi^{\dag}(\bm{q}) \rho^y \otimes \bigg[ B^{\pi}_x(\textbf{p})    \tau^x - B^{\pi}_y(\textbf{p})    \tau^y \bigg] \Psi(\bm{q}+\bm{p})\,\,\,\,t \in i\mathbb{R} \\
             \end{cases}
  \end{equation*}

\end{document}